\documentclass[10 pt,a4paper]{article}

\usepackage{subfig}
\usepackage{color}
\usepackage{amsmath}
\usepackage{amsfonts}
\usepackage{verbatim}
\usepackage{amssymb}
\usepackage{graphicx}
\usepackage{mathrsfs}
\usepackage{overpic}
\usepackage{pdfpages}

\usepackage{hyperref}

\hypersetup{%
	colorlinks=true,
	allcolors=blue,
	pdftitle={Extended phase space thermodynamics for hairy black holes},
	pdfauthor   = {Dumitru Astefanesei, Paulina Cabrera, Robert B. Mann and Ra\'ul Rojas},
	pdfkeywords = {black hole, string theory, supergravity, self-interaction, scalar field, dilaton, thermodynamics, extended phase space, black hole chemistry, Hawking-Page, Smarr, reentrant phase transitions, phase transition, pressure, cosmological constant}
}

\providecommand{\U}[1]{\protect\rule{.1in}{.1in}}
\def\pa{\partial}

\renewcommand{\(}{\left(}
\renewcommand{\)}{\right)}
\renewcommand{\[}{\left[}
\renewcommand{\]}{\right]}
\begin{document}

%%%%%%%%%%%%%%%%%%%%%%%%%%%%%%%%%%%%%%%%%%%%%%%%%%%%%%%%%%%%%%%%%%%%%%%%%%%%%%%%%%%%%%%%%%%%%%%%

\title{\huge{\bf Extended phase space thermodynamics \\ for hairy black holes}}

\author{Dumitru Astefanesei$^{1}$\footnote{Email: {\tt dumitru.astefanesei@pucv.cl}}~,~ Paulina Cabrera$^{1,2}$\footnote{Email: {\tt paulina.cabramirez@gmail.com}}~,~ Robert B. Mann$^{3,4}$\footnote{Email: {\tt bmann@uwaterloo.ca}} ~ and Ra\'ul Rojas$^{5}$\footnote{Email: {\tt raulox.2012@gmail.com}} \\
\\\textit{$^{1}$Pontificia Universidad Cat\'olica de
Valpara\'\i so, Instituto de F\'\i sica,} \\\textit{Av. Brasil 2950, Valpara\'{\i}so, Chile}\\
\\\textit{$^{2}$Universidad T\'ecnica Federico Santa Mar\'{\i}a, Departamento de F\'{\i}sica,} \\\textit{Av. Espa\~na 1680, Valpara\'{\i}so, Chile}\\
\\\textit{$^{3}$Department of Physics and Astronomy, University of Waterloo,} \\\textit{Waterloo, Ontario, N2L 3G1, Canada}\\
\\\textit{$^{4}$Perimeter Institute, 31 Caroline Street North, } \\\textit{Waterloo, ON, N2L 2Y5, Canada}\\
\\\textit{$^{5}$Departamento de F\'{\i}sica, Universidad de Concepci\'on,} \\\textit{Casilla 160-C, Concepci\'on, Chile}}

\maketitle

\begin{abstract} 
We expand our results in \cite{Astefanesei:2019ehu} to investigate a general class of exact hairy black hole solutions  in Einstein-Maxwell-dilaton gravity. The dilaton is endowed with a potential that originates from an 
electromagnetic Fayet-Iliopoulos  term in $\mathcal{N} = 2$ extended supergravity in four spacetime dimensions. We present the usual thermodynamics by using the counterterm method supplemented with boundary terms for a scalar field with mixed boundary conditions. We then extend our analysis by considering a dynamical cosmological constant and verify the isoperimetric inequality. We obtain a very rich phase diagram and criticality in both the canonical and grand canonical ensembles. Within string theory, the cosmological constant is related to the radius of the external sphere (of the compactification) and can be interpreted as a modulus. In this context, the existence of a critical value hints to the fact that the thermodynamic properties of black holes in lower dimensions depend on the size of the compactification.
\end{abstract}

\newpage
\tableofcontents

\section{Introduction} 
\label{intro}

In the context of string theory, scalar fields arise naturally as moduli  when considering specific compactifications. These moduli will appear as fields in the lower dimensional
effective field theory (see, e.g., \cite{Green:1987sp} and references therein). One such modulus is the cosmological constant that appears when considering gauged supergravity. Once embedded in string/M theory \cite{Cvetic:1999xp}, the radius of anti-de Sitter ($AdS$) spacetime is related to the radius of the external sphere. Therefore, the cosmological constant becomes dynamical if the radius of the sphere evolves in time \cite{Cvetic:2010jb}. 

In string theory, a sufficiently heavy compactified wrapped object will effectively
give rise to a lower dimensional black hole; examples  in $AdS$ gravity include the well known $R$-charged black hole solutions \cite{Cvetic:1999xp,Duff:1999gh}. Here we consider a different general class of exact hairy black hole solutions in $AdS_4$ \cite{Anabalon:2013sra}\footnote{Similar or more general exact hairy black hole solutions in $AdS$ were obtained in \cite{Lu:2013ura,Lu:2014maa,Feng:2013tza,Anabalon:2012ta,Acena:2013jya,Anabalon:2013eaa,Canfora:2021nca,Acena:2012mr}. It is also worth mentioning that similar solutions \cite{Anabalon:2013qua}, which are thermodynamically and dynamically stable \cite{Astefanesei:2019mds,Astefanesei:2020xvn,Astefanesei:2019qsg}, exist in flat spacetime.} that can be embedded in a supergravity model with dyonic
Fayet-Iliopoulos terms \cite{Anabalon:2017yhv,Anabalon:2020pez} (see also \cite{Andrianopoli:2013jra,Gallerati:2019mzs,Anabalon:2021smx,Anabalon:2020qux,Gallerati:2021cty}). Interestingly, one can study the thermodynamics of black holes in $AdS$ when the corresponding cosmological constant can vary \cite{Henneaux:1985tv,Creighton:1995au} and is taken to be a thermodynamic variable similar to `pressure' \cite{Kastor:2010gq,Kastor:2011qp,Kubiznak:2014zwa,Kubiznak:2016qmn}. In one of the pioneering works on this subject \cite{Kubiznak:2012wp}, it was understood that there is a deep analogy between charged $AdS$ black holes and Van der Waals fluids. What is important from a physical point of view is that, since there is a critical value for the `pressure', the thermodynamic properties of the black holes in string theory depend on the size of the compactification (external sphere) \cite{Cvetic:2010jb,Astefanesei:2019ehu}. It is also important to emphasize that the cosmological constant represents a relevant thermodynamic quantity in   black hole thermodynamics, as follows from the fact that it is required for the consistency of the Smarr formula, even as a fixed quantity \cite{Dolan:2012jh,Kastor:2009wy}.

Previously  we have provided a detailed analysis of the criticality phenomena for a particular exact hairy black hole solution and compared its properties with the Reissner-Nordstr\"om asymptotically anti-de Sitter (RN-$AdS$) black hole \cite{Astefanesei:2019ehu}. Interestingly, the scalar field drastically changes its properties. For example, in the grand canonical ensemble, the scalar field allows critical phenomena and, unlike the RN-$AdS$ black hole,  there is  double criticality in the canonical ensemble. 

In this paper, we carry out  a complete analysis of the criticality for a general family of exact hairy black hole solutions \cite{Anabalon:2013sra}. In this case, the potential of the scalar field  contains two extra parameters besides the cosmological constant that makes the thermodynamics in the extended phase space even richer.  In the canonical ensemble we previously observed the  particular interesting aspect  of reentrant phase transitions \cite{Astefanesei:2021vcp}. However, in this paper we shall not only provide the technical details and complete the thermodynamic analysis in the canonical ensemble, but  shall also investigate the grand canonical ensemble, which also has interesting properties that are distinct from the particular case presented in \cite{Astefanesei:2019ehu}. We use the counterterm method developed in \cite{Anabalon:2015xvl} (that is consistent with the Hamiltonian formalism \cite{Hertog:2004ns, Anabalon:2014fla}) for the scalar field to regularize the Euclidean action and quasilocal stress tensor of Brown and York \cite{Brown:1992br}. Armed with these results, we present the usual thermodynamics of hairy black holes before considering the extended phase space thermodynamics. We obtain the Smarr formula and provide a concrete check of the isoperimetric relation and provide its physical interpretation in this context. Unlike the RN-$AdS$ black hole, which has only a single critical point in
the canonical ensemble, and no interesting phase behaviour in the grand canonical ensemble, for the class of hairy black holes we consider, there exist two critical points in each ensemble, along with reentrant phase transitions in some range of the electric charge and its conjugate potential. For the grand canonical ensemble, one critical point corresponds to the termination of a sequence of standard first order phase transitions in which large black holes `condense' to small ones. The other corresponds to the beginning of a sequence of first order phase transitions exhibiting novel behaviour, in which the specific volume increases in a large-to-small phase transition instead of decreases. We shall consider these  new interesting thermodynamic properties in great detail.

The paper is organized as follows: as a set-up, in Section \ref{sec2} we briefly review the main results we have obtained in \cite{Astefanesei:2019ehu} for a particular charged hairy $AdS$ black hole solution. In Section \ref{sec3}, we present a detailed analysis of the usual and, also, extended phase space thermodynamics for the general charged hairy $AdS$ black hole solution in both, canonical and grand canonical, ensembles. Particularly, we shall consider in detail the novel first order transitions that appear above the second critical point. In the last section, we conclude with a brief review of our results.

%%%%%%%%%%%%%%%%%%%%%%%%%%%%%%%%%%%%%%%%%%%%%%%%%%%%%%%%%%%
\section{Hairy black hole chemistry framework}
\label{sec2}%%%%%%%%%%%%%%%%%%%%%%%%%%%%%%%%%%%%%%%%

In this section, we review the thermodynamics of the exact asymptotically $AdS$ charged hairy black hole solution found in \cite{Anabalon:2013sra}, corresponding to the limit $\gamma\rightarrow 1$ (when the `hair parameter\footnote{In the original papers, the hair parameter is denoted by $\nu$. In order to avoid confusion with the notation for the specific volume $v$, we shall use $\sigma$ instead of $\nu$.}' $\sigma\rightarrow \infty$) for  constant coupling in the exponential between the scalar field $\phi$ and the Maxwell invariant $F^2\equiv F_{\mu\nu}F^{\mu\nu}$, as shown in the gravitational action below. This is done in the extended phase space where the (negative) cosmological constant $\Lambda$ is a pressure term, allowing us to explore the thermodynamic behaviour for the whole set of $AdS$ theories. We use this example as a set-up for the complete analysis of the entire family that we shall consider in the next section.

Let us consider the Einstein-Maxwell-scalar theory described by the action
\begin{equation}
I=\frac{1}{2\kappa}
\int_{\mathcal{M}}
{d^4x\sqrt{-g} \[R-\frac{1}{2}(\pa\phi)^2 -U(\phi)-e^{\phi}F^2\]}
\label{g1}
\end{equation}
where $F_{\mu\nu}=\pa_{\mu}A_{\nu}-\pa_{\nu}A_{\mu}$ is the gauge field and $A_\mu$ the gauge potential, $\phi$ is the scalar field (dilaton) and $(\pa\phi)^2\equiv g^{\mu\nu}\pa_\mu\phi\pa_\nu\phi$. We adopt the unit system where the numerical values of the fundamental constants are set to unity: $G=1$, $c=1$ (so that $\kappa=8\pi$), $\hbar=1$ and, for the electromagnetic sector, we fix $\mu_0=4\pi$. 

This theory is known to support a spherically symmetric exact solution for the following scalar field potential
\begin{equation}
	\label{potinfty}
	U(\phi)=2\alpha\(2\phi+\phi\cosh\phi-3\sinh\phi\)+\frac{2\Lambda}{3}\(\cosh\phi+2\)
\end{equation}
where $\alpha$ is an arbitrary dimensionful parameter that has its origin in extended SUGRA \cite{Anabalon:2017yhv, Anabalon:2020pez}, and $\Lambda\equiv-3/\ell^2$ is the cosmological constant, with $\ell$ being the $AdS$ radius. The self-interacting potential (\ref{potinfty}), for small $\phi$, decays as $U(\phi)=-{6}/{\ell^2}-{\phi^2}/{\ell^2}+\mathcal{O}(\phi^{4})$, as expected for the $AdS$ asymptotics, and the solution to the corresponding equations of motion is
\begin{align}
	\label{metric}
	ds^2=\Omega(x)
	\[-f(x)dt^2+\frac{\eta^2dx^2}{x^2f(x)}+
	d\theta^2+\sin^2\theta d\varphi^2\]\,,\quad A_\mu=\left(-\frac{q}{x}+c\right)\delta^t_\mu\,,\quad
	\phi=\ln(x)
\end{align}
where
\begin{align}
f(x)=\alpha\(\frac{x^2-1}{2x}-\ln{x}\)
+\frac{1}{\Omega(x)}\(1-2q^2\frac{x-1}{x}\)
-\frac{\Lambda}{3}\,, \qquad \Omega(x)=\frac{x}{\eta^2(x-1)^2}
\label{conf}
\end{align}
are the metric functions, $\eta$ and $q$ are the two constants of integration and $c$ in the gauge field is an additive constant that will be used to fixed the gauge $A_t(x_+)=0$, where $x_+$ is the location of the black hole horizon, $f(x_+)=0$. The radial $x$-coordinate has the range  $1<x\leq\infty$, where $x=\infty$ is the location of the central singularity and $x\rightarrow 1$ is the boundary\footnote{This is known as the `positive branch' of solutions. There is also a negative branch, which is not studied in this paper.}. The relation to the canonical (Schwarzschild-like) coordinate, at least near the boundary, is given by $r=\sqrt{\Omega(x)}$.

For this solution, the conserved energy $E$, the Hawking temperature $T$, the Hawking-Bekenstein entropy $S$, the electric charge $Q$ and its conjugate potential $\Phi$, the pressure $P$ and thermodynamic volume $V$ are \cite{Anabalon:2013sra}
\begin{equation}
\label{mass1}	E=\frac{q^2}{\eta}-\frac{\alpha}{12\eta^3}\,, \qquad
T=-\frac{x_+f'(x_+)}{4\pi\eta}
=\frac{(x_{+}-1)^2}{8\pi\eta x_+}\[
-\alpha-2\eta^2\(\frac{x_{+}+1}{x_{+}-1}\)
+4\eta^2q^2\(\frac{x_{+}+2}{x_+}\)\]
\end{equation}
\begin{equation}
	\label{quantities2}
S=\frac{\pi x_+}{\eta^2(x_+-1)^2}\,, \qquad \Phi=\frac{q(x_+-1)}{x_+}\,, \qquad Q=\frac{q}{\eta}\,, \qquad P=-\frac{\Lambda}{8\pi}\,, \qquad V=\frac{2\pi x_+(x_++1)}{3\eta^3(x_+-1)^3}
\end{equation}
and they satisfy the extended first law $dE=TdS+\Phi dQ+VdP$.

In order to work with dimensionless quantities, in the remainder of this section we will consider the rescaled thermodynamic variables,
\begin{equation}
	\label{rescaled}
	\eta\rightarrow\sqrt{\alpha} \eta\,, \quad E\rightarrow\frac{E}{\sqrt{\alpha}}\,, \quad T\rightarrow\sqrt{\alpha}T\,, \quad
	S\rightarrow \frac{S}{\alpha}\,, \quad Q\rightarrow\frac{Q}{\sqrt{\alpha}}\,, \quad P\rightarrow \alpha P\,, \quad V\rightarrow\alpha^{-\frac{3}{2}}V
\end{equation}
This makes explicit the assumption that $\alpha>0$.

%%%%%%%%%%%%%%%%%%%%%%%%%%%%%%%%%%%%%%%%%%%%%%%
\subsection{The canonical ensemble: Fixed electric charge} %%%%%%%%%%%%%%%%%%%%%

The ensemble with $T$ and $Q$ kept fixed is achieved by imposing the boundary condition $\left.\delta(e^{\phi}\star F)\right|_{\pa\mathcal{M}}=0$. The on-shell Euclidean action is now $\tilde I^E=I^E+I^E_A$, where $I_A=(2/\kappa)\int{d^3x\sqrt{-h}e^\phi n_\mu F^{\mu\nu}A_\nu}$ is the boundary term for the gauge field, and satisfies the quantum-statistical relation $\beta^{-1}\tilde I^E=\mathcal{F}(T,Q)=E-TS$  \cite{Astefanesei:2019ehu}.

We begin with the equation of state, given parametrically by the expressions
\begin{equation}
P=\frac{3(x_++1)^2} {8\pi x_+} \[\frac{2(x_++1)^2}{x_+(x_+-1)}\frac{Q^2}{v^4} -\frac{1}{v^2} -\frac{x_+^2-1-2x_+\ln{x_+}}{2(x_++1)^2}\]
\end{equation}
\begin{equation}
\label{state1}
T=\frac{(x_++1)^2} {4\pi x_+}\[\frac{2(x_++1)(x_++2)}{(x_+-1)x_+}\frac{Q^2}{v^3} -\frac{1}{v}-\frac{v}{2}\frac{(x_+-1)^3}{(x_++1)^3} \]
\end{equation}
where  $v\equiv 3V/2S$ is the specific volume  that measures the thermodynamic volume per degree of freedom \cite{Gunasekaran:2012dq}. 
It can be straightforwardly shown that, in the limit $x_+\rightarrow 1$, these expressions reduce to the RN-$AdS$ equation of state, $P=T/v-1/(2\pi v^2)+2Q^2/(\pi v^4)+O(v^{-5})$, as expected. In the opposite limit, $x_+\rightarrow\infty$, we obtain
\begin{equation}
\label{specificvol}
v=\frac{1}{\eta}+2\eta r_+^2+\mathcal{O}(r_+^4)\, \qquad \eta(r_+\rightarrow 0)=\frac{\(1+\sqrt{1+4Q^2}\)^{\frac{1}{2}}}{2Q}
\end{equation}
where $r_+\equiv\sqrt{\Omega(x_+)}$. This indicates that in the limit $x_+\rightarrow\infty$ ($S\rightarrow 0$), the specific volume tends to a constant value, i.e., $V\propto S$.

Critical points in this ensemble satisfy the conditions
\begin{equation}
\(\frac{\pa P}{\pa v}\)_{T_c,Q}=0\,, \qquad \(\frac{\pa^2 P}{\pa v^2}\)_{T_c,Q}=0
\end{equation}
for different critical temperatures. We find two solutions to these equations, corresponding to
two critical points,  for all  fixed $|Q|>0$. 

\begin{figure}[t!]
\centering
\includegraphics[scale=0.21]{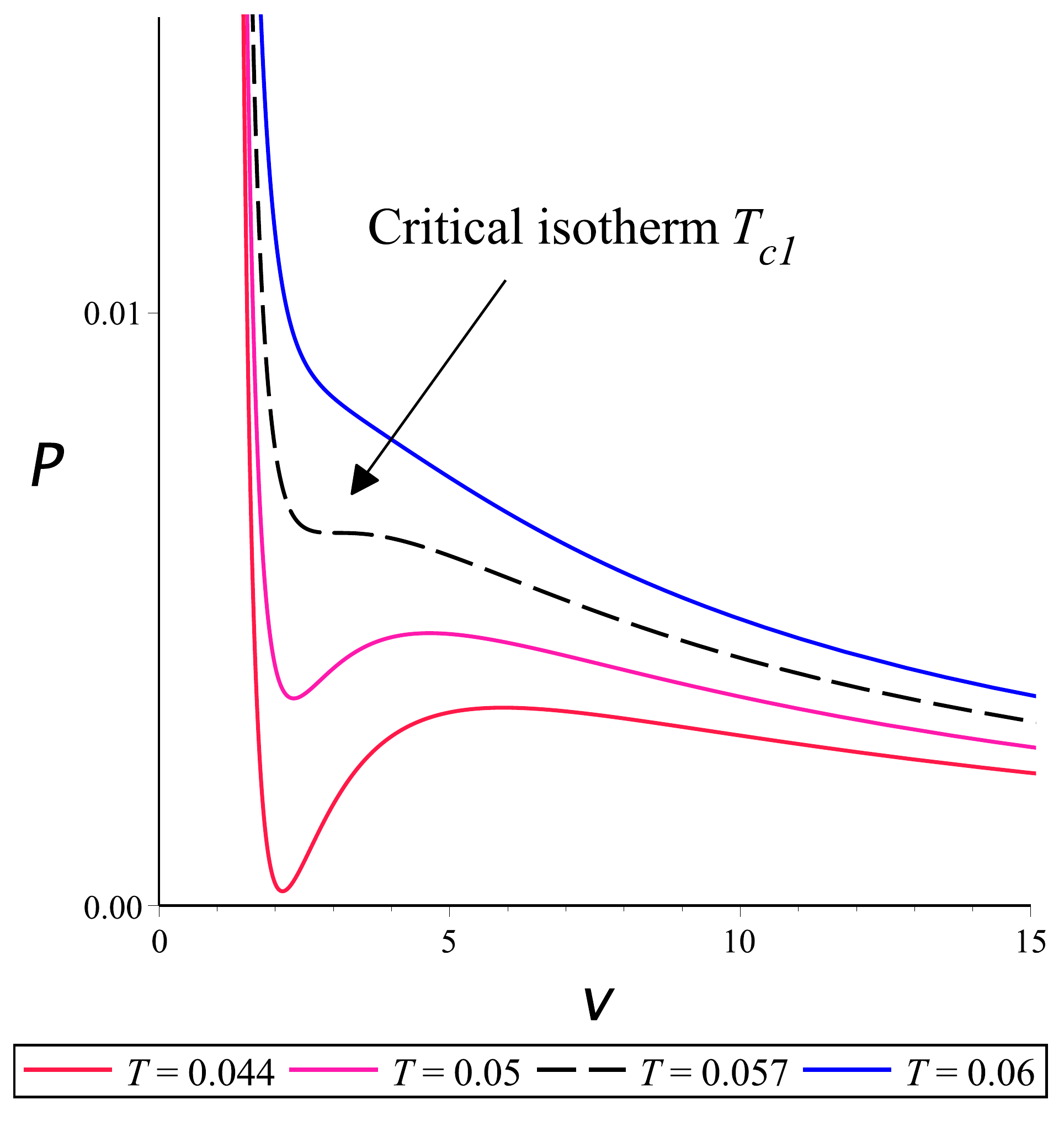}
\includegraphics[scale=0.21]{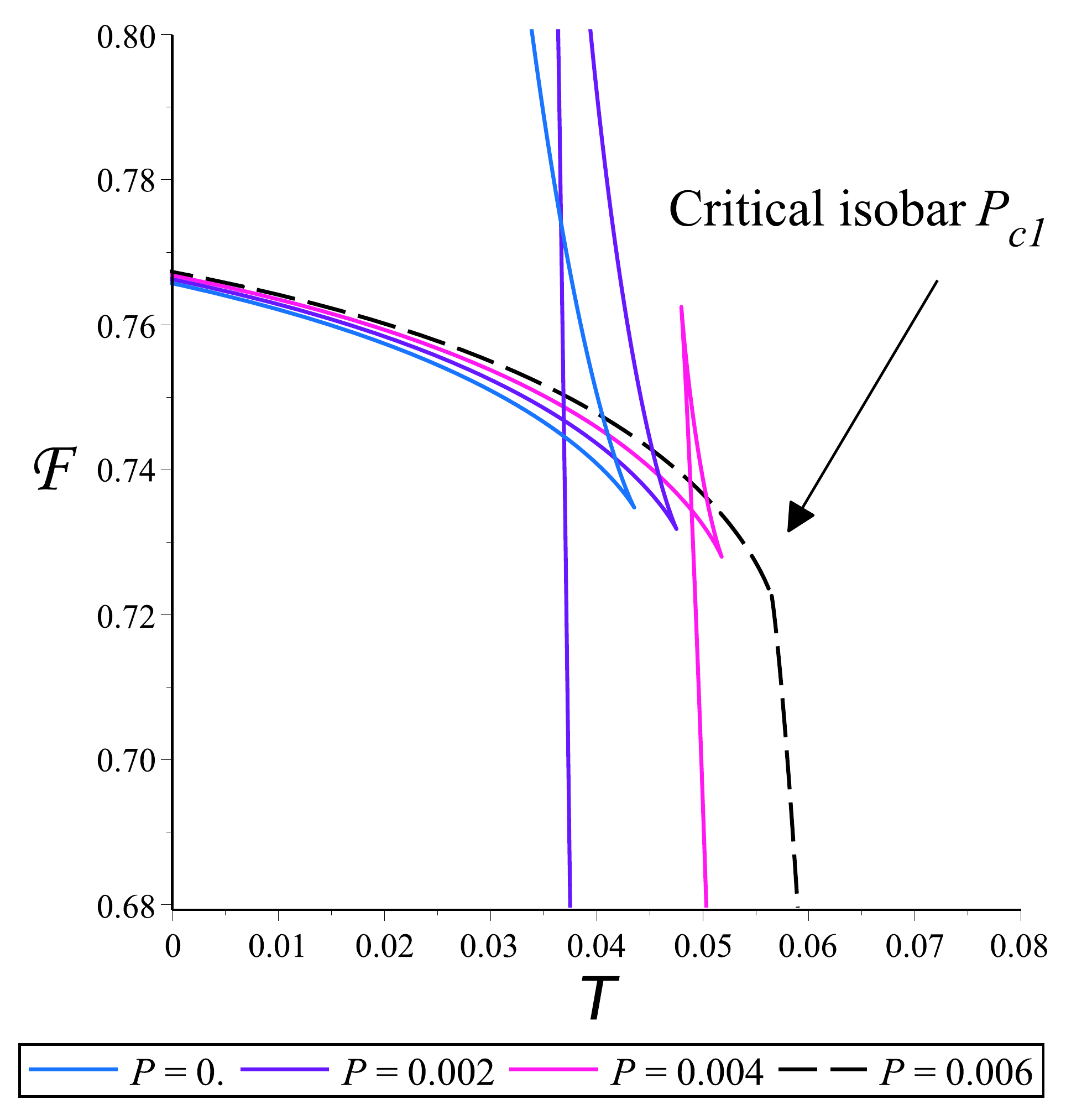}
\includegraphics[scale=0.21]{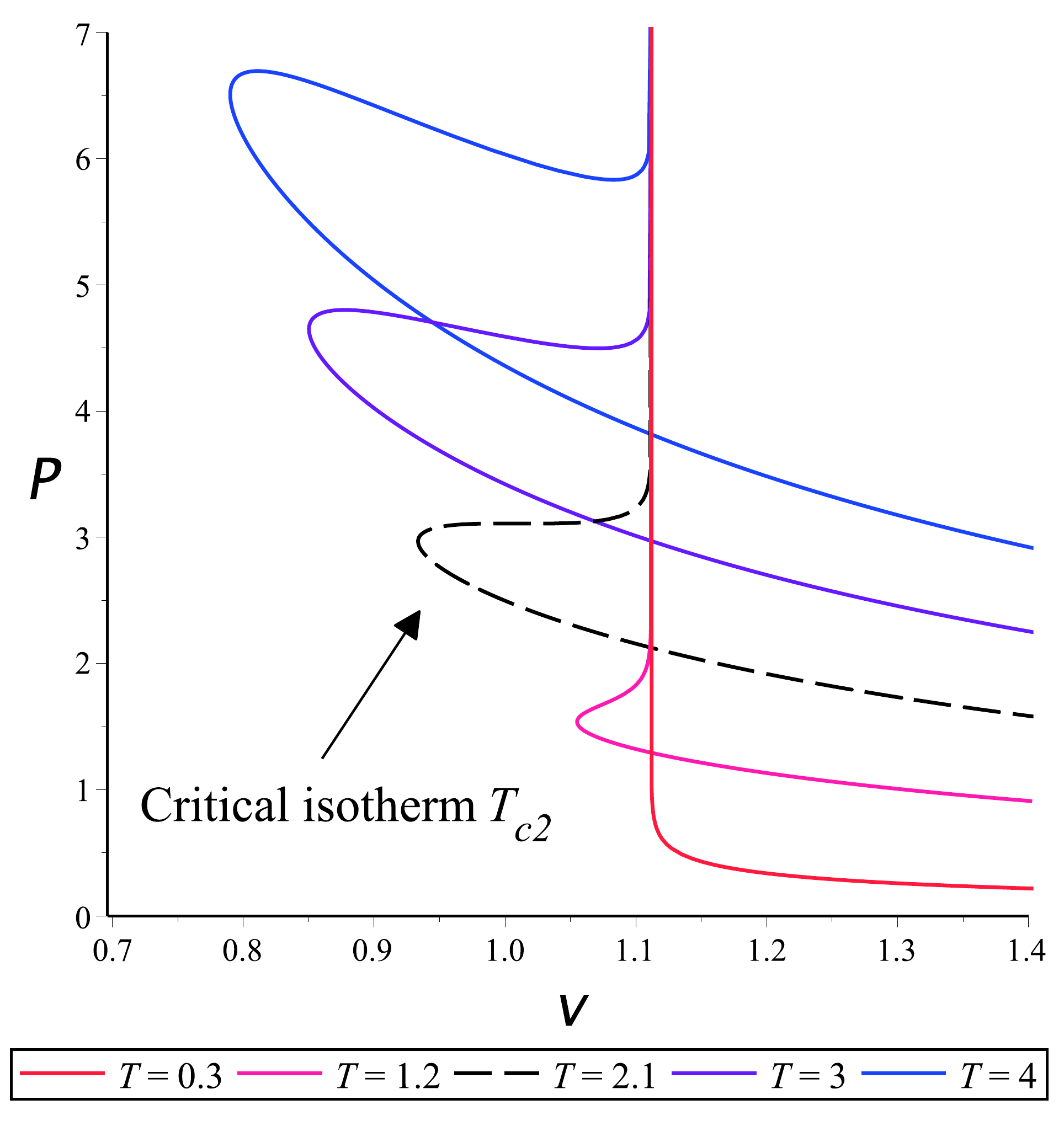}
\includegraphics[scale=0.21]{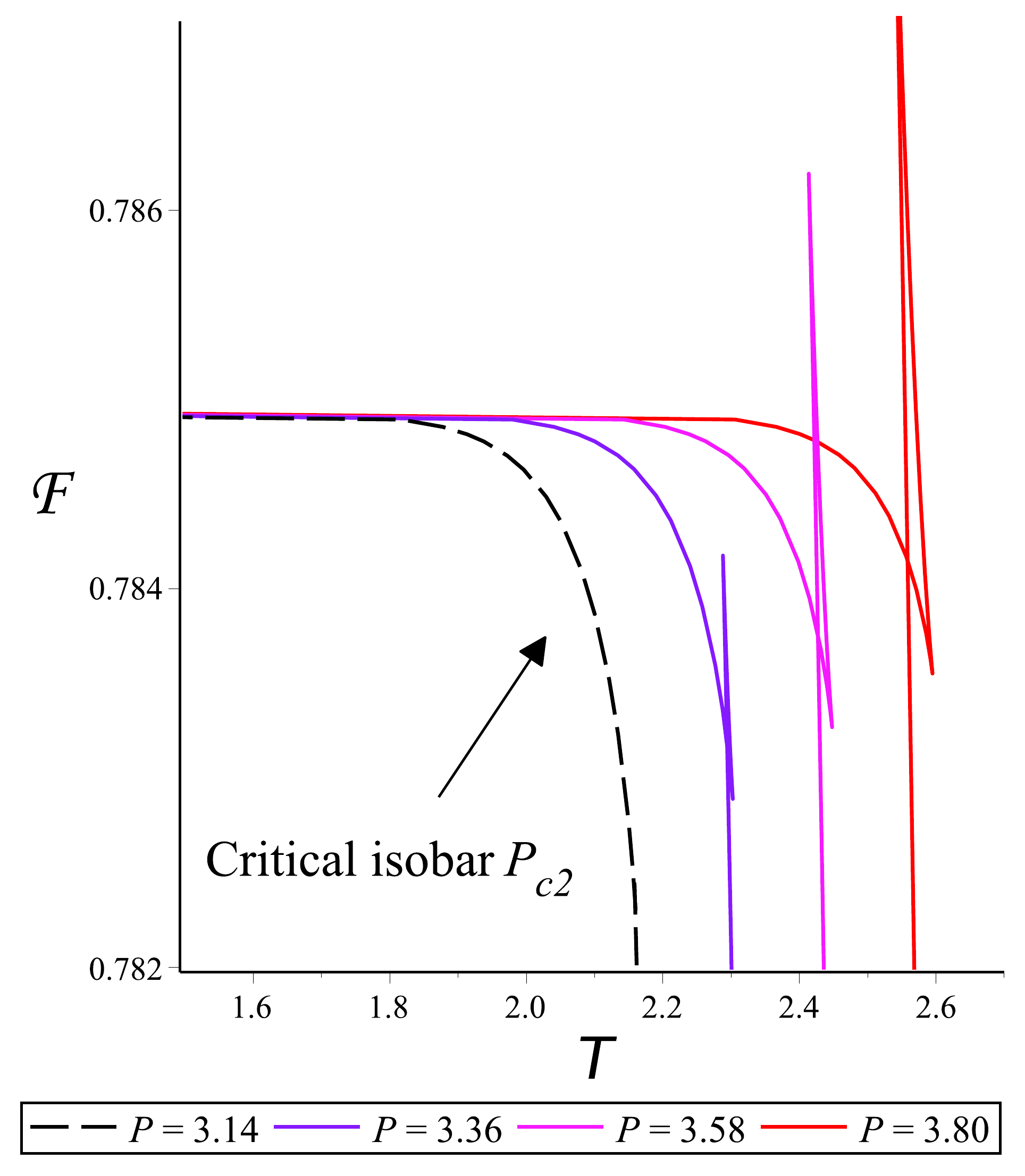}
\caption{Equation of state and $\mathcal{F}-T$ diagram in the canonical ensemble for $Q=1$. The left-hand (right-hand) panels show the behaviour around the first (second) critical point. Critical isobars/isotherms are given by dashed lines.
}
\label{F1}
\end{figure}

In Fig.~\ref{F1}, we depict the equation of state $P-v$ (fixed $T$) and free energy $\mathcal{F}-T$ (fixed $P$) diagrams for $Q=1$. The first critical point ($c1$) is reminiscent of RN-$AdS$ criticality, whereas the second critical point ($c2$) yields new features due to the scalar hair. This one  is quite novel. Below the critical temperature $T_{c2}$, $v(P)$ is a single-valued function, whereas above this temperature it is multivalued; the critical point corresponds to the point at which the  local maximum and minimum of $P(v)$ become coincident, shown in the dashed line in the panel second from right in Fig.~\ref{F1}. 
According to the $\mathcal{F}-T$ diagrams, both critical points are associated with first-order 
phase transitions between thermally stable phases, as follows from the fact that, for the coexisting phases, $C_Q\equiv T(\pa S/\pa T)_Q=-T(\pa^2\mathcal{F}/\pa{T}^2)_Q>0$. 

%%%%%%%%%%%%%%%%%%%%%%%%%%%%%%%%%
%%%%%%%%%%%%%%%%%%%%%%%%%%%%%%%%%%%%%%%%%%%%%%%%%%%%%%

%%%%%%%%%%%%%%%%%%%%%%%%%%%%%%%%%%%%%%%%%%%%%%%%%%%%%%%%%%%%%%%%%%

\subsection{The grand canonical ensemble: Fixed conjugate potential} %%%%%%%%%%%%%%%%%%%%%%%%%%%%%%%%%%%%%%

The thermodynamic ensemble with $T$ and $\Phi$   kept fixed is compatible with the boundary condition $\left.\delta A_\mu\right|_{\pa\mathcal{M}}=0$. The on-shell action computed in the Euclidean section, $I^E$, satisfies the quantum-statistical relation $\beta^{-1}I^E=\mathcal{G}(T,\Phi)=E-TS-\Phi Q$ \cite{Astefanesei:2019ehu}, where $\beta=T^{-1}$ is the periodicity in the Euclidean time and $\mathcal{G}$ is the grand canonical thermodynamic potential.

Let us first consider the equation of state $P-T-v$. Parametrically, we have
\begin{equation}
P(v,x_+)=\frac{3(x_{+}+1)^2\Phi^2}{4\pi(x_{+}-1)v^2} +\frac{3(x_{+}+1)^2}{8\pi x_+} \[\frac{2x_+\ln{x_+}-x_+^2+1}{2(x_{+}+1)^2} -\frac{1}{v^2}\]
\end{equation}
\begin{equation}
T(v,x_+)=\frac{(x_{+}+1)(x_{+}+2)\Phi^2}{2\pi(x_{+}-1)v} -\frac{(x_{+}-1)^3}{4\pi x_+(x_{+}+1)v} \[\(\frac{x_{+}+1}{x_{+}-1}\)^3+\frac{1}{2}v^2\]
\end{equation}
where $1<x_+\leq\infty$.
It is straightforward to show that, in the large black hole limit $x_+\rightarrow 1$, the equation of state reduces to the RN-$AdS$ equation of state, namely, $P={T}/{v}+(\Phi^2-1)/(2\pi v^2)+\mathcal{O}(1/v^3)$, as expected.

\begin{figure}[t!]
\centering
\includegraphics[scale=0.21]{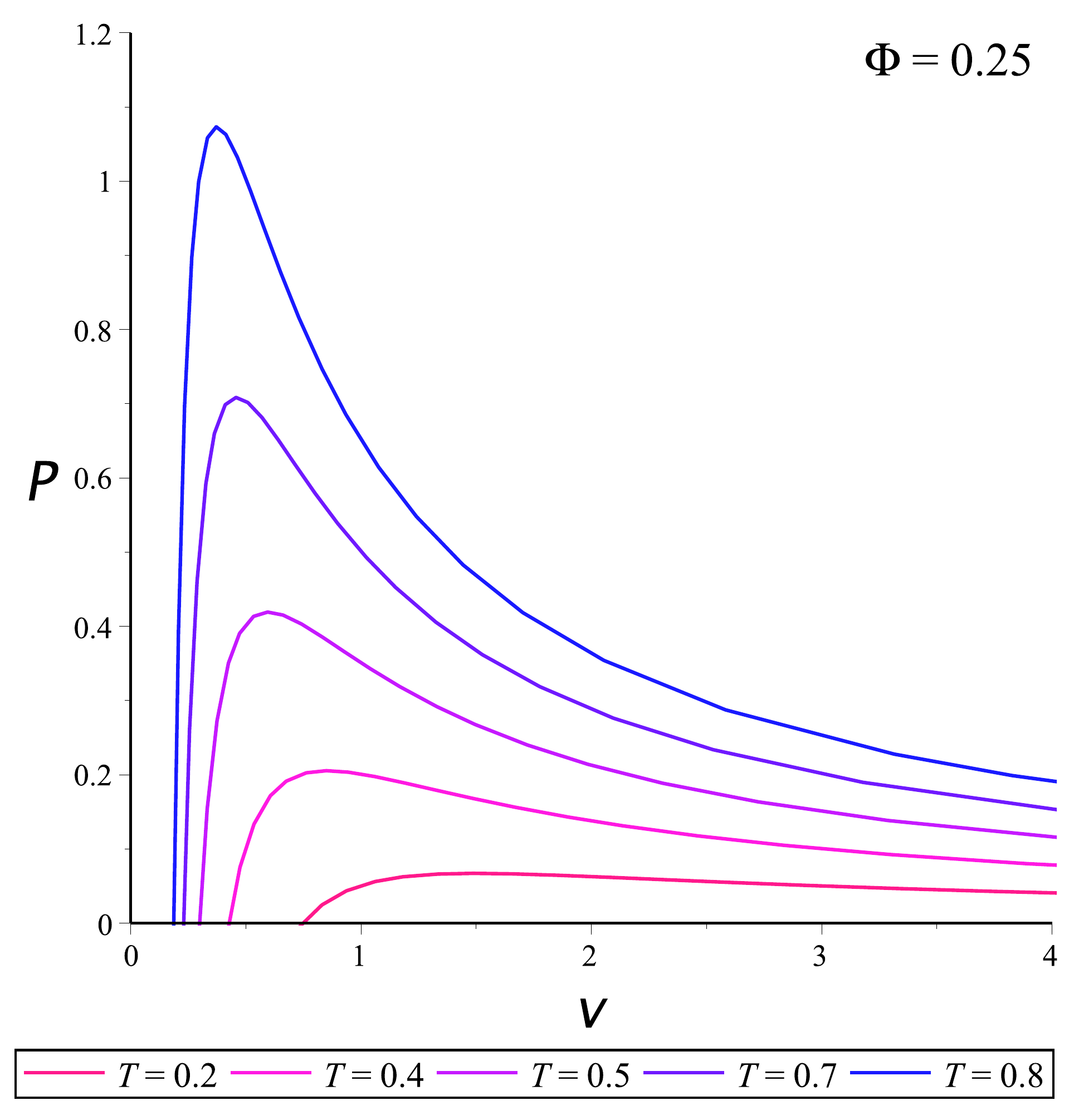}\; 
\includegraphics[scale=0.21]{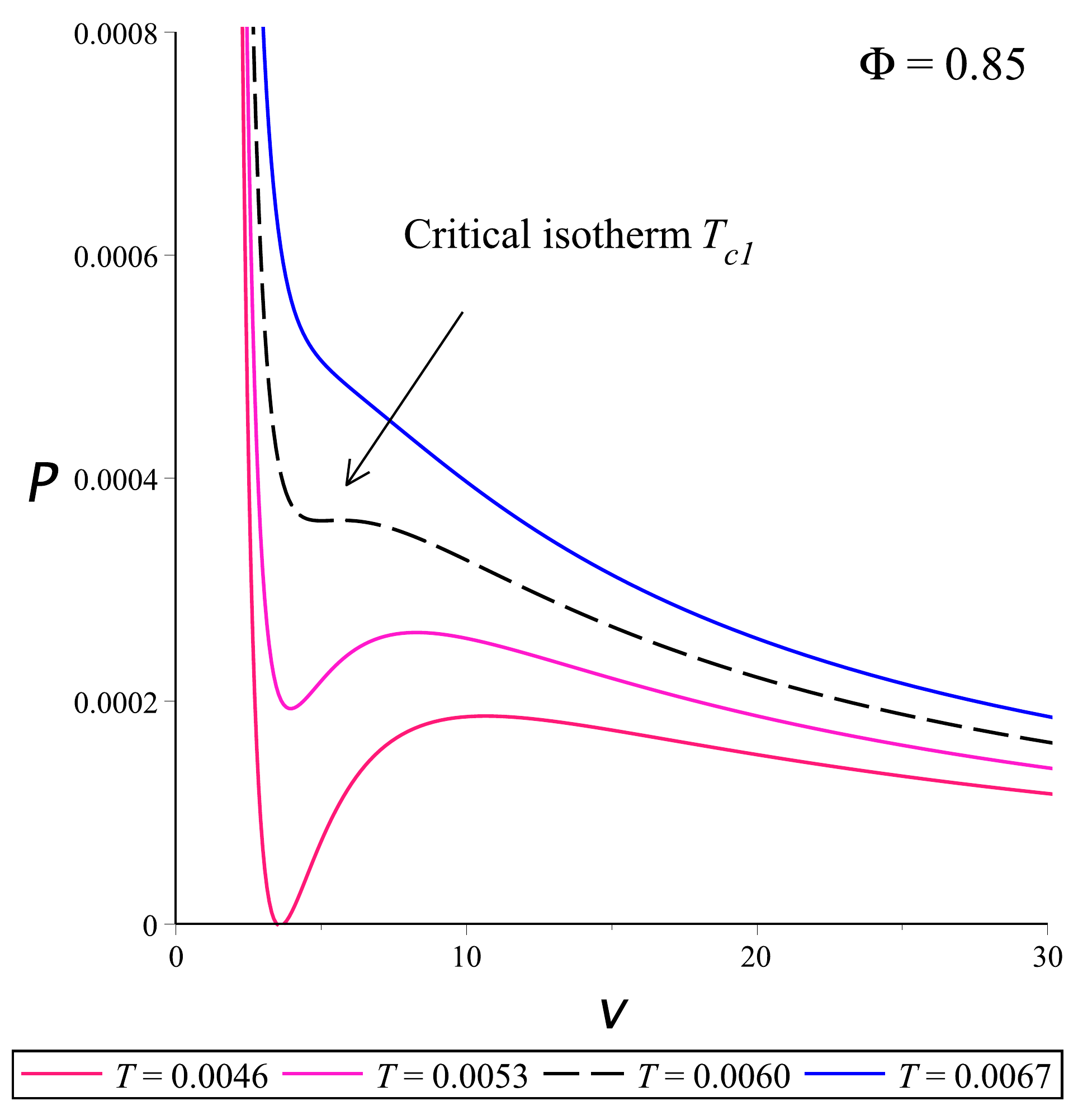}
\includegraphics[scale=0.21]{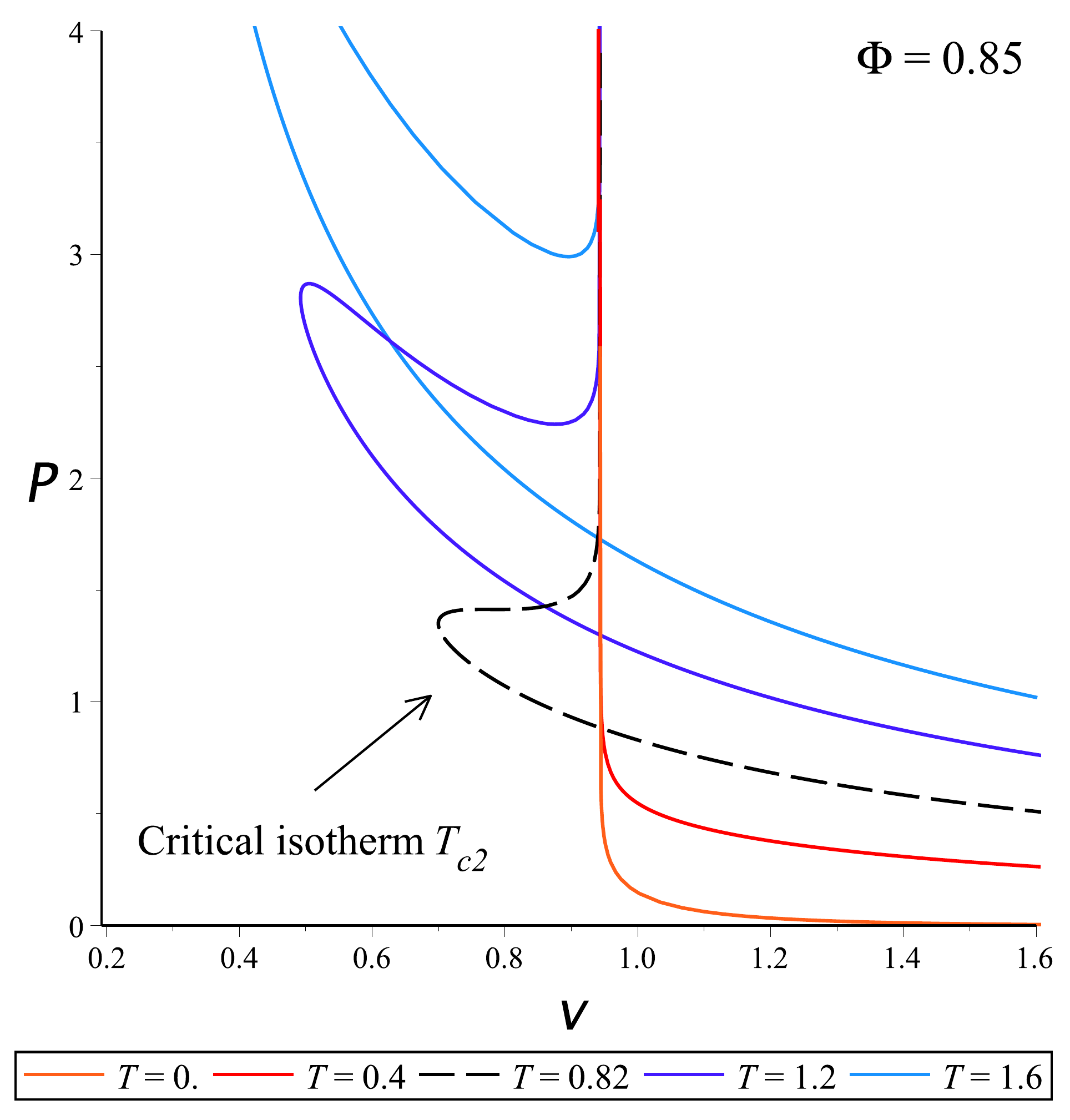}\;
\includegraphics[scale=0.21]{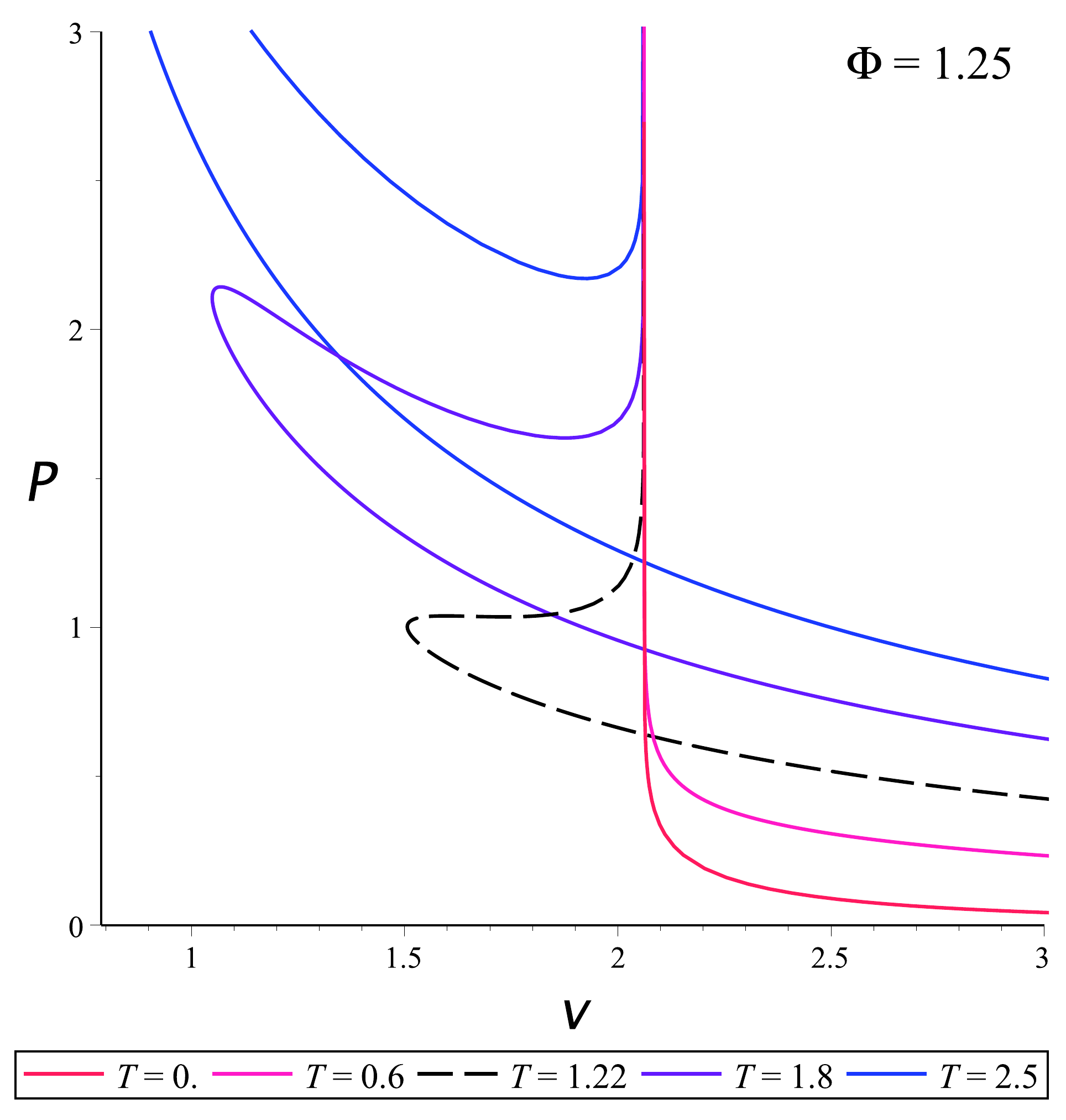}
\caption{Equation of state. The left-hand panel, for $\Phi=0.25<1/\sqrt{2}$, shows no criticality. The middle panels, for $1/\sqrt{2}<\Phi=0.85<1$, show two different critical isotherms, $T_{c1}$ and $T_{c2}>T_{c1}$, given by the dashed lines. The right-hand panel, for $\Phi=1.25$, shows one critical isotherm, again given by the dashed line.}
\label{F2}
\end{figure}

While there is no critical phenomena for the RN-$AdS$ black hole in the grand canonical ensemble, the situation here is more interesting. We have previously reported one critical point in this ensemble  \cite{Astefanesei:2019ehu}. Upon further investigating this case we  find at most two critical points, each satisfying the conditions
\begin{equation}
	\label{cond1}
\(\frac{\pa P}{\pa v}\)_{T_c,\Phi}=0\;, \qquad \(\frac{\pa^2 P}{\pa v^2}\)_{T_c,\Phi}=0
\end{equation}
at different critical temperatures $T_c$. The equations in (\ref{cond1}) have two solutions if ${1/\sqrt{2}}<\Phi<1$, one solution if $\Phi>1$, and no solution if $\Phi<1/\sqrt{2}$. This is illustrated in Fig.~\ref{F2}, where these three situations are shown. We see that there are two kinds of critical isotherms
for the intermediate values of $\Phi$. One corresponds to the standard Van der Waals case, where the $P-v$ curve has a point of inflection. The other has the same 
novel features as in the canonical ensemble, corresponding to the coincidence of the local maximum and minimum of $P(v)$. This novel point is the only critical point for large values of $\Phi>1$.

\begin{figure}[t!]
	\centering
	\includegraphics[scale=0.21]{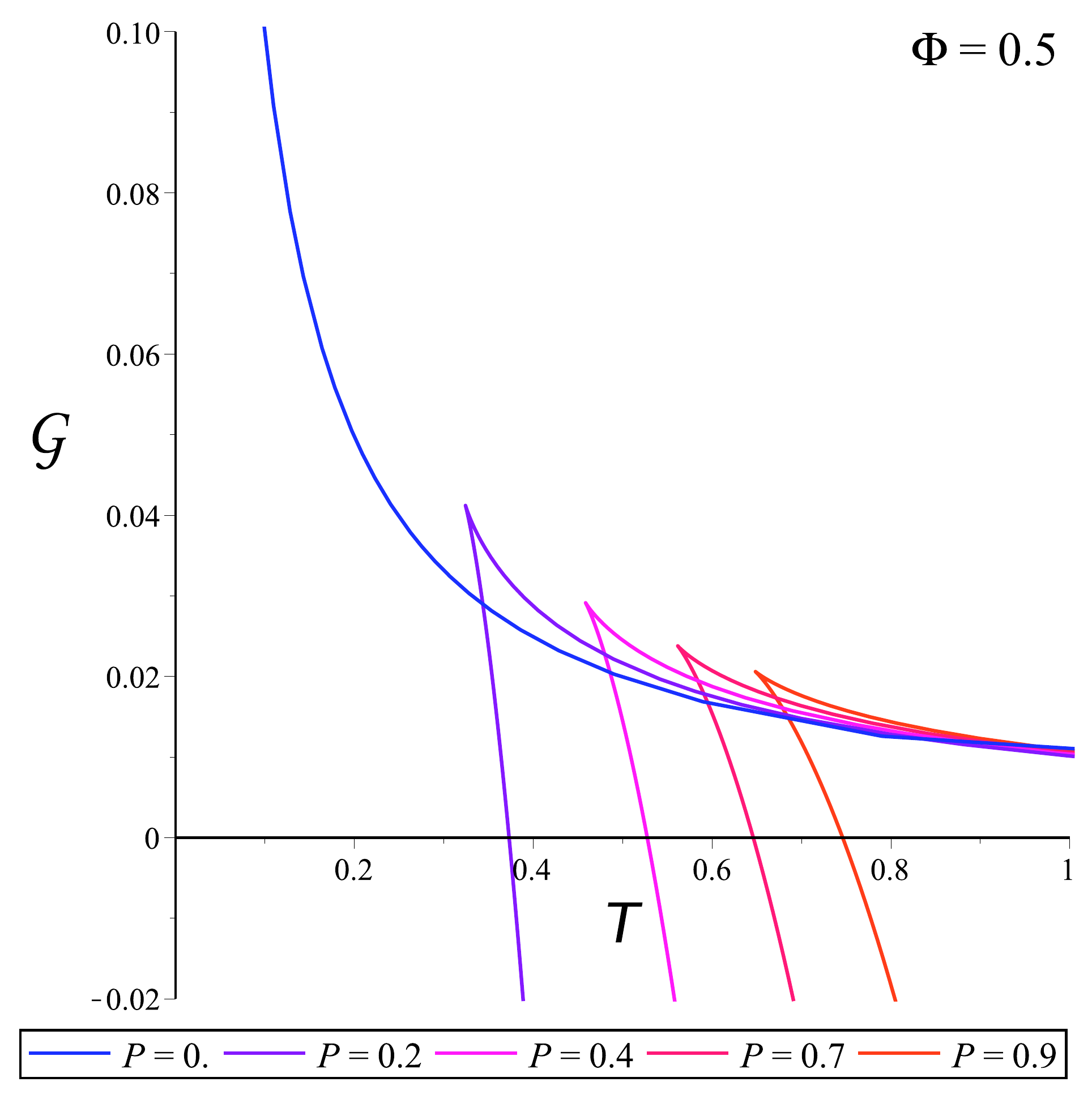}\; 
	\includegraphics[scale=0.21]{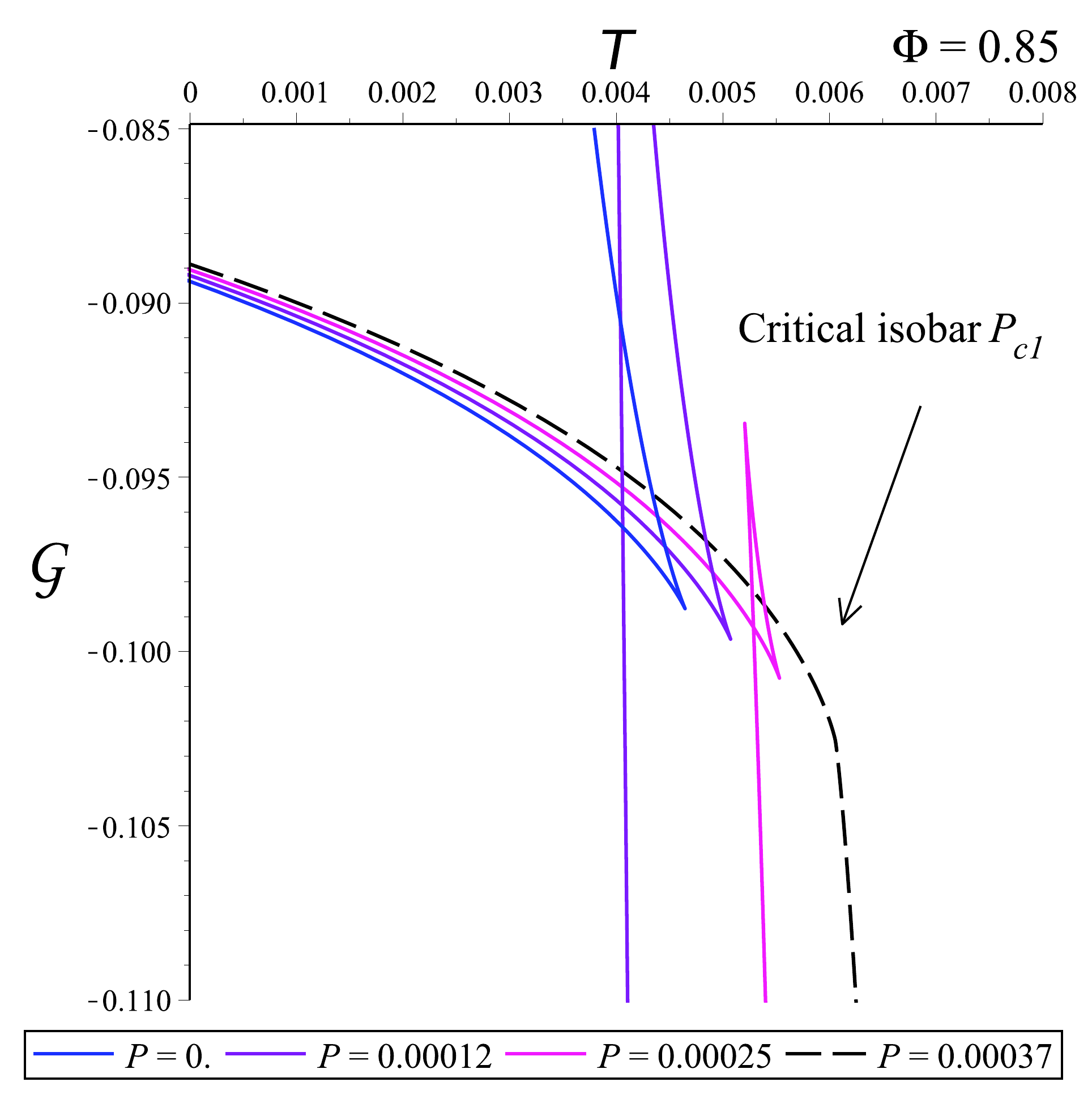}
	\includegraphics[scale=0.21]{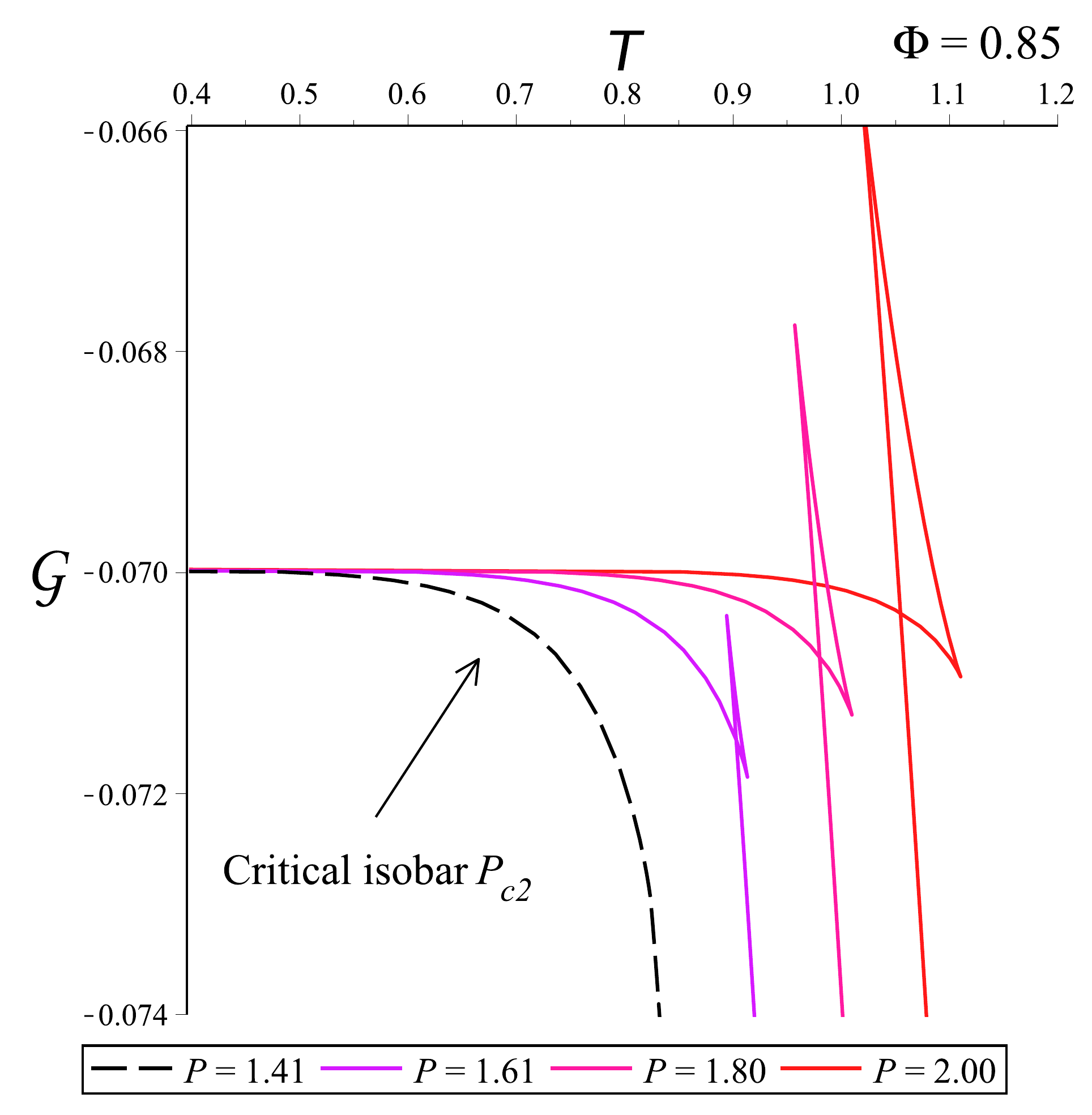}\;
	\includegraphics[scale=0.21]{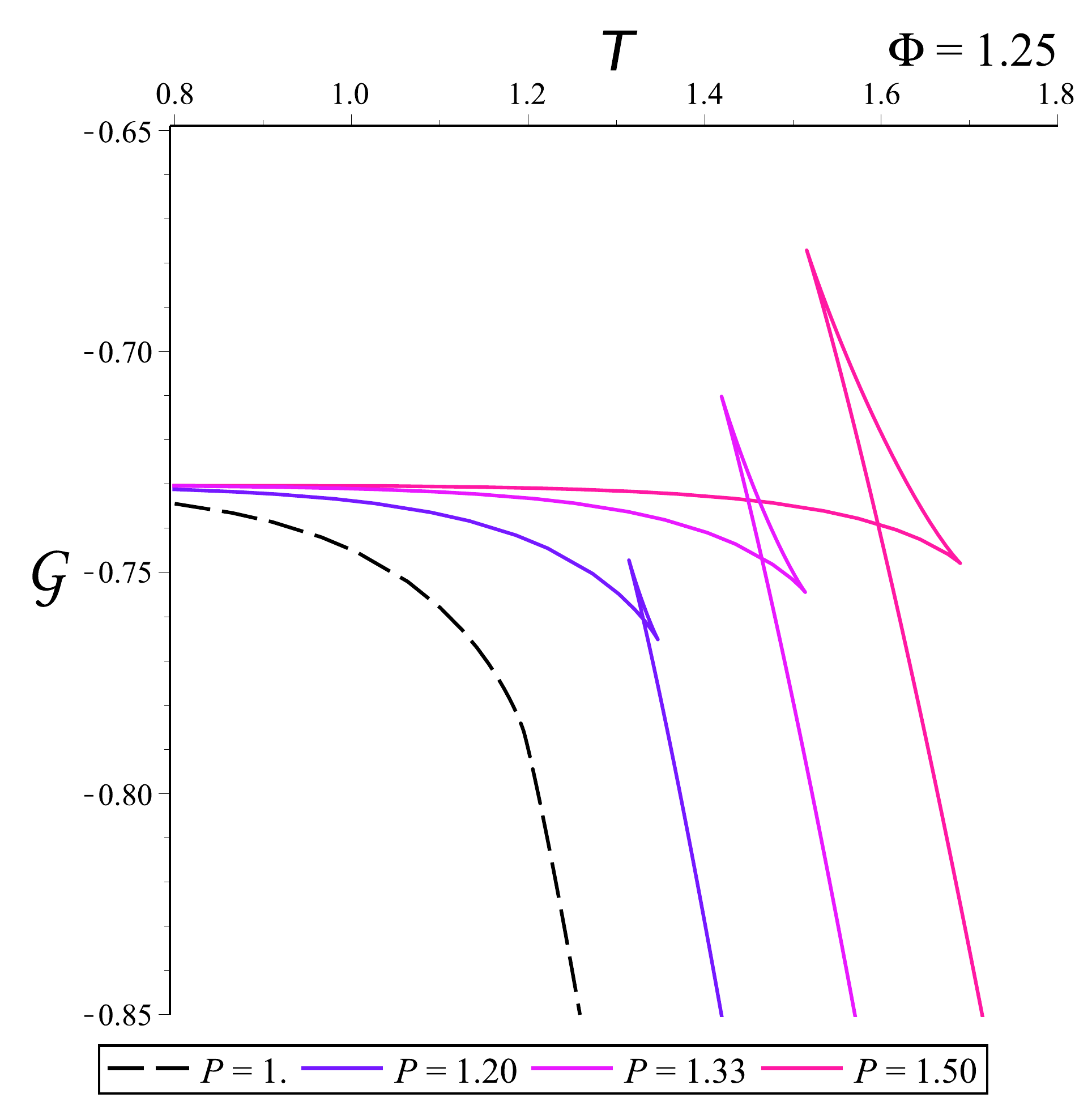}
	\caption{$\mathcal{G}-T$ diagrams. The left-hand panel, for  $\Phi=0.5<1/\sqrt{2}$, shows no criticality, but exhibits a Hawking-Page-type phase transition at $\mathcal{G}=0$. The middle panels, for $1/\sqrt{2}<\Phi=0.85<1$, show two different critical isobars, $P_{c1}$ and $P_{c2}>P_{c1}$, given by the dashed lines. The right-hand panel, for $\Phi=1.25$, shows one critical isobar, again given by the dashed line.}
	\label{F3}
\end{figure}

The existence of critical isotherms is indicative of  phase transitions, which we examine by studying the thermodynamic potential $\mathcal{G}=E-TS-\Phi Q$. In Fig.~\ref{F3}, we depict the thermodynamic potential for the three situations. It is remarkable that all the critical points are associated with large-to-small first order phase transitions between two thermally stable phases. Hawking-Page phase transitions\footnote{The two phases involved in the first order Hawking-Page transitions should be a large black hole and the ground state of the theory. The fact that the solutions can be embedded in SUGRA is a sufficient condition for the existence of a stable ground state of the theory. While  explicit construction of the ground state of the theory  is outside the scope of this paper, we would like to point out the reference \cite{Anabalon:2022aig} where exact hairy soliton solutions were constructed in supergravity.} (in which the hairy black hole discharges to thermal AdS) only take place for $\Phi<1/\sqrt{2}$. 

\section{Extended phase space thermodynamics: the general case}
\label{sec3}

In this section, we investigate the thermodynamics and critical behaviour of the general class of solutions for which the scalar potential contains an extra parameter, $\sigma$, that controls the coupling between the scalar and gauge field, as well as the self-interaction scalar field potential. The action is
\begin{equation}
I=\frac{1}{2\kappa}
\int_{\mathcal{M}}
{d^4x\sqrt{-g} \[R-e^{\gamma\phi}F^2 -\frac{1}{2}(\partial\phi)^2 -U(\phi)\]}
\label{sqrt3action}
\end{equation}
where
\begin{equation}
\gamma\equiv\(\frac{\sigma+1}{\sigma-1}\)^\frac{1}{2}
\end{equation}
and the scalar field potential is \cite{Anabalon:2013sra}
\begin{align}
U(\phi)&=\frac{2\alpha}{\sigma^2} \[\frac{\sigma-1}{\sigma+2} \sinh\(\frac{\sigma+1}{l_\sigma}\phi\) -\frac{\sigma+1}{\sigma-2} \sinh\(\frac{\sigma-1}{l_\sigma}\phi\) +\frac{4\sigma^2-4}{\sigma^2-4} \sinh\(\frac{\phi}{l_\sigma}\)\]
\notag \\ 
& \quad +\frac{\sigma^2-4}{3\sigma^2} \[\frac{\sigma-1}{\sigma+2} \exp\(-\frac{\sigma+1}{l_\sigma}\phi\) + \frac{\sigma+1}{\sigma-2} \exp\(\frac{\sigma-1}{l_\sigma}\phi\) +\frac{4\sigma^2-4}{\sigma^2-4} \exp\(-\frac{\phi}{l_\sigma}\)\] \Lambda
\label{V0}
\end{align}
where $l_\sigma\equiv \sqrt{\sigma^2-1}$. We refer to the parameter $\sigma$ as the `hair parameter' in the sense that we recover the usual RN-$AdS$ black hole for the specific value $\sigma=-1$. When $\sigma\leq -1$, the coupling exponent is $0\leq \gamma<1$. We are interested in the cases $\sigma>1$ for which the coupling is stronger, $\gamma>1$, and so the contribution from the scalar field becomes non-trivial in its capacity for inducing relevant changes on the thermodynamic properties we propose to explore. The limit $\sigma=1$ corresponds to the Schwarzschild black hole and the limit $\sigma\rightarrow\infty$ corresponds to the case studied in the previous section\footnote{The limit $\sigma\rightarrow\infty$ should be carefully taken \cite{Anabalon:2013qua}.}.

\begin{figure}[t!]
\centering
\includegraphics[scale=0.4]{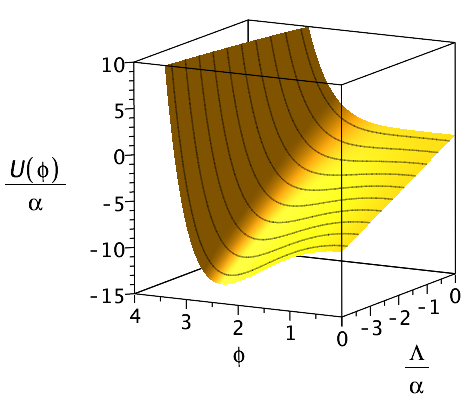}
\qquad\includegraphics[scale=0.4]{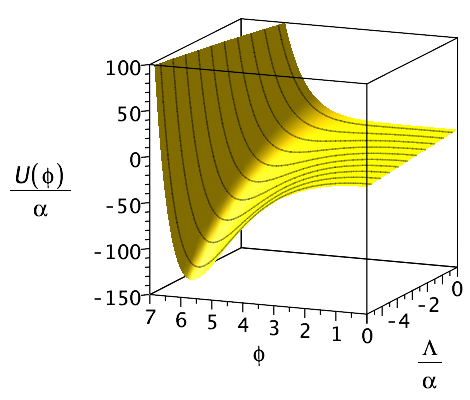}
\caption{The scalar field potential $U(\phi)$ vs $\phi$. Left-hand panel: $\sigma=\sqrt{3}$. Right-hand panel $\sigma=3$.}
\label{F4}\end{figure}

 For small $\phi$, the  potential decays in accordance with the $AdS$ asymptotics,
\begin{equation}
U(\phi)=-\frac{6}{\ell^2} -\frac{\phi^2}{\ell^2} -\frac{1}{12} \(\frac{\sigma^2-3}{\sigma^2-1}\) \cdot\frac{\phi^4}{\ell^2} %+\frac{2\alpha\ell^2+\nu^2-4} {(\nu^2-1)^\frac{3}{2}} \cdot\frac{\phi^5}{\ell^2} 
+\mathcal{O}(\phi^5)
\end{equation}
where $\Lambda=-3/\ell^2$. As shown in Fig.~\ref{F4}, the potential is bounded from below and has a global minimum at a finite value of $\phi$, for any (negative) value of $\Lambda$, provided $\sigma>1$ and $\alpha>0$. We will assume $\sigma>1$ and $\alpha>0$ from now on.

The equations of motion are
\begin{equation}
R_{\mu\nu}-\frac{1}{2}\pa_\mu\phi\pa_\nu\phi -\frac{1}{2}g_{\mu\nu}U(\phi)-T_{\mu\nu}^{EM}=0\,, \quad \Box\phi-\gamma e^{\gamma\phi}F^2-
\frac{dU(\phi)}{d\phi}=0\,, \quad \nabla_\mu\(e^{\gamma\phi}F^{\mu\nu}\)=0
\end{equation}
where $T_{\mu\nu}^{EM}=2e^{\gamma\phi}
(F_{\mu\alpha}F_{\nu}{}^{\alpha}-\frac{1}{4}g_{\mu\nu}F^2)$ is the energy-momentum tensor for the electromagnetic field. The exact solution to the equations of motion, provided the potential (\ref{V0}), is
\begin{equation}
	ds^2=\Omega(x)\[-f(x)dt^2
	+\frac{\eta^2dx^2}{f(x)}
	+d\Sigma^2\], \quad
	A_\mu =\(-\frac{q}{\sigma x^\sigma}+\frac{q}{\sigma x_+^\sigma}\)\delta_\mu^t, \quad
	\phi=l_\sigma\ln(x)
	\label{scalar}
\end{equation}
where $\eta$ and $q$ are the constants of integration related to the conserved charges, namely the mass and electric charge of the black holes\footnote{The scalar field is `secondary hair' that is present outside the horizon that has no associated conserved charge.}, and  $d\Sigma:=d\theta^2+\sin^2\theta d\varphi^2$. The metric functions $f(x)$ and $\Omega(x)$ are
\begin{equation}
f(x)=\frac{1}{\ell^2} +\alpha\[\frac{1}{\sigma^2-4} -\frac{x^2}{\sigma^2} \(1+\frac{x^{-\sigma}}{\sigma-2} -\frac{x^\sigma}{\sigma+2}\)\] +\frac{x}{\Omega(x)} \[1-\frac{2q^2(x^\sigma-1)}{\sigma(\sigma-1)x^\sigma}\]\,, \quad 	\Omega(x) 
=\frac{\sigma^2x^{\sigma-1}}{\eta^2\(x^\sigma-1\)^2}
\end{equation}
The black hole horizon is located at $x=x_+$, where $f(x_+)=0$. There is no loss in generality in assuming $\eta\geq 0$. The canonical radial coordinate is given by the change $r^2=\Omega(x)$. In the asymptotic region ($x\rightarrow 1$),
\begin{equation}
x=1+\frac{1}{\eta{r}}-\frac{\sigma^2-1}{24}\(\frac{1}{\eta^3r^3} -\frac{1}{\eta^4r^4}+\cdots\)
\label{coord}
\end{equation}
Since the mass of the scalar field potential is $m^2=-2\ell^{-2}$, it should be expected that
% in the static regime 
the scalar field falls off as $\phi(r)=\frac{A}{r}+\frac{B}{r^2}+\mathcal{O}(r^{-3})$. It turns out that, indeed, $A={l_\sigma}{\eta}^{-1}$ and $B=-(1/2){l_\sigma}{\eta}^{-2}$. So in this case, the scalar field obeys a mixed boundary condition since both modes $A$ and $B$ are non vanishing. It is convenient to introduce a function $W=W(A)$ that relates both $A$ and $B=B(A)$, by means of $B(A)\equiv\frac{dW(A)}{dA}$. It follows straightforwardly that
\begin{equation}
\label{W(A)}
W(A)=-\frac{A^3}{6l_\sigma}
\end{equation}
This expression for $W(A)$ is relevant for the computation of the contribution to the on-shell Euclidean action coming from the scalar field.

%%%%%%%%%%%%%%%%%%%%%%%%%%%%%%%%%%%%
\subsection{Euclidean action and the usual thermodynamics}
\label{euclidean}

In this section, we use counterterms consistent with the mixed boundary condition of the scalar field \cite{Marolf:2006nd,Anabalon:2015xvl} to compute the on-shell Euclidean action. We do the computation using the boundary condition $\left.\delta A_{\mu}\right|_{\pa\mathcal{M}}=0$ for the gauge field. Once the action is regularized, we use the Brown-York formalism \cite{Brown:1992br} to compute the quasilocal boundary stress tensor and the conserved energy. Finally, we verify the first law of black hole thermodynamics and the quantum-statistical relation. We would like to emphasize that the hair parameter $\sigma$ is kept arbitrary and, therefore, this analysis is more general than the one presented in \cite{Astefanesei:2019ehu}.

The full regularized action $I^E$ in the Euclidean section is composed of the bulk part $I^E_{\text{bulk}}$ given by Eq. (\ref{sqrt3action}), the Gibbons-Hawking boundary term $I^E_{GH}$ \cite{Gibbons:1976ue}, the gravitational counterterm for asymptotically $AdS$ spacetime $I^E_g$ \cite{Balasubramanian:1999re}, and the boundary term for the scalar field $I^E_\phi$ \cite{Anabalon:2015xvl},
\begin{equation}
\label{full}
I^E=I^E_{\text{bulk}} -\frac{1}{\kappa}\int_{\pa\mathcal{M}} {d^3x\sqrt{-h}K} +\frac{1}{\kappa}\int_{\pa\mathcal{M}} {d^3x\sqrt{-h}} \(\frac{2}{\ell}+\frac{\ell\mathcal{R}}{2}\) +\frac{1}{2\kappa} \int_{\pa\mathcal{M}}d^3x\sqrt{-h} \[\frac{\phi^2}{2\ell}+\frac{W(A)}{\ell A^3}\phi^3\]
\end{equation}
respectively, where $\mathcal{R}$ is the Ricci scalar for the foliation $x=x_B=const$, where $B$ stands for boundary, and $h_{ab}$ is the induced metric on the boundary $\pa\mathcal{M}$. At the end, we shall consider the limit $x_B\rightarrow 1$. 

Let us proceed by computing the terms in (\ref{full}) one by one. By using the equations of motion, the bulk part of the action in the Euclidean section can be reduced to $	I_{\text{bulk}}^E=-(1/4){\beta}{\eta}^{-1}[f(x_B)\Omega'(x_B) -2\eta^2\int_{x_+}^{x_B}\Omega(x)\,dx]$. 
Now, by expanding this first result in powers of $(x_B-1)$, we get
\begin{equation}
I_{\text{bulk}}^E=
\beta\[
\frac{1}{4\eta}
+\frac{\alpha}{6\eta^3}
-\frac{q^2}{\eta(\sigma-1)}
+\frac{\sigma(x_+^\sigma+1)}{4\eta(x_+^\sigma-1)} -\frac{\sigma^2-1}{48\eta^3\ell^2}\]+\frac{\beta}{2\eta^3\ell^2(x_B-1)^3}+\mathcal{O}\(x_B-1\)
\end{equation}
The next terms are computed on the hypersurface $x=x_B$, where $h_{ab}dx^adx^b =\Omega(x)\[-f(x)dt^2+d\Sigma\]$. For the Gibbons-Hawking boundary term, let us first write down the extrinsic curvature tensor\footnote{The normal unit to the hypersurface $x=const$ is $n_\mu=-\delta_\mu^{r}(g^{xx})^{-\frac{1}{2}}$, and $K_{\mu\nu}=\nabla_\mu n_{\nu}$, $K_{ab}=(\frac{dx^\mu}{dy^a})(\frac{dx^\nu}{dy^b})K_{\mu\nu}$.} and its trace,
\begin{equation}
	K_{ab}=\frac{\Omega'}{2\eta}\(\frac{f}{\Omega}\)^\frac{1}{2} \[\frac{(f\Omega)'}{\Omega'}\delta^t_a\delta^t_b -\delta^\theta_a\delta^\theta_b -\sin^2\theta\delta^\phi_a\delta^\phi_b\]\,,\quad K=-\frac{1}{2\eta} \(\frac{f}{\Omega}\)^\frac{1}{2} \(\frac{f'}{f}+\frac{3\Omega'}{\Omega}\)
\end{equation}
It follows that
\begin{equation}
I_{GH}^E
=\beta\[{\frac{3q^2}{2\eta\(\sigma-1\) }}-\frac{3}{4\eta}-\frac{\alpha}{4\eta^3} +\frac{\sigma^2-1}{16\eta^3\ell^2}\] -\frac{3\beta}{2\eta^3\ell^2(x_B-1)^3} -\frac{\beta}{\eta(x_B-1)} +\mathcal{O}\(x_B-1\)
\end{equation}
The Ricci scalar on the boundary is $\mathcal{R}=2/\Omega(x_B)$ and, thus, the gravitational counterterm is
\begin{equation}
I_g^E
=\beta\[\frac{1}{2\eta}+\frac{\alpha}{6\eta^3} -\frac{q^2}{\eta(\sigma-1)}
+\frac{\sigma^2-1}{8\eta^3\ell^2}\] +\frac{\beta}{\eta^3\ell^2(x_B-1)^3} +\frac{\beta(\frac{1}{\eta}-\frac{\sigma^2-1}{8\eta^3\ell^2})}{x_B-1}+\mathcal{O}\(x_B-1\)
\end{equation}
For the boundary term of the scalar field, we consider the expression $W$ obtained in (\ref{W(A)}). The result is
\begin{equation}
I_\phi^E
=-\frac{\beta(\sigma^2-1)}{6\eta^3\ell^2} +\frac{\beta(\sigma^2-1)}{8\eta^3\ell^2(x_B-1)} +\mathcal{O}(x_B-1)
\end{equation}
Now, by adding up all the contributions, we find that the divergent terms $\propto(x_B-1)^{-1}$ and $\propto(x_B-1)^{-3}$ cancel each other and, in the limit $x_B\rightarrow 1$, the final result is finite
\begin{equation}
\label{qsrelation}
I^E=I_{\text{bulk}}^E+I_{GH}^E+I_g^E+I_\phi^E
=\beta\[\frac{1}{4\eta}+\frac{\alpha}{12\eta^3} -\frac{q^2}{2\eta(\sigma-1)} +\frac{\sigma+1+(\sigma-1)x_+^\sigma}{4\eta(x_+^\sigma-1)}\]
\end{equation}

The total action $I^E$ satisfies the quantum-statistical relation, as we shall show. Let us compute the thermodynamic quantities for this solution. We start by computing the conserved energy $E$. We use the Brown-York formalism \cite{Brown:1992br}, which requires the quasilocal boundary stress tensor $\tau_{ab}$. For the full action (\ref{full}), we have
\begin{equation}
\tau_{ab}\equiv-\frac{2}{\sqrt{-h}}\frac{\delta I}{\delta h^{ab}} =-\frac{1}{\kappa}\(K_{ab}-h_{ab}K +\frac{2}{l}h_{ab}-lG_{ab}\)-\frac{h_{ab}}{2\kappa\ell} \[\frac{\phi^2}{2}+ \frac{W(A)}{A^3}\phi^3\]
\end{equation}
and, according to the Brown-York formalism, the conserved energy is
\begin{equation}
E=\oint_{s_\infty^2}{d^2x\sqrt{\Sigma}n^a\tau_{ab}\xi^b}
\end{equation}
where $\Sigma=\Omega^2(x_B)\sin^2\theta$ is the determinant of the metric on the 2-sphere $ds_\Sigma^2=\Omega(x_B)d\Sigma$, 
$\xi^a=\delta^a_t$ is the time-like Killing vector and $n_a$ the normal unit to $t=const$, given by
\begin{equation}
	n_a=\frac{\delta_a^t}{\sqrt{-g^{tt}}}=\sqrt{\Omega f}\delta_a^t
\end{equation}
To compute the conserved energy, we only need the leading terms in powers of $(x_B-1)$ of $\tau_{tt}$, that is,
\begin{equation}
\tau_{tt}=\frac{1}{4\pi \ell}\(\frac{q^2}{\sigma-1} -\frac{\alpha+3\eta^2}{6\eta^2}\)(x_B-1) +\mathcal{O}\[(x_B-1)^2\]
\end{equation}
Using this result,   the conserved energy of the system is
\begin{equation}
E=\frac{q^2}{\eta(\sigma-1)}-\frac{\alpha+3\eta^2}{6\eta^3}
\end{equation}
Let us now compute the remaining thermodynamic quantities. The electric charge $Q$ can be obtained, as usual, by using the Gauss law at the boundary 
\begin{equation}
Q=\frac{1}{4\pi}\oint_{s^2_\infty}{e^{\gamma\phi}\star F}=\frac{q}{\eta}\,, \qquad \Phi
%=\frac{q}{\nu x_+^\nu}\(x_+^\nu-1\)
=\frac{q}{\sigma x_+^\sigma}\(x_+^\sigma-1\)
\end{equation}
where its conjugate potential is defined as $\Phi\equiv A_t(x=x_+)-A_t(x=1)$. 
The Hawking-Bekenstein entropy is $S=A/4$, where $A=4\pi\Omega(x_+)$ is the area of the event horizon, and the expression for the Hawking temperature
\begin{equation}
S=\frac{\pi\sigma^2 x_+^{\sigma-1}}{\eta^2\(x_+^\sigma-1\)^2}\,, \qquad	T=-\frac{f'(x_+)}{4\pi\eta} =\frac{3E}{2S}- \frac{x_+^{-\frac{\sigma-1}{2}}}{4\sigma\sqrt{\pi{S}}} \[\frac{4\pi\sigma Q^2}{x_+S}+x_+^\sigma(\sigma-1)+\sigma+1\]
\end{equation}
 is obtained by removing the conical singularity in the Euclidean metric.
It is straightforward to show that first law for charged black holes $dE=TdS+\Phi dQ$ is satisfied\footnote{Since the scalar field is secondary hair, no scalar charge appears in the first law \cite{Astefanesei:2018vga}.}. Now, we can easily verify that the Euclidean action, given by Eq. (\ref{qsrelation}), satisfies the quantum-statistical relation,
\begin{equation}\label{IE-grand}
\frac{I^E}{\beta}=E-TS-\Phi Q\equiv\mathcal{G}(T,\Phi)
\end{equation}
where $\mathcal{G}$ is the grand canonical thermodynamic potential. The first law can be written as $d\mathcal{G}=-SdT-Qd\Phi$, from where it follows that $\mathcal{G}=\mathcal{G}(T,\Phi)$.

%%%%%%%%%%%
%%%%%%%%%%%%%%%
\subsection{Smarr formula and the reverse isoperimetric inequality}

Let us now consider the extended phase space where the cosmological constant represents the pressure of a perfect fluid of density $\rho=-P$, with $P=-\Lambda/(8\pi)$. The extended first law is
\begin{equation}
dE=TdS+VdP+\Phi dQ
\end{equation}
where $V$ is the thermodynamic volume given by
\begin{equation}
V\equiv\(\frac{\pa{E}}{\pa{P}}\)_{Q,S}
	=\frac{2\pi\sigma^2}{3\eta^3} \frac{(\sigma+1)x_+^{2(\sigma-1)}+(\sigma-1)x_+^{\sigma-2}}{(x_+^\sigma-1)^3}
\end{equation}

The thermodynamic variables satisfy a simple relation known as the Smarr formula. It can be obtained by scaling arguments, i.e., by looking at the dimensions of the corresponding thermodynamic variables \cite{Kastor:2009wy}. Since the theory is given by two dimensionful constants ($\Lambda$ and $\alpha$), besides the standard thermodynamic variables $(E,T,S,Q,\Phi)$, the Smarr formula incorporates the extra pairs $PV$ and $\alpha\mathcal{B}$, and it reads
\begin{equation}
E=2TS+\Phi Q-2PV-2\alpha\mathcal{B} 
\end{equation}
where $\mathcal{B}\equiv\({\pa E}/{\pa\alpha}\)_{S,Q,P}$ measures how $E$ changes due to variations in $\alpha$. This suggests that the first law can be further extended to 
\begin{equation}
dE=TdS+VdP+\Phi dQ+\mathcal{B}d\alpha
\end{equation}
provided we can find a concrete physical interpretation for $\alpha$. Nevertheless, we treat the parameter $\alpha$ as a constant without variation and so the quantity $\mathcal{B}$ is not going to be relevant in our analysis.

Returning to thermodynamic volume $V$, by using (\ref{coord})  we observe that $V=4\pi r_+^3/3+\mathcal{O}(r_+)$, or in other words the 
  leading contribution  is the Euclidean volume, as could be expected from the fact that for large black holes the scalar field and its self-interaction are negligible at the event horizon. In general, the thermodynamic volume $V$ is conjectured to satisfy the so-called Reverse Isoperimetric Inequality \cite{Cvetic:2010jb}:
\begin{equation}
\label{RII}
\mathcal{R}\equiv\[
\frac{(d-1)V}{\omega_{d-2}}\]^{\frac{1}{d-1}}
\(\frac{\omega_{d-2}}{A}\)^{\frac{1}{d-2}} \,\,\, \geq \,\,\, 1
\end{equation}
where $d$ is the number of dimensions of the spacetime, $\omega_{d-2}$ is the area of the unit cross section, and $A$ is the area of the black hole event horizon. In our case, with $d=4$ and $\omega_2=4\pi$, we find that 
\begin{equation}
\mathcal{R}_{\sigma}(x_+)
={x_+^{-\frac{1}{6}(\sigma+1)}}\[ \frac{\sigma+1}{2\sigma}x_+^\sigma+\frac{\sigma-1}{2\sigma}\]^\frac{1}{3}
\end{equation}
is the general expression\footnote{Since the value of the scalar field on the horizon is $\phi_+\equiv l_\sigma\ln(x_+)$, the ratio $\mathcal{R}$ depends, alternatively, on $\sigma$ and $\phi_+$.} for the ratio $\mathcal{R}$.
From the following considerations
\begin{equation}
\lim_{x_+\rightarrow 1}\mathcal{R}_{\sigma}(x_+)=1 \,, \qquad \lim_{x_+\rightarrow\infty}\mathcal{R}_\sigma(x_+)\rightarrow \infty \,, \qquad \frac{d\mathcal{R}_\sigma(x_+)}{dx_+} =\frac{
2^\frac{2}{3}
x_+^{-\frac{1}{6}\sigma-\frac{7}{6}}
(x_+^\sigma-1)(\sigma-1)}{12\sigma^\frac{1}{3}\[x_+^\sigma(\sigma+1)+\sigma-1\]^\frac{2}{3}} \geq 0
\end{equation}
it is straightforward to show that the inequality   (\ref{RII})  is satisfied.
The physical interpretation is that, for a fixed thermodynamic volume, the charged hairy $AdS$ black hole carries less entropy than its RN-$AdS$ counterpart, for which $\mathcal{R}=1$. This is expected, because the remaining entropy is carried by the scalar field (`hairy' degrees of freedom) outside the event horizon.

Henceforth we shall   rescale the thermodynamic quantities as in \eqref{rescaled}; consequently  $\alpha > 0$ will not appear explicitly in any further expression.

\subsection{Extended thermodynamics}

With the on-shell Euclidean action  properly computed, we have the tools to study the thermodynamics. 
The hair parameter $\sigma$ introduces a  new phenomenon of  reentrant phase transitions in both the canonical and grand canonical ensembles, in addition
 to the novel transition behaviour seen in the previous section for $\sigma\to\infty$.

%%%%%%%%%%%%%%%%%%%%%%%%%%%%%%%%%%%%%%%
\subsubsection{The canonical ensemble}
%%%%%%%%%%%%%%%%%%%%%%%%%%%%%%%%%%%%%%

We first consider  thermodynamics in the extended phase space with the electric charge of the black hole being kept fixed. The boundary condition for the gauge field is   $\delta(e^{\gamma\phi}\star F)|_{\pa\mathcal{M}}=0$.  The thermodynamic potential in this ensemble
is $\mathcal{F}(T,Q) = E-TS$ and can be obtained by a Legendre transformation of the thermodynamic potential from the grand canonical ensemble
\eqref{IE-grand}. This is equivalent to adding a boundary term \cite{Hawking:1995ap}
\begin{equation}
I_A^E=-\frac{2}{\kappa}\int_{\pa\mathcal{M}} {d^3x\sqrt{h}e^{\gamma\phi}n_\mu F^{\mu\nu}A_\nu}=\beta\, Q\Phi
\end{equation}
to the action, yielding   $\mathcal{F}(T,Q)=\beta^{-1}\tilde I^E=E-TS$, where $\tilde I^E=I^E+I^E_A$.

Let us first study the equation of state, given parametrically by 
\begin{equation}
T=\frac{1}{4\pi\eta^3\Omega(x_+)} \[\frac{2\eta^4 \((\sigma+2)x_+^\sigma+2\sigma-2\)Q^2} {\sigma(\sigma-1)x_+^{\sigma}} 
-1\]
-\frac{(\sigma+2)x_+^{\frac{1}{2}(\sigma+1)}+(\sigma-2)x_+^{-\frac{1}{2}(\sigma-1)}}{4\pi\sigma\sqrt{\Omega(x_+)}}
\end{equation}
\begin{equation}
v=\frac{x_+^\sigma(\sigma+1)+\sigma-1}{\eta(x_+^\sigma-1)x_+}
\end{equation}
where $\eta=\eta(x_+,P,Q)$ is obtained from\footnote{Since $\eta>0$, Eq. (\ref{etaP}) has two solutions of interest. One of these solutions for $\eta$ is positive only for $\sigma>2$, while the other one is positive only for $1<\sigma<2$. For the particular case $\sigma=2$, note that  $$\lim_{\sigma\rightarrow 2}\[\frac{x_+^{2-\sigma}}{\sigma^2(\sigma-2)}-\frac{1}{\sigma^2-4}\]=-\frac{1}{4}\[\ln(x_+)+\frac{3}{4}\].$$} $f(x_+)=0$,
\begin{equation}
\label{etaP}
\frac{2Q^2x_+^{2-2\sigma}(x_+^\sigma-1)^3\eta^4}{\sigma^3(\sigma-1)}-\frac{(x_+^\sigma-1)^2\eta^2}{\sigma^2x_+^{\sigma-2}}+\frac{x_+^2}{\sigma^2}\(1+\frac{x_+^{-\sigma}}{\sigma-2}-\frac{x_+^\sigma}{\sigma+2}\)-\frac{1}{\sigma^2-4}-\frac{8\pi P}{3}=0
\end{equation}
For large black holes, the equation of state can be put in the form $T=Pv+1/(2\pi v)-2Q^2/(\pi v^3)+\mathcal{O}(v^{-5})$, regardless of the value of the hair parameter $\sigma$. This is   because the scalar field and its self-interaction are negligible at the event horizon of a large black hole, and thus the corrections to the equation of state due to the scalar are  subleading.

\begin{figure}[t!]
\centering
\includegraphics[scale=0.36]{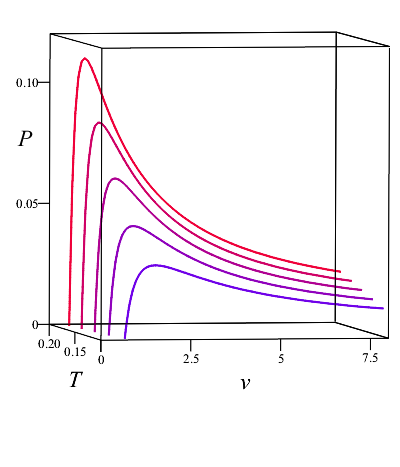}
\includegraphics[scale=0.37]{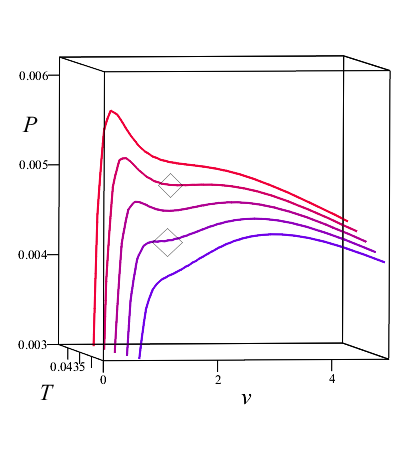}
\includegraphics[scale=0.37]{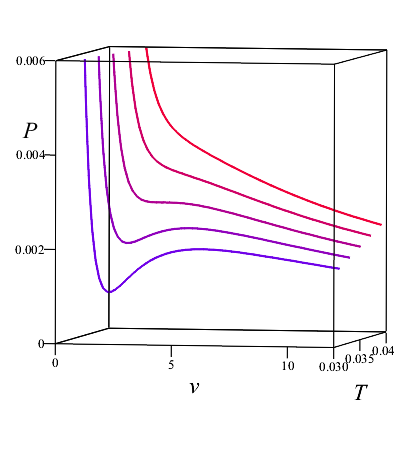}
\caption{Equation of state for $Q=1.50$ (left-hand panel), $Q=2.65$ (middle) and $Q=2.90$ (right-hand panel), for the theory with $\sigma=2$. For $Q_{min}<Q<Q_0$, two critical points are observed.}
\label{F5}
\end{figure}

When the scalar field and its self-interaction become important near the event horizon, which is the case for intermediate and small black holes, the thermodynamic behaviour becomes more interesting. Two values for the electric charge are relevant in our analysis: $Q_{\text{min}}$ and $Q_{0}$. Let us, for concreteness, fix $\sigma=2$, for which $Q_{\text{min}}\approx 2.622$ and $Q_0\approx 2.712$. In Fig.~\ref{F5}, we  illustrate the equation of state for three representative values of $Q$. For $Q<Q_{\text{min}}$ (left-hand panel in Fig.~\ref{F5}), there is no critical behaviour. For $Q>Q_0$ (right-hand panel in Fig.~\ref{F5}), the critical behaviour is qualitatively the same as that of the RN-$AdS$. For $Q_{\text{min}}<Q<Q_0$ (middle plot in Fig.~\ref{F5}), two critical isotherms are observed. These two critical isotherms are related to  reentrant phase behaviour.

\begin{figure}[t!]
\centering
\includegraphics[scale=0.40]{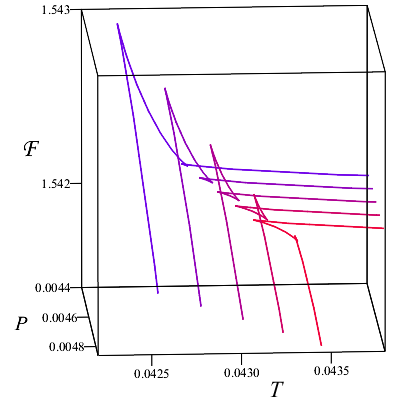}\qquad
\includegraphics[scale=0.35]{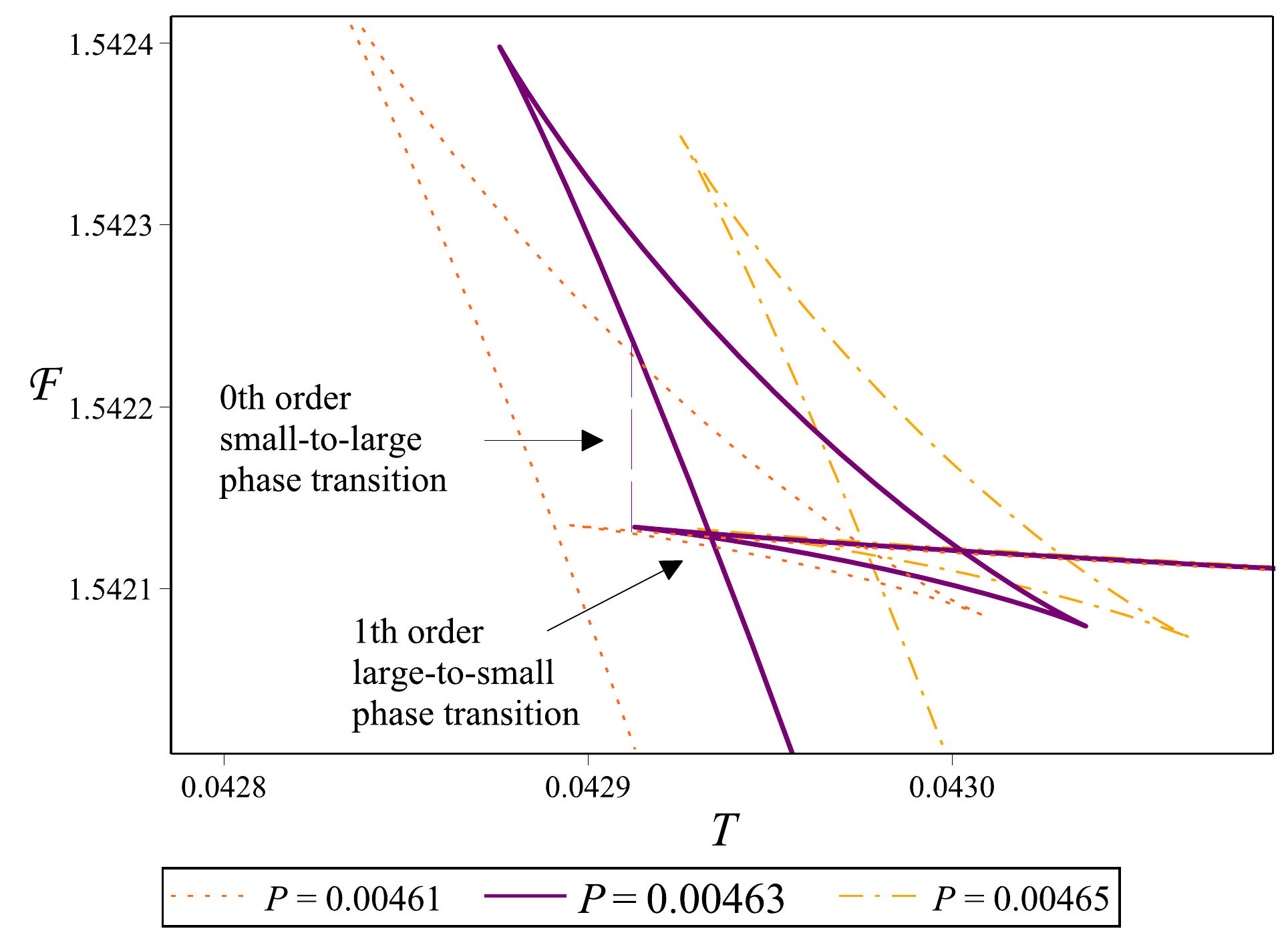}
\caption{$\mathcal{F}-T-P$ for $Q=2.65$ and $\sigma=2$. As pressure increases, the inverted swallowtail moves leftward with respect to the leftmost part of the curve, and a second (standard) swallowtail with a first order phase transition appears in addition to a zeroth order phase transition.}
\label{F6}
\end{figure}

To appreciate better the nature of this double criticality within $Q_{\text{min}}<Q<Q_0$, consider the $\mathcal{F}-T$ diagram, depicted in Fig.~\ref{F6}. As pressure increases from small values (the bluer curve in Fig.~\ref{F6}), an `inverted' swallowtail appears. There is no first order phase transition in this case because the curves that intersect themselves in the inverted swallowtail are not at the global minimum of $\mathcal{F}$. However, as the pressure further increases, the inverted swallowtail moves leftward with respect to the lower part of the curve, eventually giving rise to a second (standard) swallowtail. This standard swallowtail results in a first order phase transition from large to small black holes in the direction of decreasing temperature. For a tiny range for $P$, there is also a zeroth order phase transition from a small black hole to a large one, characterized by a jump discontinuity in $\mathcal{F}$, as well as in its first derivative $(\pa\mathcal{F}/\pa T)_P$, as it is shown in the second panel of Fig.~\ref{F6}.

This `reentrant phase behaviour', reported also in \cite{Astefanesei:2021vcp}, is new when compared to the behaviour for hairy black holes in the theory  $\sigma=\infty$. It happens within the interval $Q_{min}<Q<Q_0$. It is also important to notice that $\Lambda$ is not required to be considered a thermodynamic variable for reentrant phase behaviour to occur, because the parameter that is being varied is temperature. Therefore, once $Q$ is conveniently fixed (within $Q_{\text{min}}<Q<Q_0$), it is only necessary to search for different (fixed) values of $P$ for which reentrant phase transitions occur. For the concrete case $Q=2.65$ in the theory $\sigma=2$, reentrant phase behaviour exists approximately within the tiny interval $4.61\cdot 10^{-3}< P< 4.65\cdot 10^{-3}$. These values of $P$ for which there is reentrant phase behavior were obtained by solving numerically the equations $(\pa P/\pa v)_{T,Q}=0$ and $(\pa^2P/\pa v^2)_{T,Q}=0$.

%%%%%%%%%%%%%%%%%%%%%%%%%%%%%%%%%%%%%%%%%
\subsubsection{The grand canonical ensemble}
\label{sec3:2}

The parametric expressions for the equation of state in this ensemble are
\begin{equation}
P=\frac{3}{8\pi\sigma^2}\[\frac{x_+^{2-\sigma}}{\sigma-2} -\frac{x_+^{\sigma+2}}{\sigma+2} +x_+^2 -\frac{\sigma^2}{\sigma^2-4} +\frac{\mathcal{X}_1^2}{x_+^{3\sigma-4}v^2}\(
\frac{\mathcal{X}_2-2\sigma+2}{\mathcal{X}_2-3\sigma} \frac{2\sigma\Phi^2}{\sigma-1}-1\)\]
\end{equation}
\begin{equation}
\label{temper}
T=\frac{x_+}{4\pi\sigma} \[-\frac{(x_+^\sigma-1)^3}{\sigma\mathcal{X}_1}v+\frac{2\mathcal{X}_1\mathcal{X}_2}{(\sigma-1)x_+^{\sigma-2}(x_+^\sigma-1)}\frac{\Phi^2}{v} -\frac{\mathcal{X}_1(\mathcal{X}_2-\sigma)}{\sigma x_+^{2\sigma-2}v}\]
\end{equation}
where $\mathcal{X}_1\equiv (\sigma+1)x_+^{2\sigma-2}+(\sigma-1)x_+^{\sigma-2}$ and $\mathcal{X}_2\equiv (\sigma+2)x_+^\sigma+2\sigma-2$ have been defined for simplicity. The specific volume $v$ is obtained as usual,
\begin{equation}
	\label{svol}
	v\equiv \frac{3V}{2S} =\frac{x_+^\sigma(\sigma+1)+\sigma-1}{\eta(x_+^\sigma-1)x_+}
\end{equation}

\begin{figure}[t!]
\centering
\includegraphics[scale=0.40]{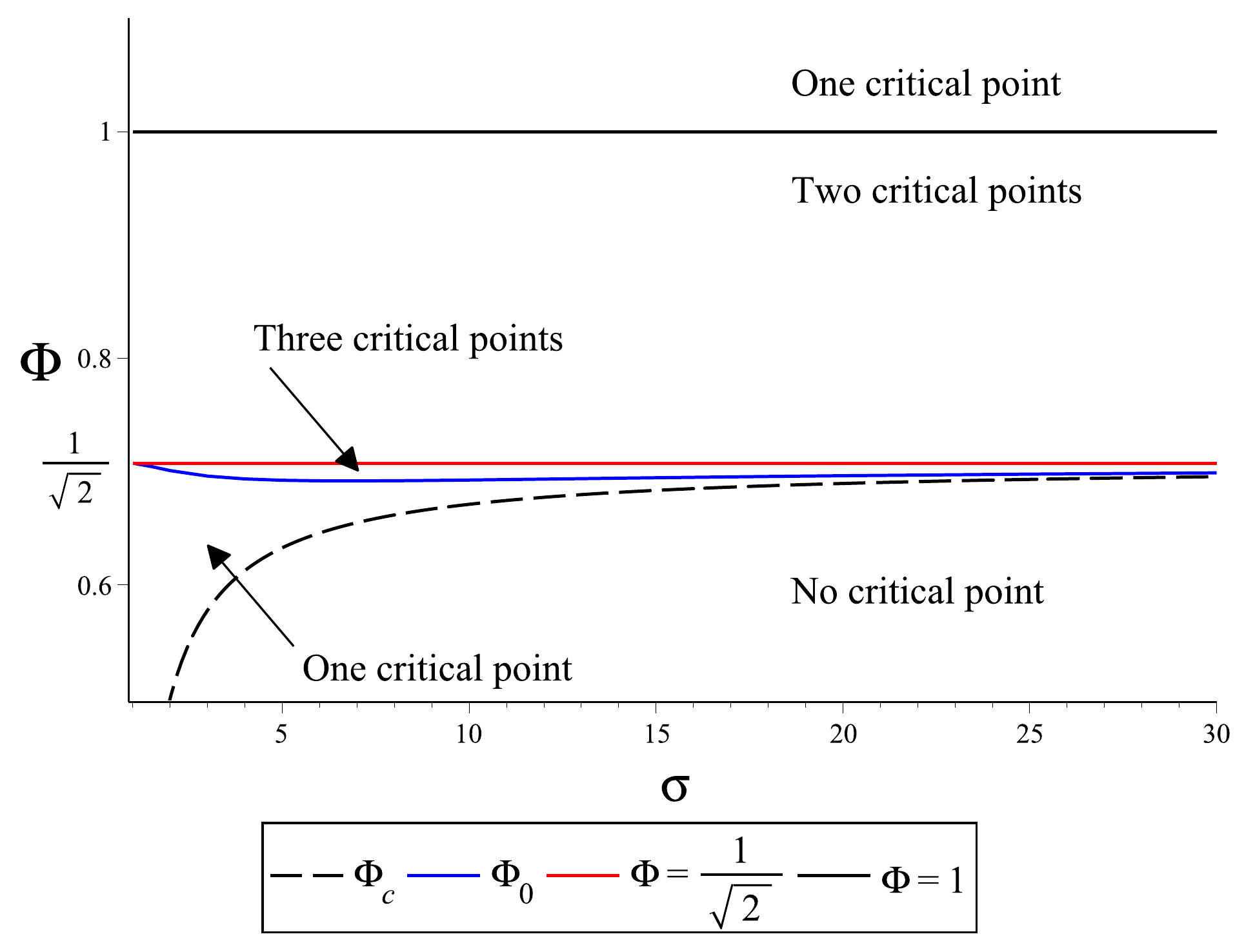}
\caption{The number of critical points depends on the value of $\Phi$. The values $\Phi_c$ and $\Phi_0$ depend on $\sigma$.} 
\label{F7}
\end{figure}

In this ensemble, the value of $\Phi$ determines in a remarkable way the thermodynamic behaviour of black holes. For $\Phi<\Phi_c\equiv\sqrt{(\sigma-1)/(2\sigma)}$, there is no criticality, i.e., the conditions for criticality given in (\ref{cond1}) are not satisfied. For $\Phi_c<\Phi<\Phi_0(\sigma)$, there is one critical point. The dependence of $\Phi_0$ on $\sigma$ is shown in Fig.~\ref{F7}. Notice that $\Phi_0(\sigma)<1/\sqrt{2}$. For $\Phi_0<\Phi<1/\sqrt{2}$, there are three critical points. For $1/\sqrt{2}<\Phi<1$, there are two critical points and, for $\Phi>1$, there is one critical point. 

\begin{figure}[t!]
\centering
\includegraphics[scale=0.23]{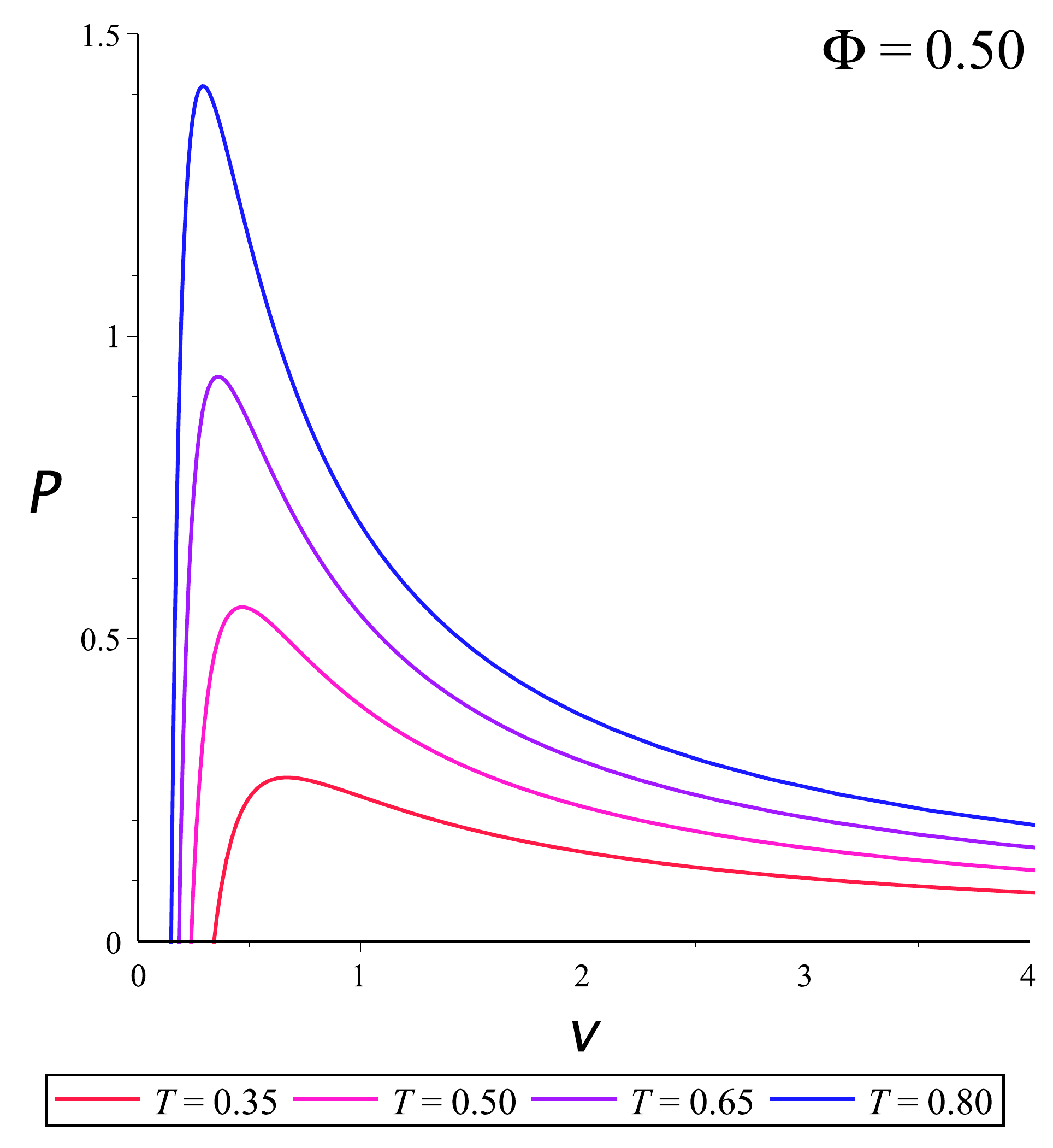}
\includegraphics[scale=0.23]{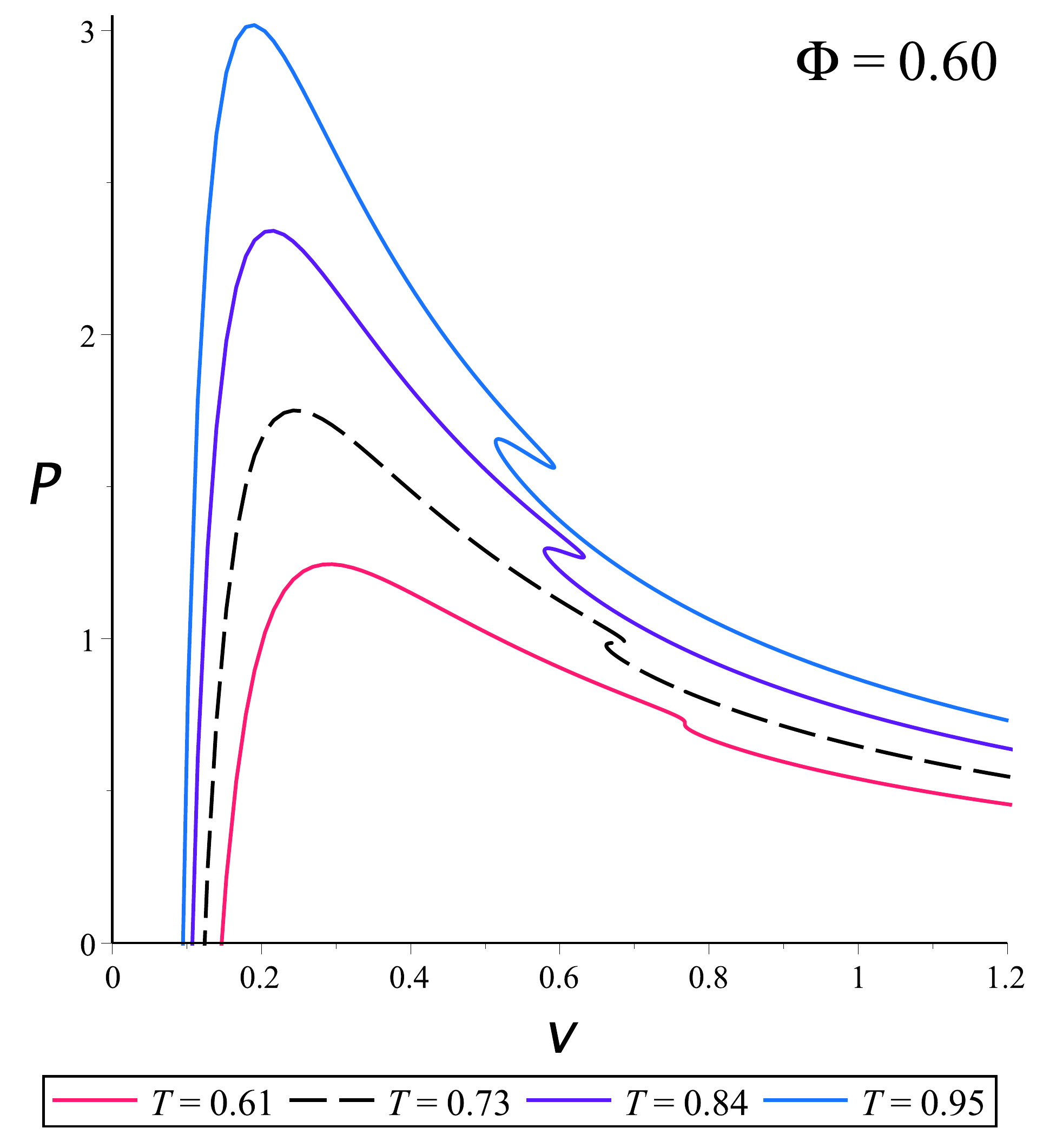}
\begin{overpic}[scale=0.38]{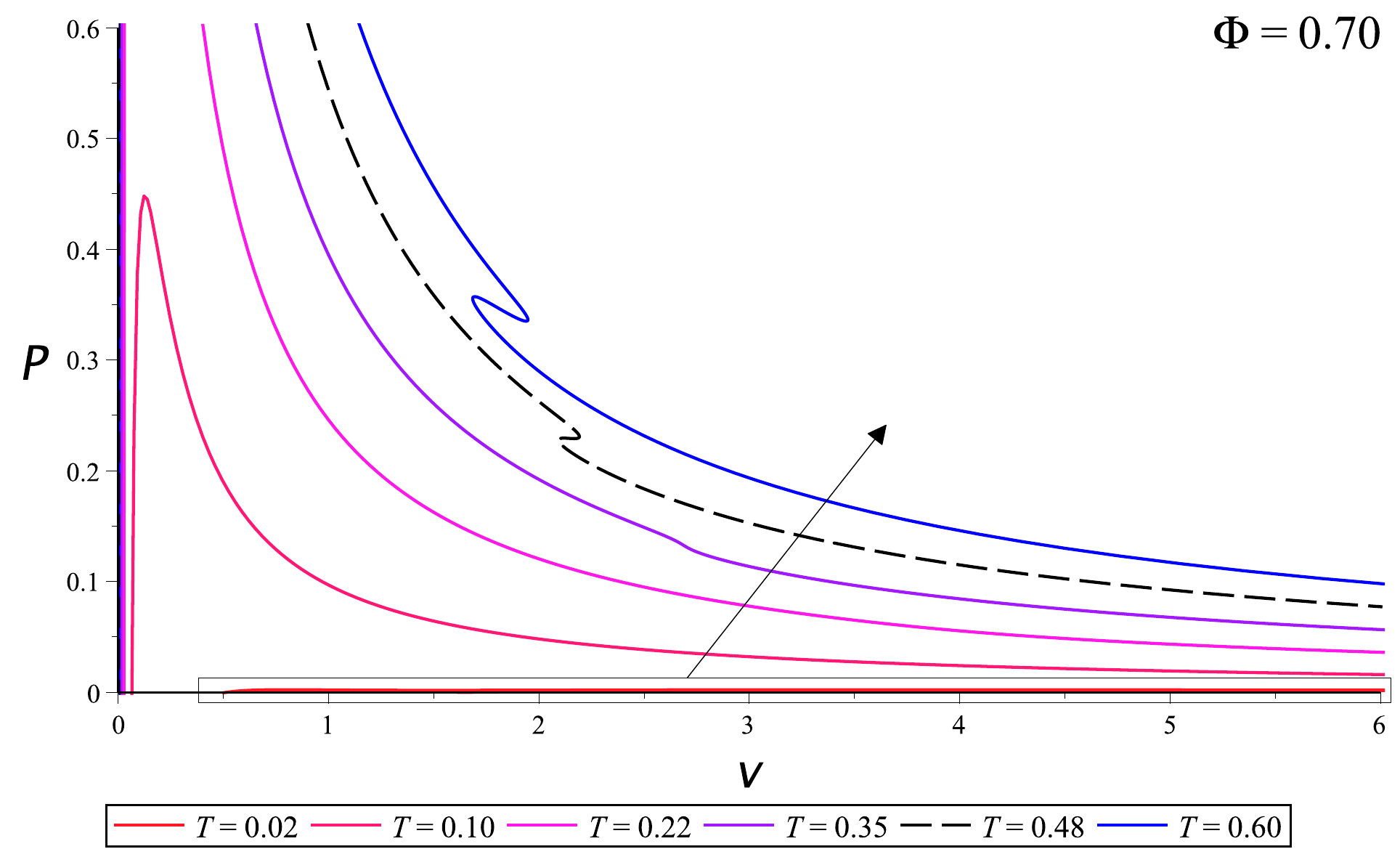}\put(54,32) {\includegraphics[scale=0.17]{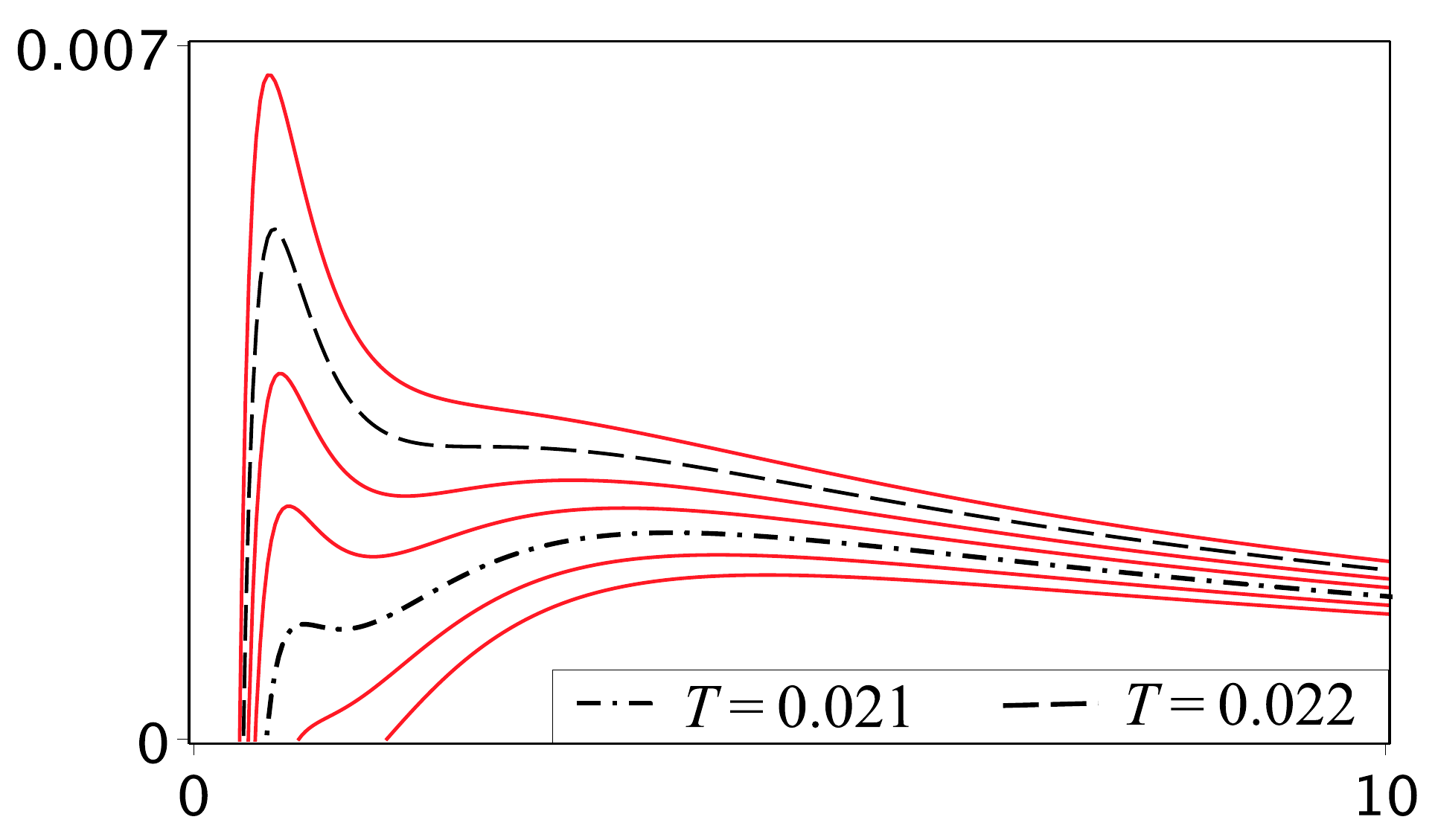}}
\end{overpic} \\
\includegraphics[scale=0.25]{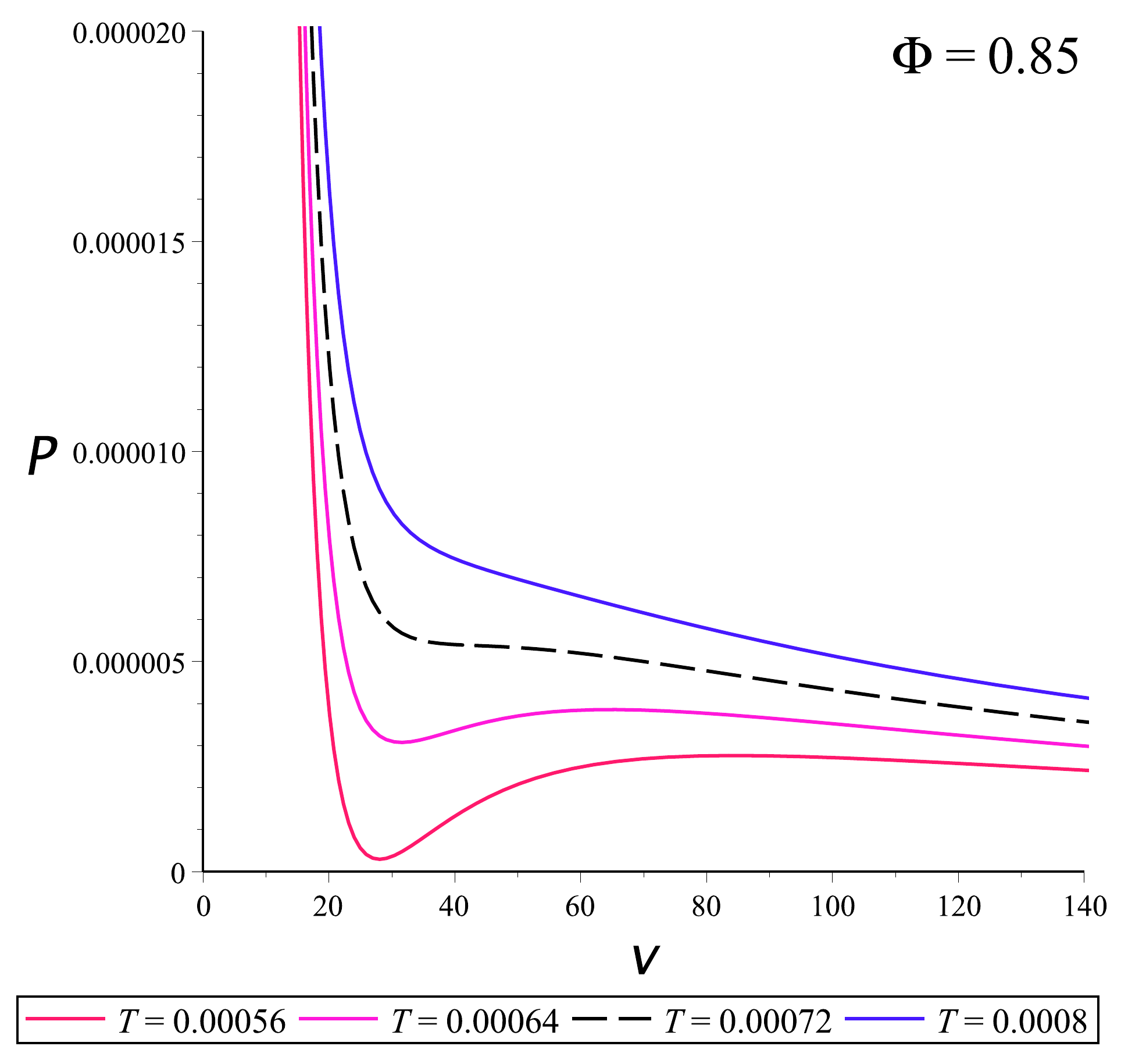}
\includegraphics[scale=0.25]{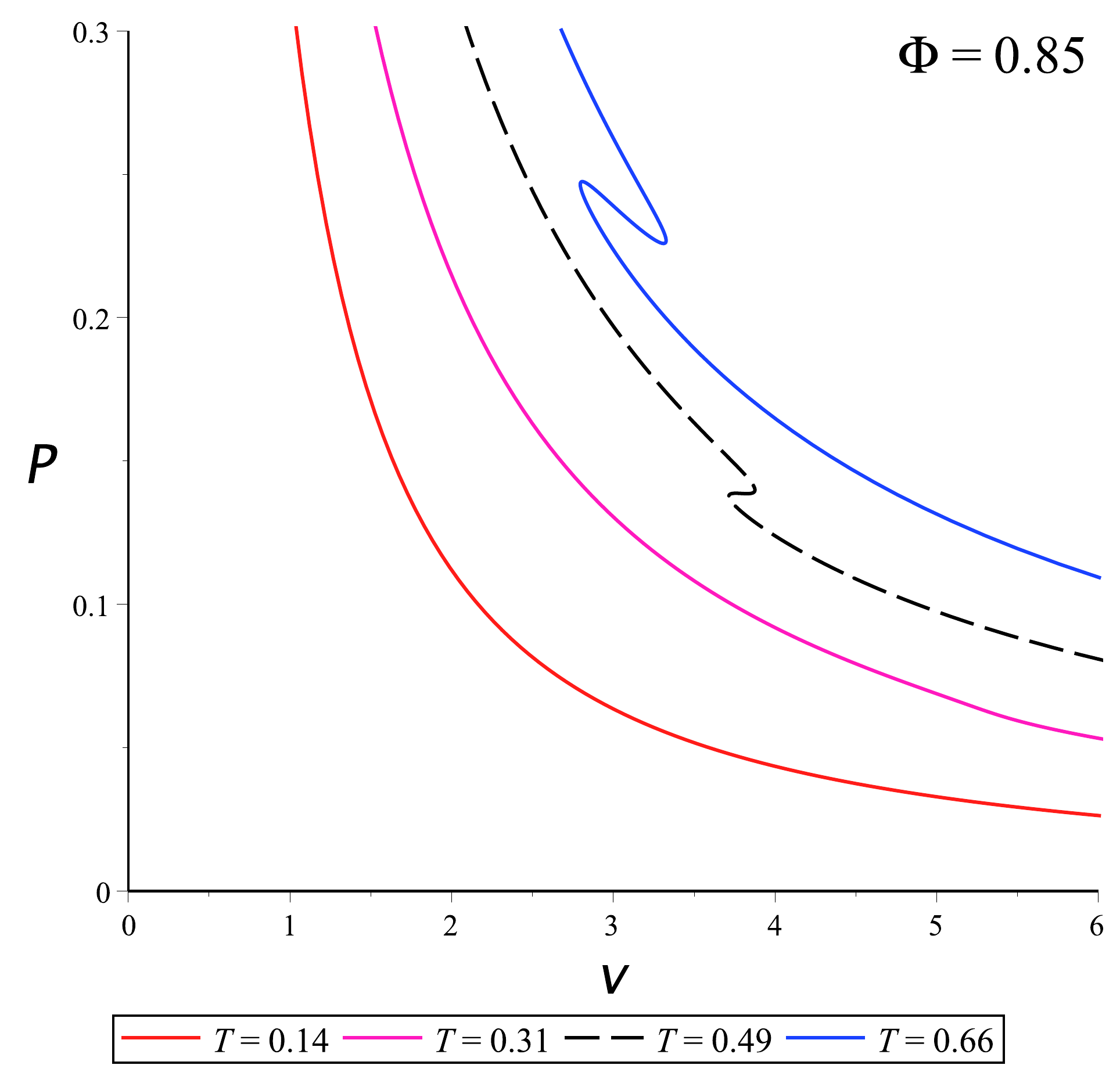}
\includegraphics[scale=0.25]{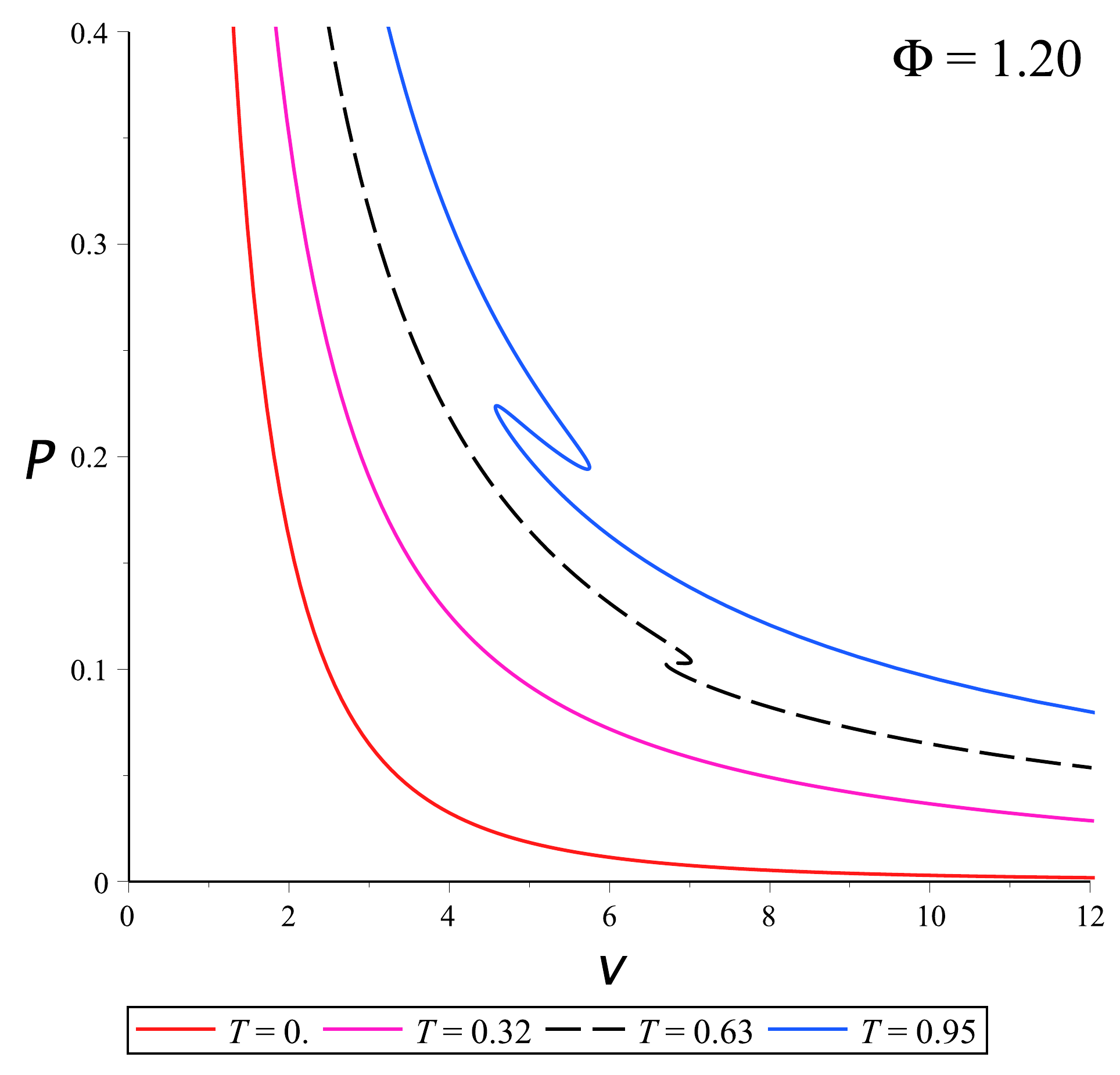}
\caption{\small Equation of state $P-v$ for five characteristic values of the conjugate potential, in the model $\sigma=3$. The panels show the cases: $\Phi=0.50<\Phi_c$, $\Phi_c<\Phi=0.60<\Phi_0$, $\Phi_0<\Phi=0.70<1/\sqrt{2}$, $1/\sqrt{2}<\Phi=0.85<1$, $\Phi>1$, respectively.
Dashed lines correspond to critical behaviour.} 
\label{F8}
\end{figure}

This rich behaviour is depicted in Fig.~\ref{F8} for different values of $\Phi$. We see from the upper left panel that for small $\Phi <  \sqrt{(\sigma-1)/(2\sigma)}$ there are no phase transitions, but once  $\Phi = \sqrt{(\sigma-1)/(2\sigma)}$ we see a new kind of criticality, in which 
 the local maximum and minimum of $P(v)$ are coincident, but $P(v)$  is not single-valued, 
noted in the previous section.  Above the critical point neither $P(v)$ nor $v(P)$ are single-valued functions.
For larger values of $\Phi > \sqrt{(\sigma-1)/(2\sigma)}$ a novel 
phase transition is present (shown in the upper central panel),  whose behaviour we shall discuss in the next section. As $\Phi$ crosses the next threshold at $\Phi = 1/\sqrt{2}$, two new critical points appear at very low pressures, shown in the inset of the upper right panel in  Fig.~\ref{F8}; this is the reentrant behaviour shown in the middle diagram of Fig.~\ref{F5}.   Over this range of values of $\Phi$, as temperature increases, we will have the rentrant behaviour discussed in the previous subsection, followed by a novel phase transition of the same type as for  $\ \sqrt{(\sigma-1)/(2\sigma)}< \Phi < 1/\sqrt{2}$. As $\Phi$ becomes larger than $1/\sqrt{2}$, the middle critical point disappears, and only the smallest one (corresponding to the standard Van der Waals case) and the largest one (corresponding to the novel case) are present.  
For $\Phi > 1$ the Van der Waals critical point vanishes, and only the one corresponding to the novel case remains.

\begin{figure}[t!]
	\centering
	\includegraphics[scale=0.21]{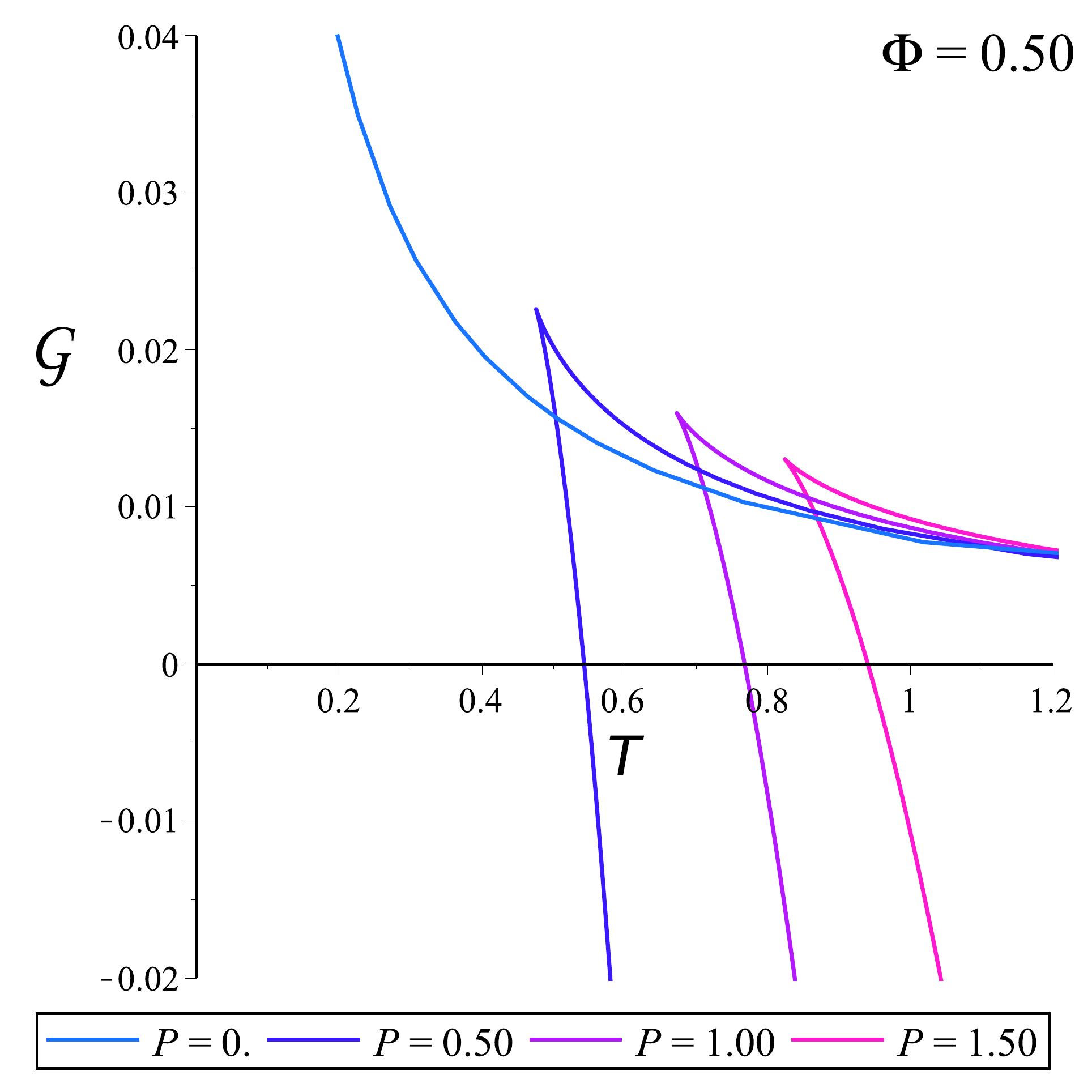}
	\includegraphics[scale=0.21]{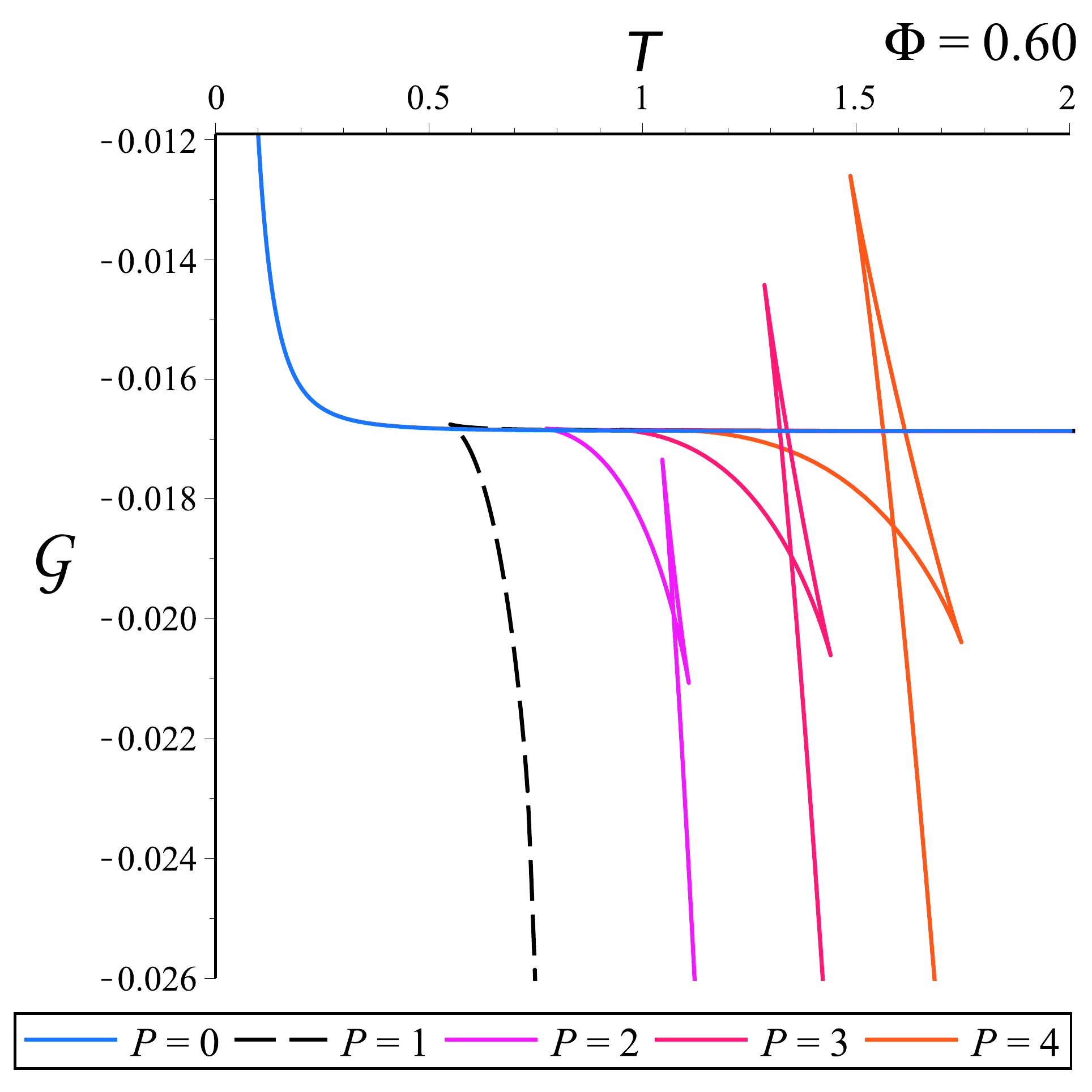}
	\begin{overpic}[scale=0.42]{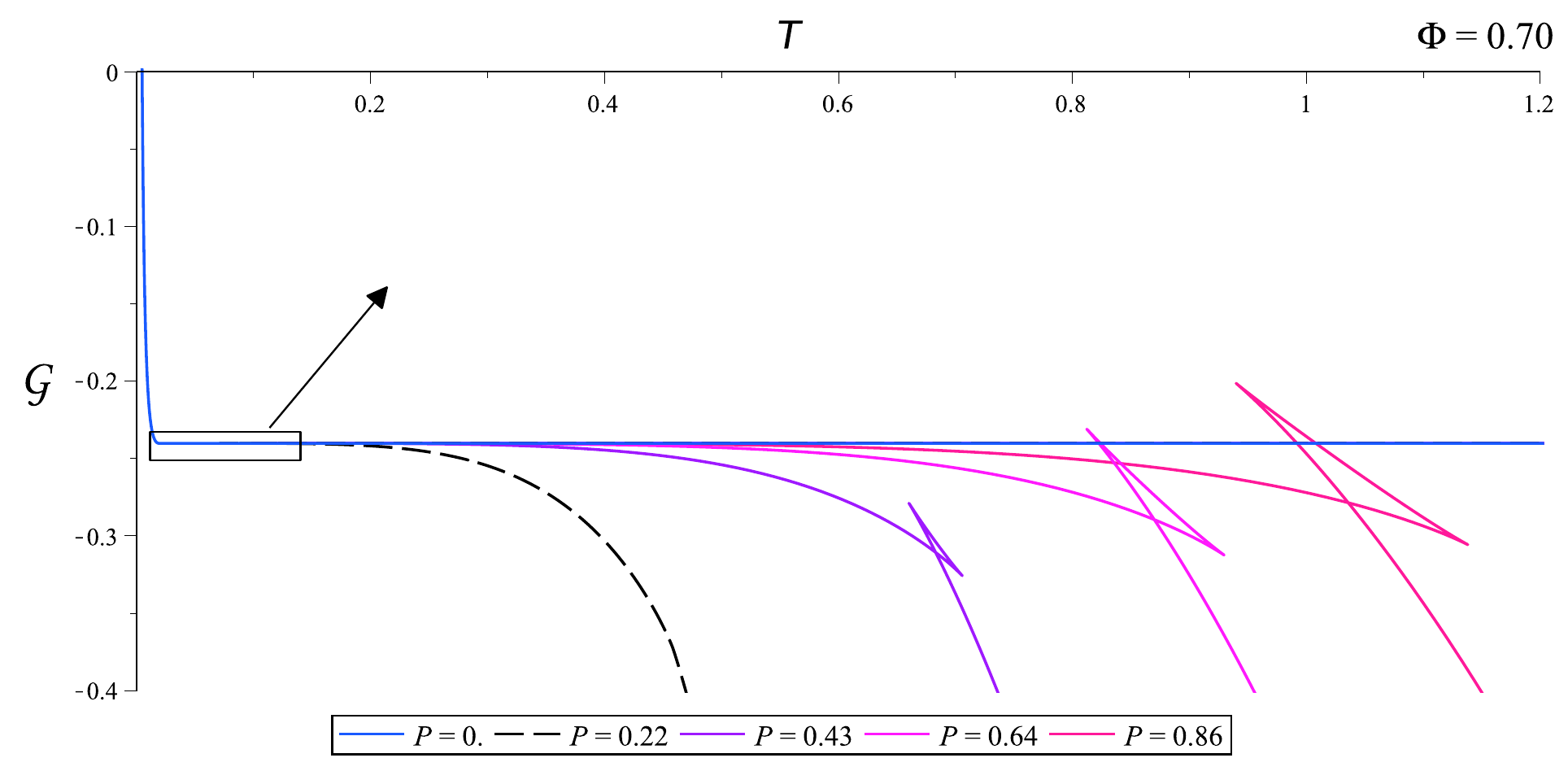}\put(25,24) {\includegraphics[scale=0.24]{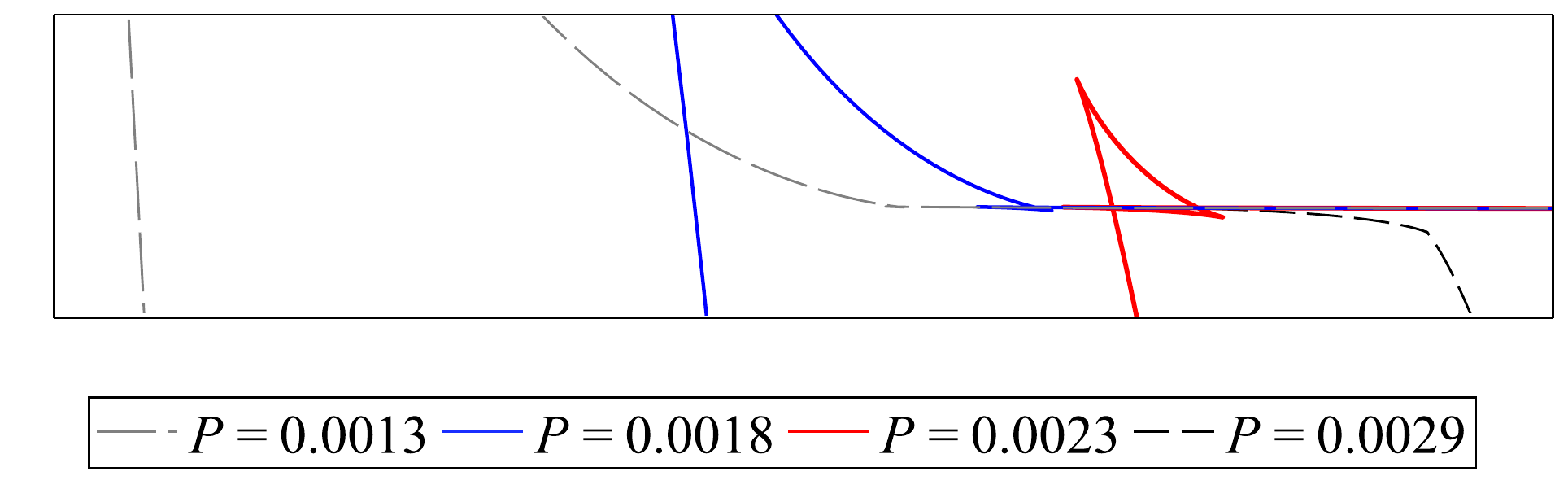}}
	\end{overpic} \\
	\includegraphics[scale=0.24]{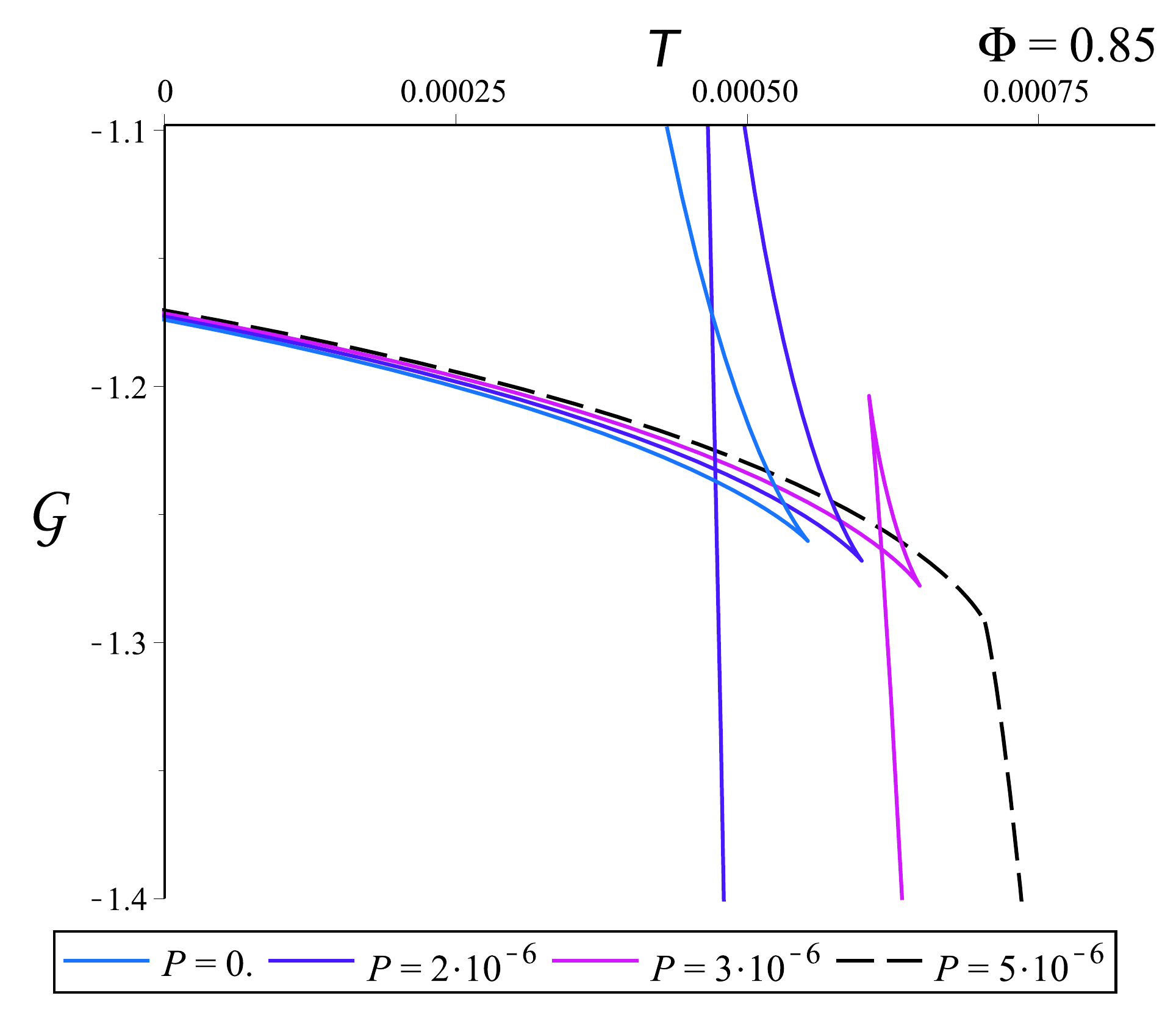}
	\includegraphics[scale=0.24]{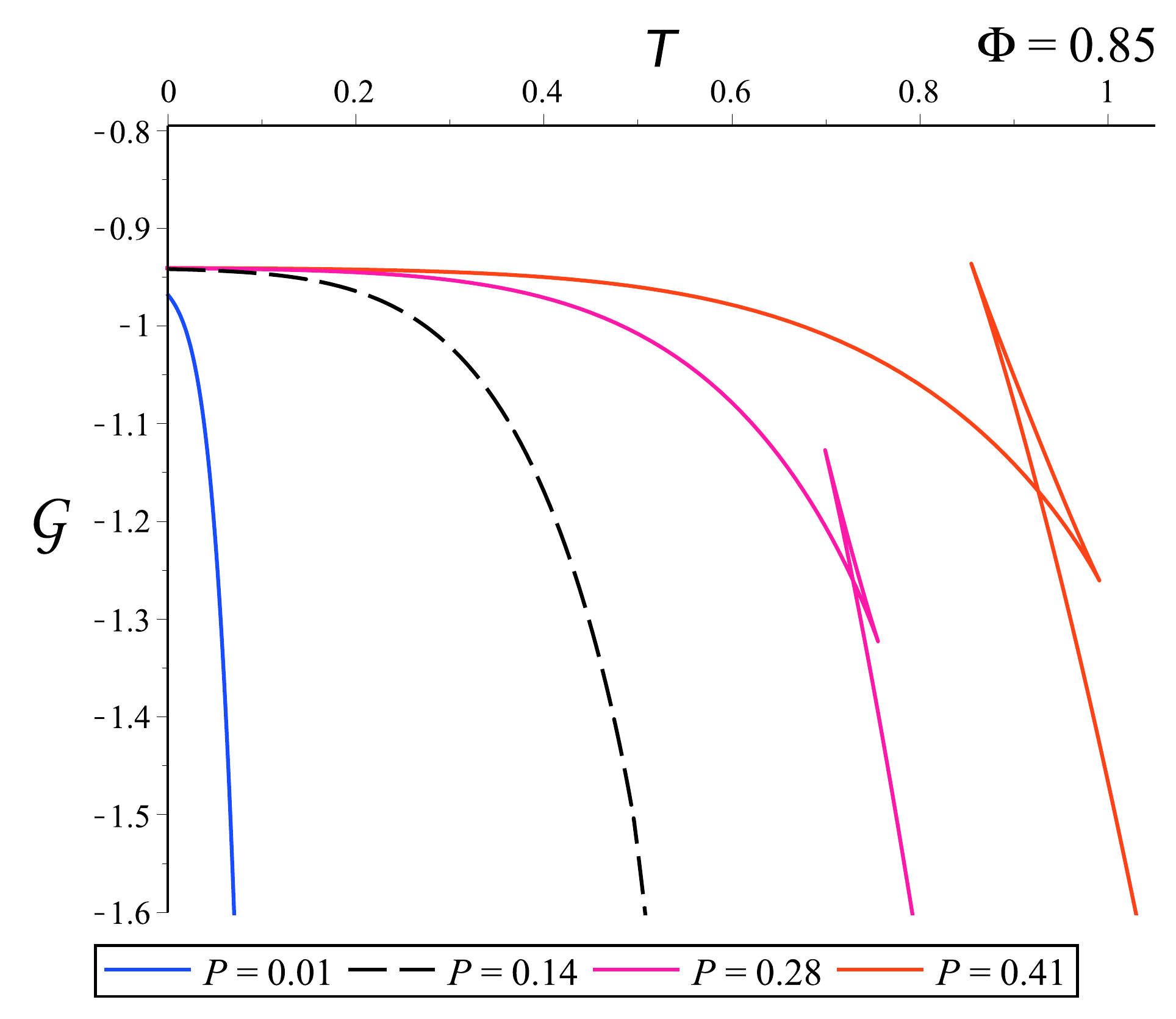}
	\includegraphics[scale=0.24]{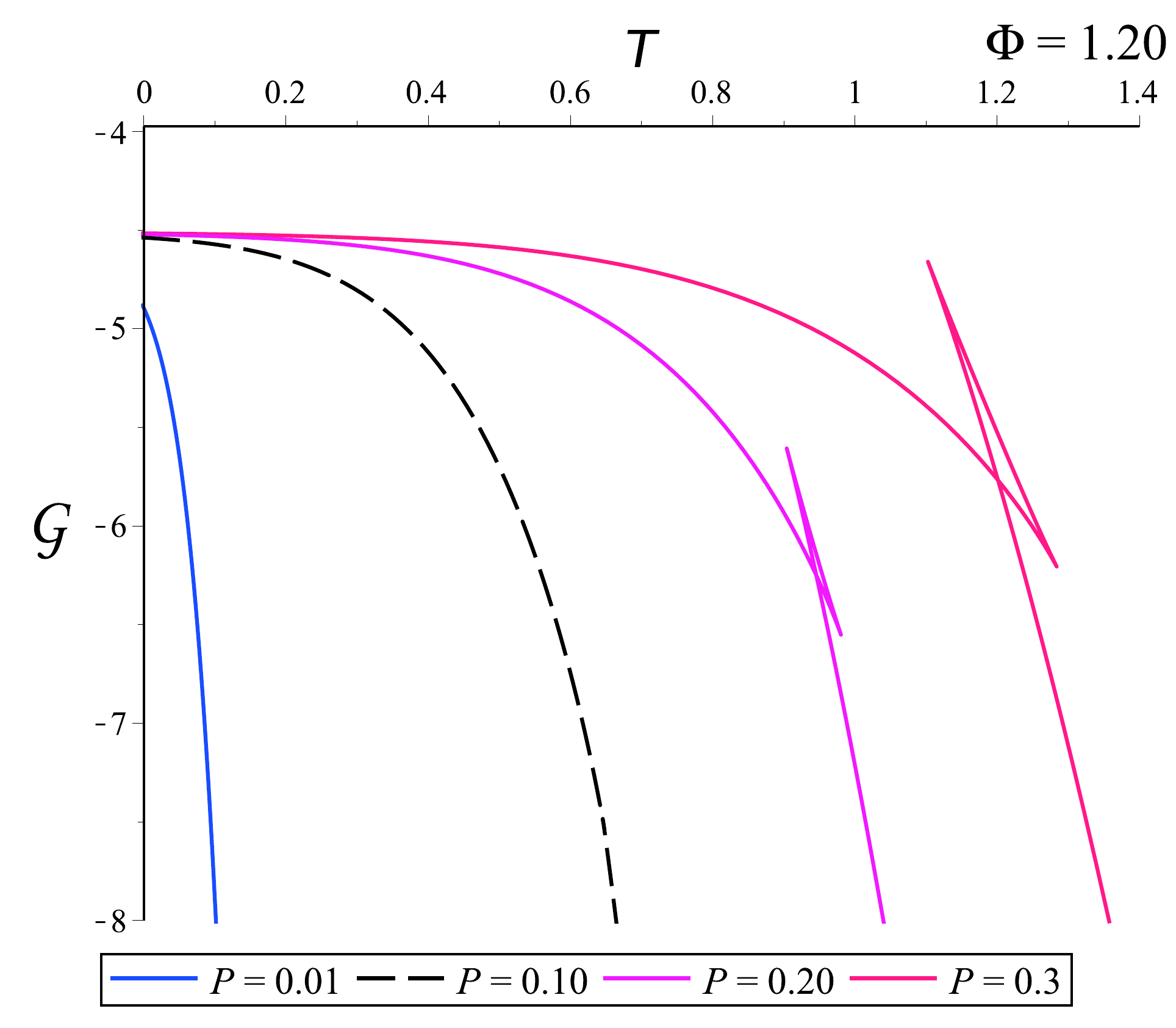}
	\caption{$\mathcal{G}$ vs $T$ diagram for five characteristic values of $\Phi$, in the model $\sigma=3$. The panels show the cases: $\Phi=0.50<\Phi_c$, $\Phi_c<\Phi=0.60<\Phi_0$, $\Phi_0<\Phi=0.70<1/\sqrt{2}$, $1/\sqrt{2}<\Phi=0.85<1$, $\Phi>1$, respectively. Dashed lines correspond to critical behaviour. These diagrams are consistent with the equation of state depicted in Fig. \ref{F8}}  
	\label{F9}
\end{figure}

\begin{figure}[t!]
\centering
\includegraphics[scale=0.36]{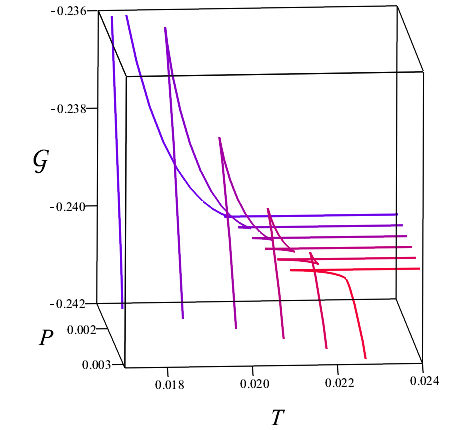}\qquad
\includegraphics[scale=0.30]{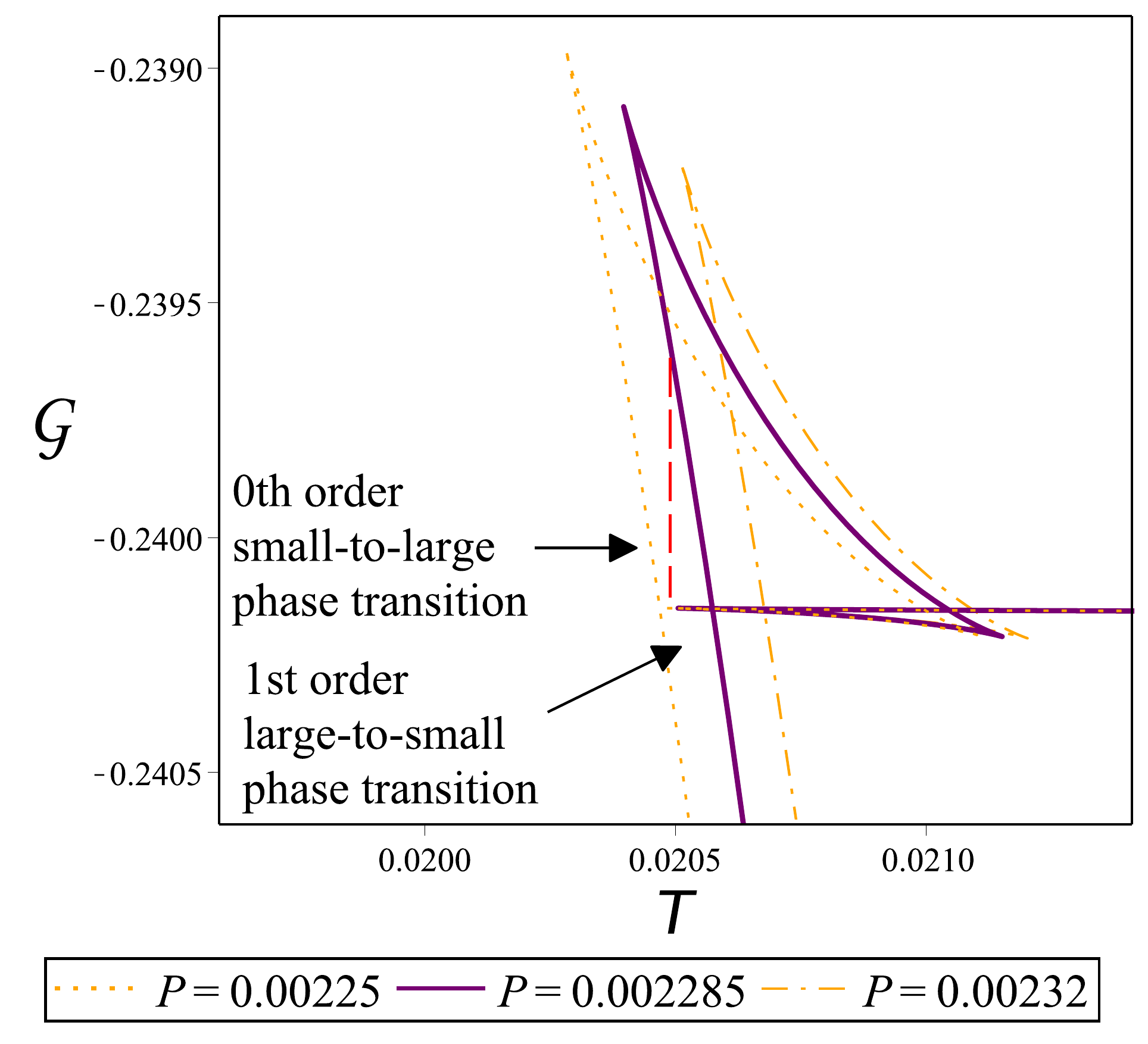}
\caption{$\mathcal{G}-T$ for the special case $\Phi_c<\Phi<\Phi_0$. We take $\Phi=0.70$ in the model $\sigma=3$. For this case, there are three critical isobars. Here we show only the reentrant phase behaviour. As pressure increases from small values,  an inverted swallowtail  appears (at the first critical point). Then the branch of large stable black holes intersects the inverted swallowtail and a second swallowtail (a standard one) forms. This is when a reentrant phase transition takes place, as detailed in the second panel. The standard swallowtail shrinks to zero at the second critical point.}  
\label{F10}
\end{figure}

In order to elucidate the nature of the critical points, we study the thermodynamic potential vs temperature. In parametric form we have 
\begin{equation}
\mathcal{G}=
\frac{1}{12{\eta}^{3}}
-\frac{\sigma} {4\eta(x_+^\sigma-1)^2}\[x_+^{2\sigma}\(1-\frac{2\sigma\Phi^2}{\sigma-1}\)-1\]
\end{equation}
\begin{equation}
T=\frac{x_+}{4\pi\eta\sigma} \[-\frac{(x_+^\sigma-1)^2} {\sigma x_+^\sigma} +\frac{2\eta^2\mathcal{X}_2\Phi^2}{\sigma-1} -\frac{\eta^2(\mathcal{X}_2-\sigma)(x_+^\sigma-1)}{\sigma x_+^\sigma}\]
\end{equation}
where
\begin{equation}
\eta=\sqrt{\frac{2\sigma^2(\sigma-1)x_+^{\sigma-2}} {(x_+^\sigma-1)\[\mathcal{X}_2-\sigma x_+^\sigma(3-4\Phi^2)\]} \[\frac{8\pi P}{3}+\frac{1}{\sigma^2-4}-\frac{x_+^2}{\sigma^2} \(1+\frac{x_+^{-\sigma}}{\sigma-2}-\frac{x_+^{\sigma}}{\sigma+2}\)\]}
\end{equation}
with $\mathcal{X}_2$ given in the line below (\ref{temper}). By studying the thermodynamic potential, we find that for almost every critical behaviour observed in $P-v$ diagrams, there is a standard swallowtail typical for large-to-small first order phase transitions between stable phases. 
However this swallowtail does not signify a standard Van der Waals transition, but rather novel phase behaviour that we shall discuss in more detail in the next subsection.
The only exception occurs for the interval $\Phi_0<\Phi<1/\sqrt{2}$, where there is  reentrant phase behaviour in addition to the novel behaviour. All the different situations are depicted in Fig.~\ref{F9} and the exceptional case for the reentrant phase behaviour within $\Phi_0<\Phi<1/\sqrt{2}$ is depicted in Fig.~\ref{F10}. In this case, there is a reentrant phase transition in the sense that there is a range of (fixed values of) $P$ for which the system can go from large to small to large black hole in the direction of decreasing temperature.

\begin{figure}[t!]
\centering
\includegraphics[scale=0.47]{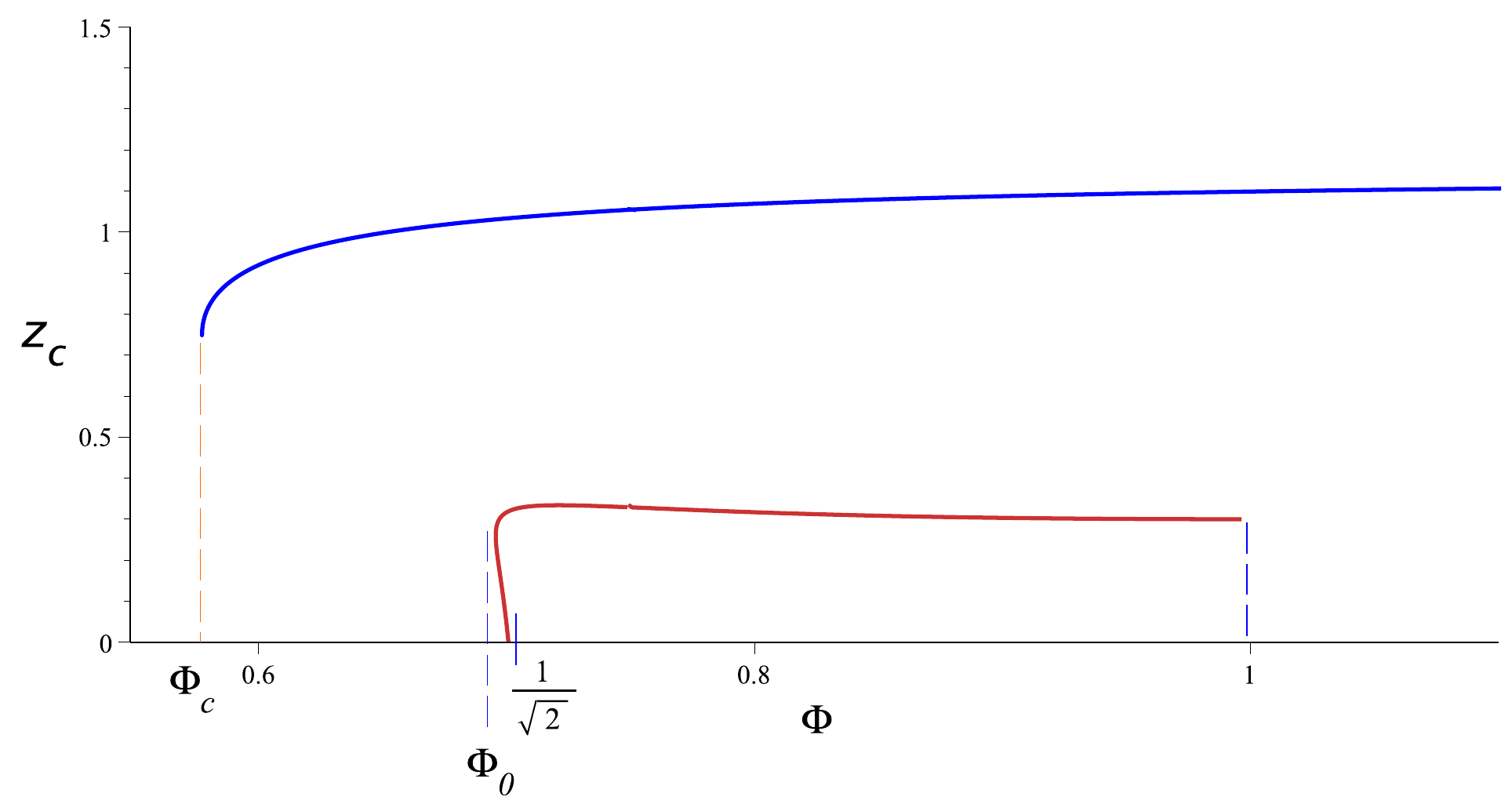}\quad
\begin{overpic}[scale=0.20]{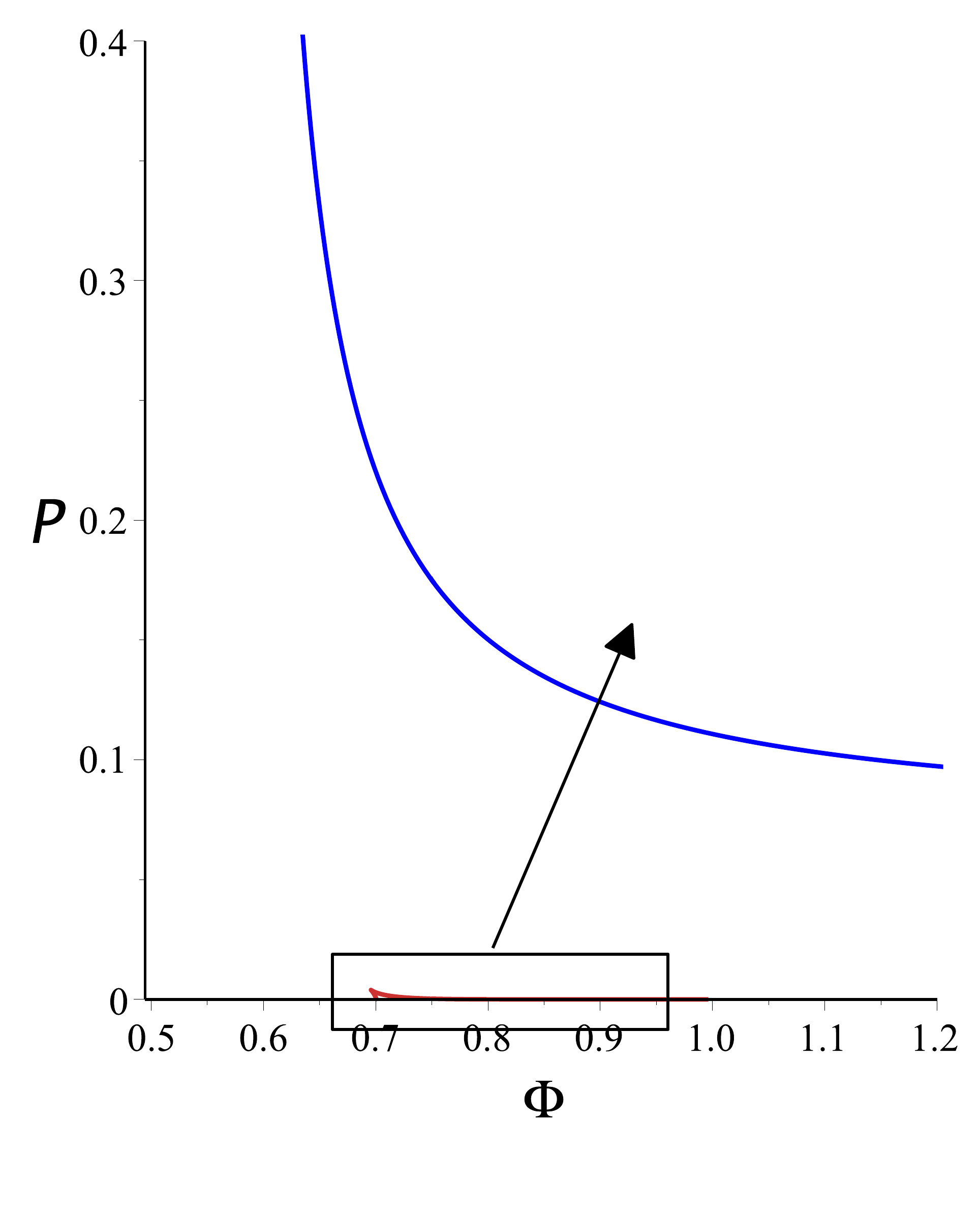}
\put(48,48) {\includegraphics[scale=0.09]{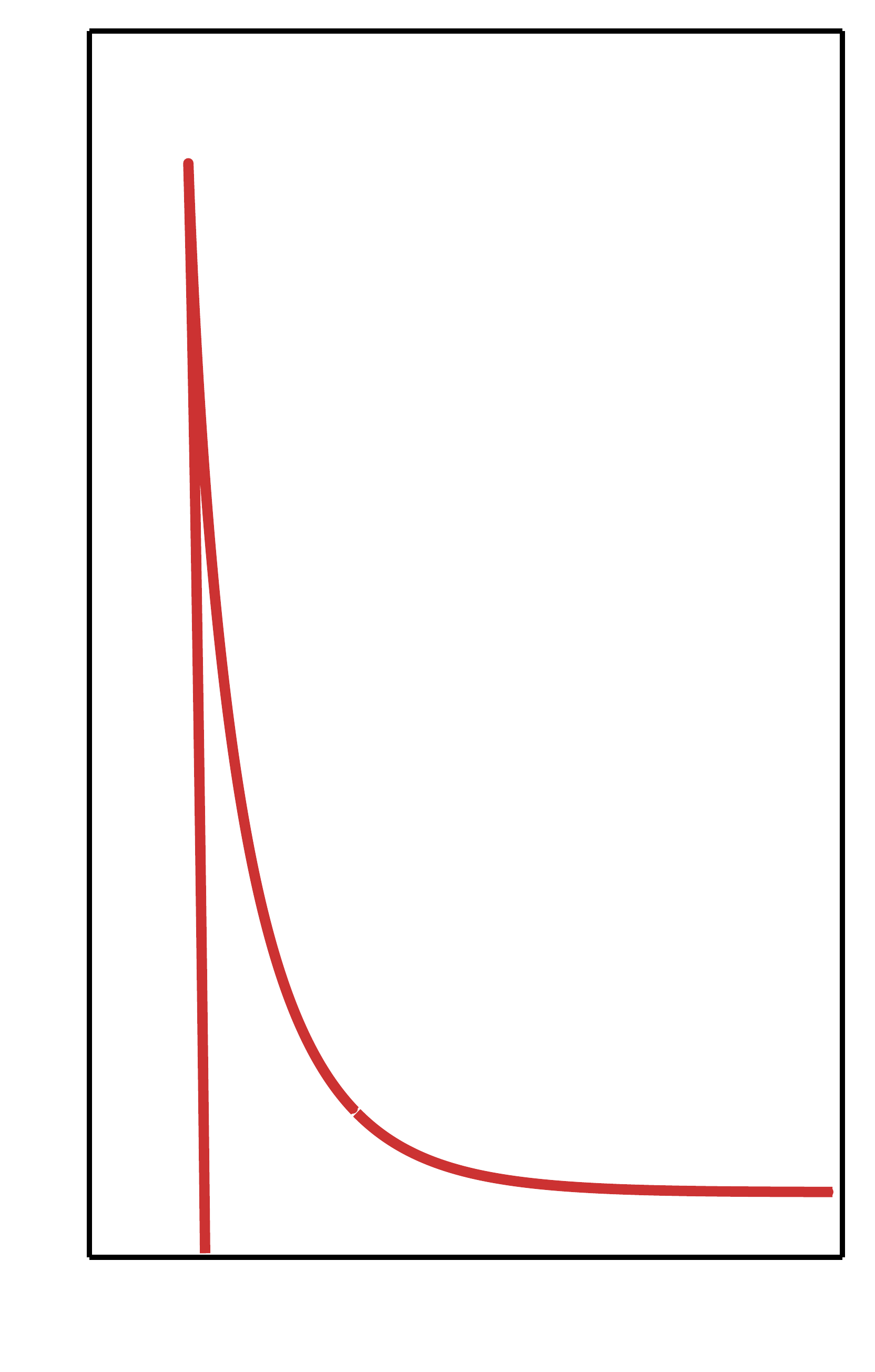}}
\end{overpic}
\caption{Left-hand panel: Critical compressibility factor $z_c$ vs $\Phi$, for $\sigma=3$, and $\Phi_c\approx 0.5774$, $\Phi_0\approx 0.6957$. Right-hand panel: Critical pressures vs $\Phi$.}
\label{F11}
\end{figure}

One (dimensionless) quantity that provides information about the criticality properties of the system is the `critical compressibility factor', defined as
\begin{equation}
z_c\equiv\frac{P_cv_c}{T_c}
\end{equation}
For  both a Van der Waals fluid and the RN-$AdS$ black hole, $z_c$ turns out to be exactly $3/8$. Unlike the critical exponents, which are expected to be universal, $z_c$ is known to differ from one substance to another in ordinary chemistry \cite{Ryuzo,Kulinskii,Herschbach}.
In this case, the critical compressibility factor depends on $\Phi$, as depicted in Fig.~\ref{F11}, for $\sigma=3$. The dependence of $z_c$ on $\Phi$ indicates that the conjugate potential plays the role of a fluid parameter that characterizes the nature the dual conformal field theory. Also in Fig.~\ref{F11}, we have depicted the critical pressures for the whole range of $\Phi$. Notice that in the limit $\Phi\rightarrow \Phi_c^{+}$, $P\rightarrow\infty$, and in the other hand, in the limit $\Phi\rightarrow\infty$, $P\rightarrow 0.0768$.

\subsection{Novel Phase Transitions}

Several of the phase transitions that appear in both the canonical and grand canonical ensembles have  rather unusual features.  First, the free-energy diagrams exhibit swallowtails whose size increases with increasing pressure, as is clear from the rightmost diagram in Fig.~\ref{F1},  the two rightmost diagrams in Fig.~\ref{F3},  and the second and fourth diagrams in Fig.~\ref{F9}.  This kind of phenomenon has been seen before in Lovelock gravity  \cite{Frassino:2014pha}, and is referred to as a reverse Van der Waals phase transition. It denotes
 a situation where condensation of large black holes into small ones takes place at increasingly higher temperatures and pressures above a critical point, instead decreasing values of these quantities, as holds for the standard case \cite{Kubiznak:2016qmn};
an example for the class of hairy black holes we are considering is shown in the middle diagram in Fig.~\ref{F5}, with the re-entrant behaviour shown in
the right panel for Fig.~\ref{F6}.

However the $P-v$ diagrams that correspond to the swallowtails  in the third diagram in Fig.~\ref{F1}, the two rightmost diagrams in 
Fig.~\ref{F2}, and all but the two leftmost diagrams in Fig.~\ref{F8} indicate that something quite different is going on in these cases. 
The phase transition for these novel cases takes place at the temperature $T_{c2}$. There is a `subcritical temperature' $T_{sc}(<T_{c2})$ at which
\begin{equation}
\(\frac{\pa v}{\pa P}\)_{T_{sc}}=0\,, \qquad \(\frac{\pa^2 v}{\pa P^2}\)_{T_{sc}}=0
\end{equation}
For $T\leq T_{sc}$, $P$ is a single-valued function of $v$, while for $T>T_{sc}$ there is a region in $v$ where $P$ is triple-valued.

For concreteness, we focus on the situation $1/\sqrt{2}<\Phi<1$ in the grand canonical ensemble, 
 for which there are two critical points. We fix $\Phi=0.85$ and $\sigma=3$. The critical points are
\begin{equation}
c1: \quad (P=5.13\cdot 10^{-6},v=42.67,T=7.09\cdot 10^{-4})\,, \qquad c2: \quad (P=0.1347,v=3.8361,T=0.4789)
\end{equation} 
This case is interesting because the critical point $c1$ taking place at the lower pressure is similar to the Van der Waals one, whereas the second critical point, $c2$, at higher pressure, has no analogue with standard thermodynamic systems. In Fig.~\ref{F11} we   depict in detail the critical isotherms for this case. Near the critical point $c1$ we observe standard Van der Waals behaviour, but around the critical point $c2$, the pressure is multi-valued: for a given $(T,v)$, there are at most three possible values of $P$, as is clear from the right diagram in Fig.~\ref{F11}. 
\begin{figure}[t!]
\centering
\includegraphics[scale=0.38]{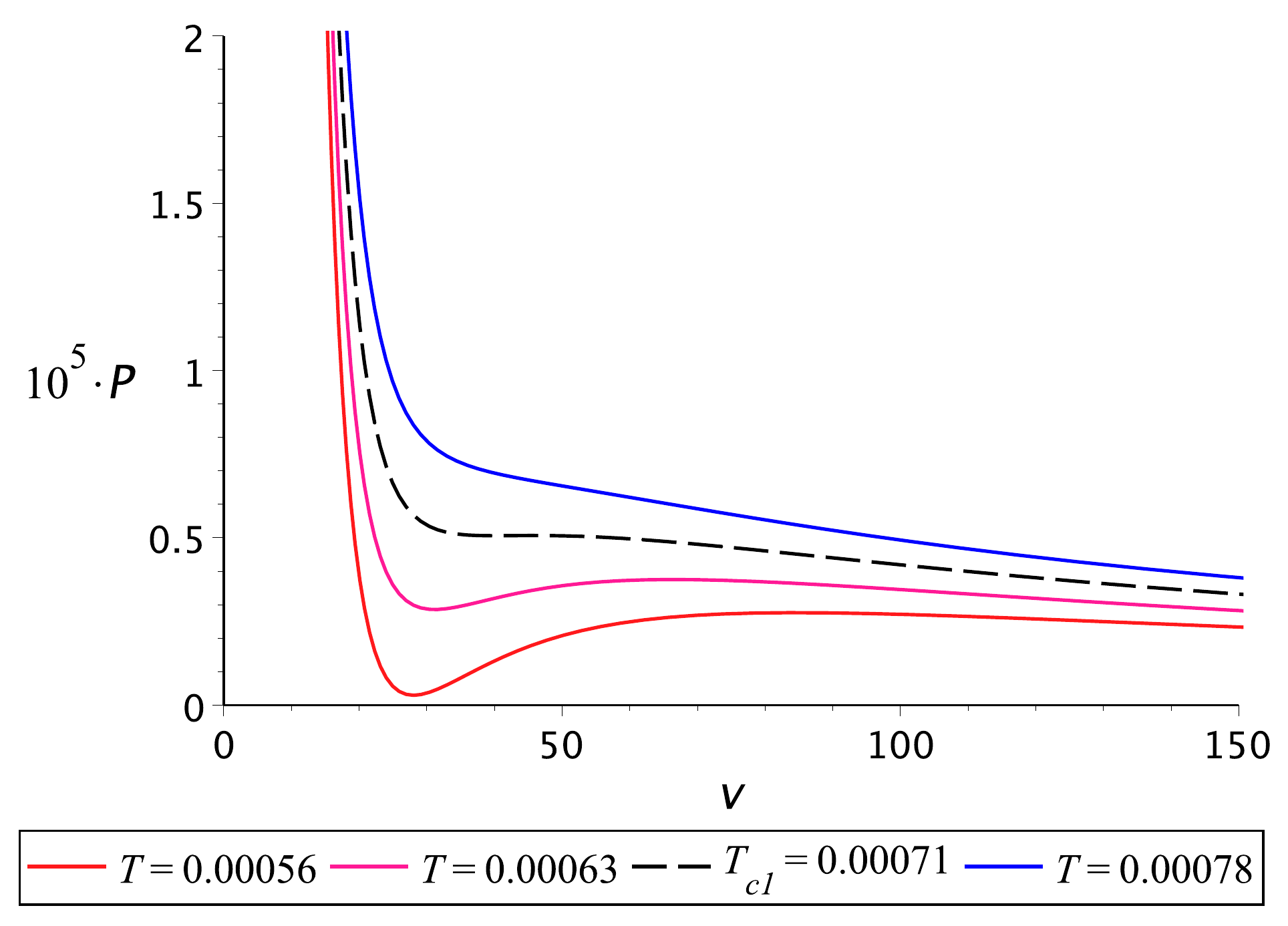}\qquad
\begin{overpic}[scale=0.38]{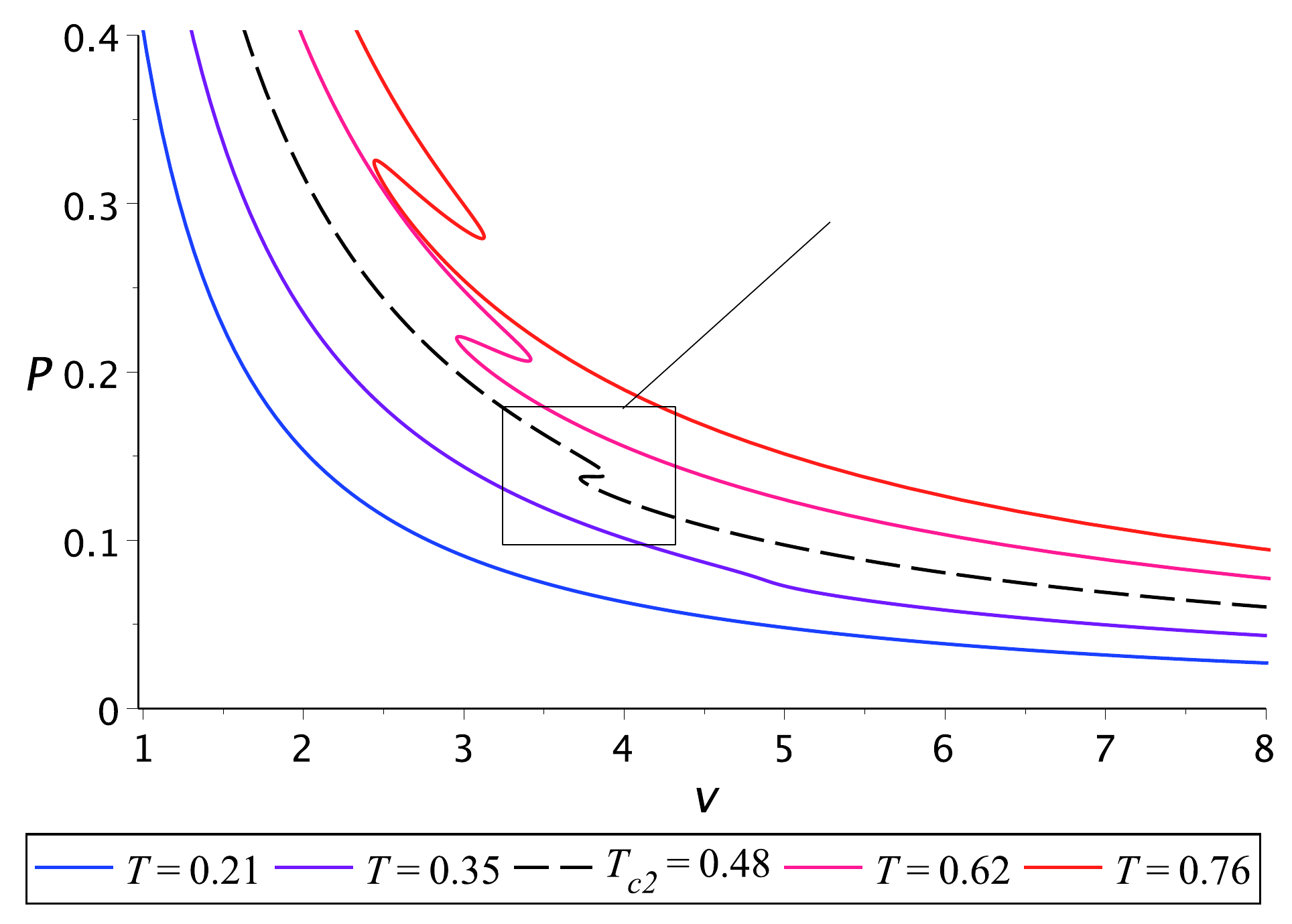}
\put(60,42){\includegraphics[scale=0.14]{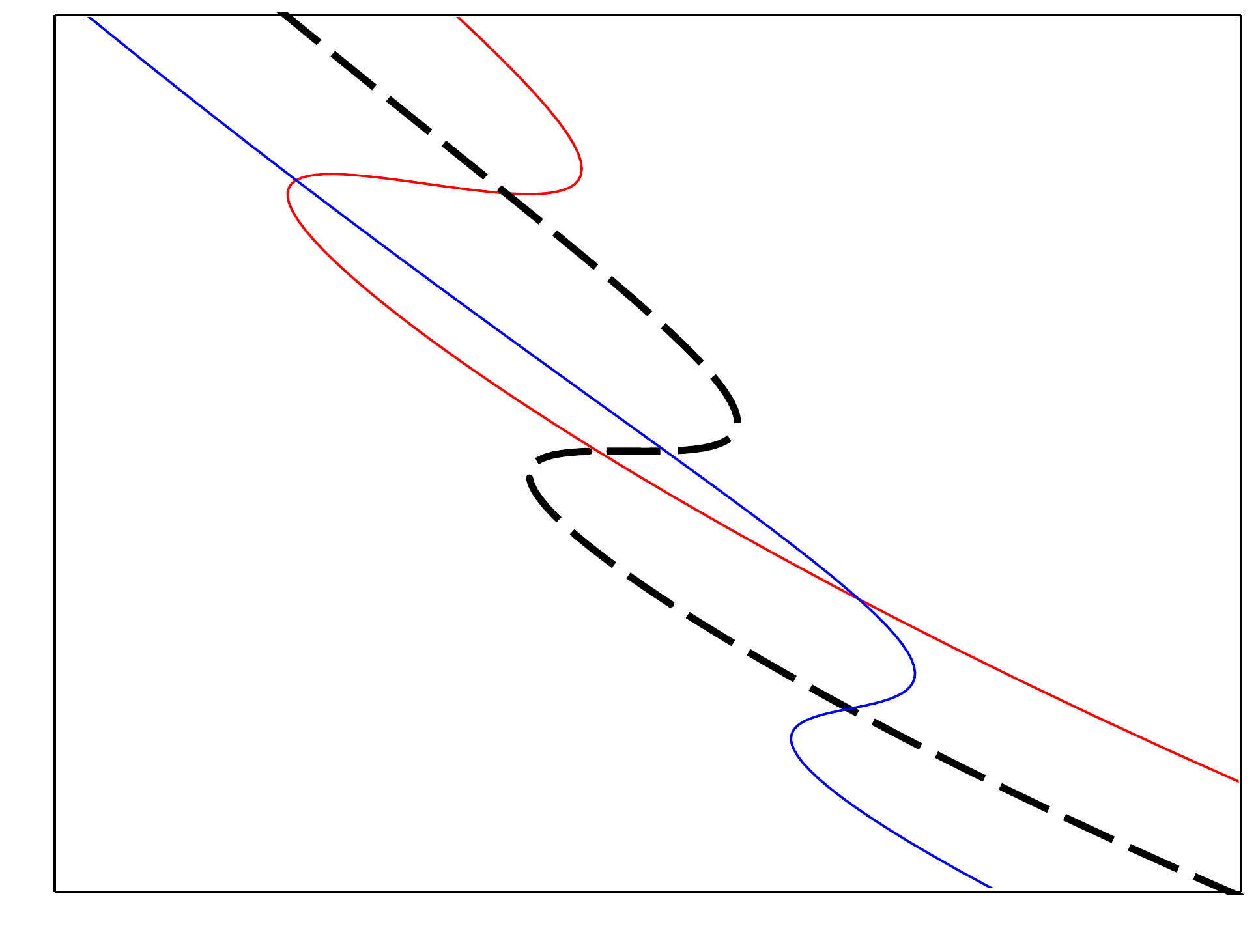}}
\end{overpic}
\caption{Equation of state for $\Phi=0.85$ ($\sigma=3$). There are two critical isotherms (black dashed curves),
one corresponding to the critical point $c1$ (left), the other to the critical point $c2$ (right).}
\label{F12}
\end{figure}

\begin{figure}[t!]
\centering
\includegraphics[scale=0.28]{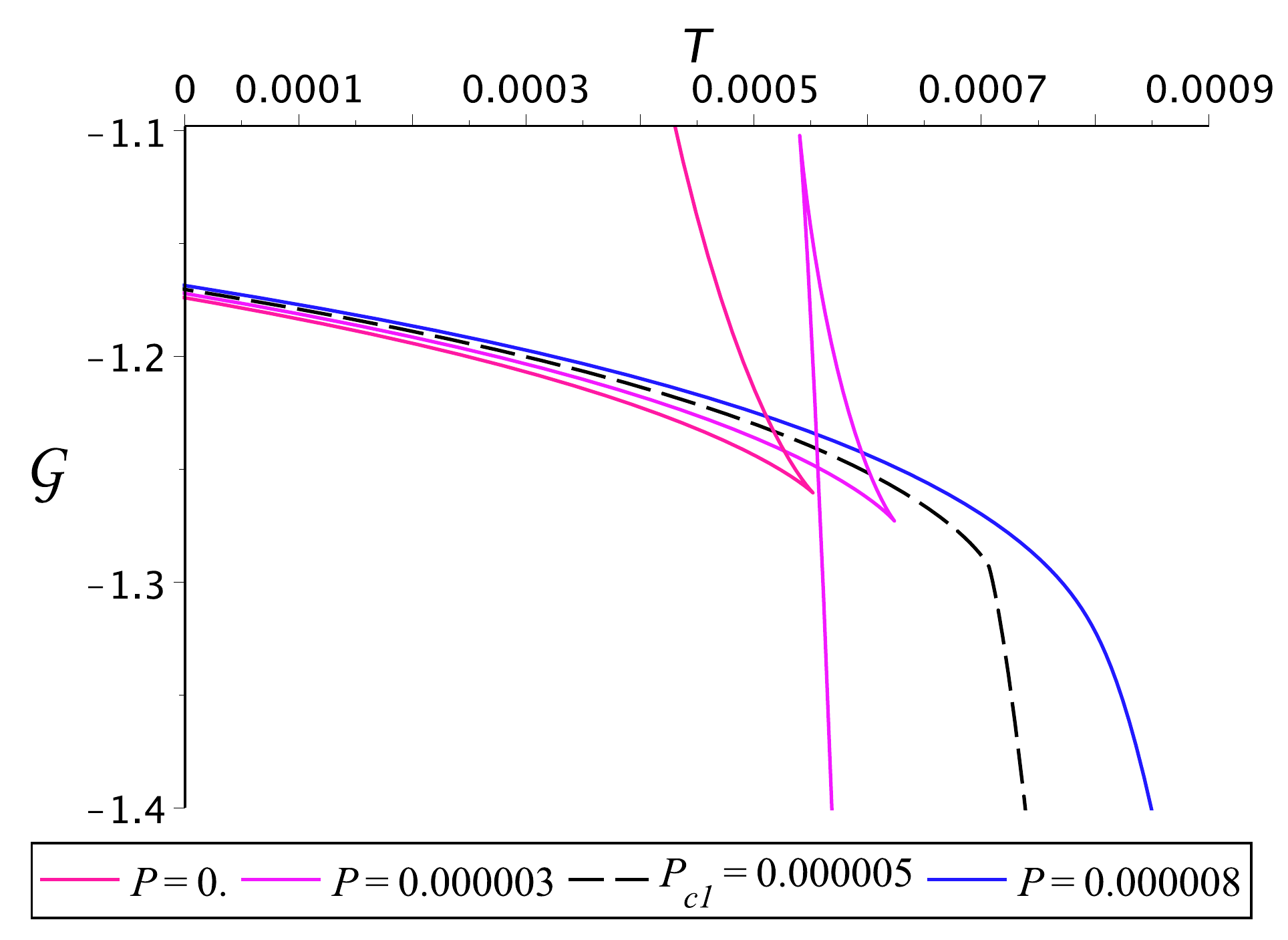}
\includegraphics[scale=0.28]{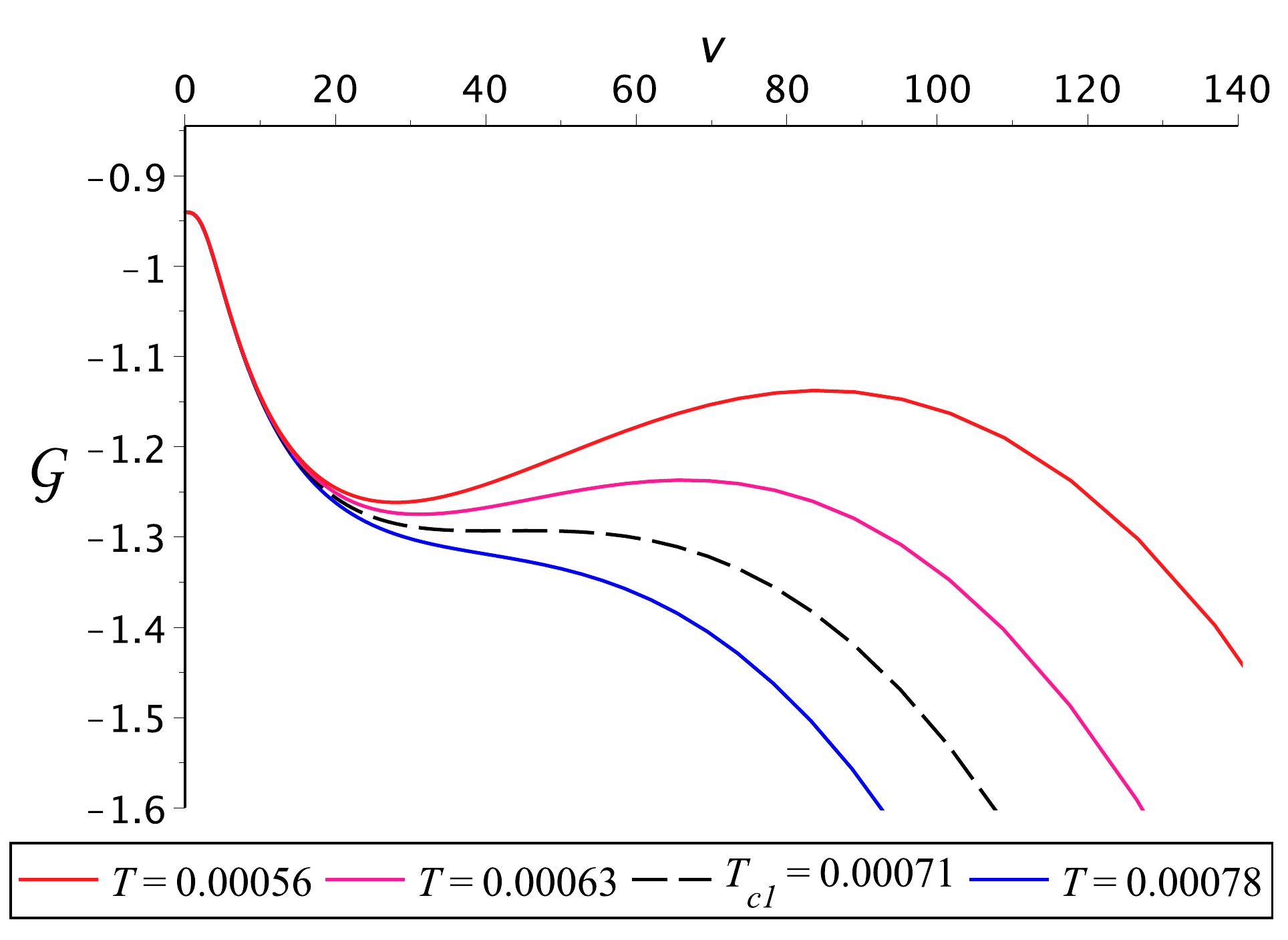}
\includegraphics[scale=0.28]{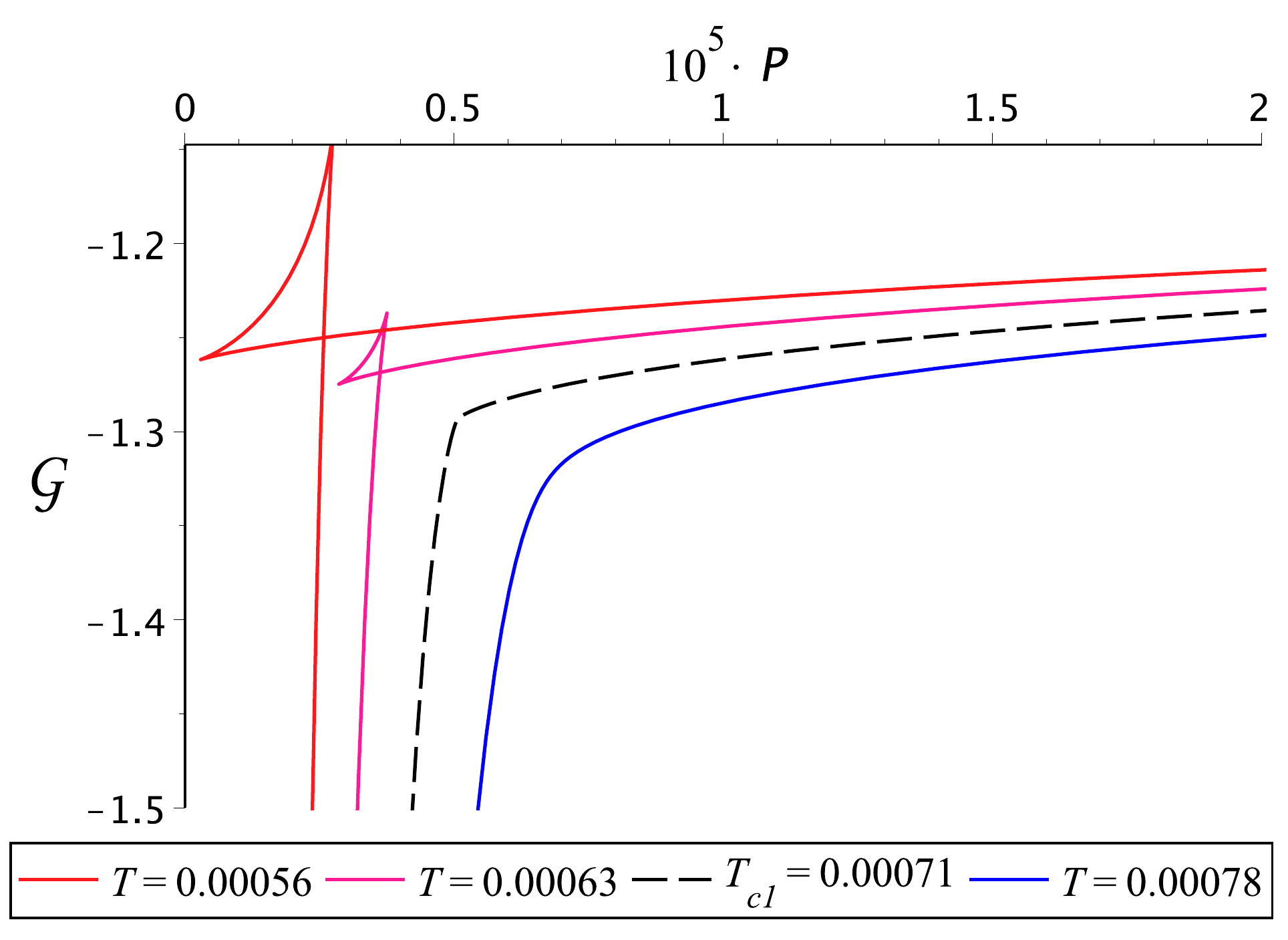}
\caption{$\mathcal{G}-T$, $\mathcal{G}-v$ and $\mathcal{G}-P$ diagrams for the first critical point. The (first order) phase transition is discontinuous in $S=-(\pa\mathcal{G}/\pa{T})_{P}$ and $V=(\pa\mathcal{G}/\pa{P})_{T}$.}
\label{F13}
\end{figure}

In order to understand the distinction between these two kinds of phase transitions, let us begin by analyzing the phase transition for 
the more familiar  critical point $c1$. Fig.~\ref{F13} depicts the $\mathcal{G}-T$, $\mathcal{G}-v$ and $\mathcal{G}-P$ diagrams\footnote{In the $\mathcal{G}-v$ diagram --with $T$ kept fixed-- the issue is that pressure is not held fixed along a given isotherm, but this is precisely the case for black hole chemistry.}.  From $\mathcal{G}-T$, we know that the large-to-small first order black hole phase transition is discontinuous in the entropy $S=-(\pa\mathcal{G}/\pa{T})_{P}$  in the direction of decreasing temperature. From the $\mathcal{G}-P$ diagram, we know that it is discontinuous in $V=(\pa\mathcal{G}/\pa{P})_{T}$ in the direction of increasing pressure, which is quite intuitive.  Essentially, for an ensemble of large black holes, as the thermodynamic volume decreases the pressure increases, until a point is reached where `condensation' begins, with the large black holes changing into small ones with no further increase in pressure or free energy, as is clear from
the central panel in Fig.~\ref{F13}.  The pressure at which this occurs is given by Maxwell's equal area law. As volume is further decreased, more and more large black holes in the ensemble will condense into small ones, until the entire ensemble consists of small black holes. As the volume decreases even more, the pressure significantly increases, since further condensation is impossible.  The situation is fully analogous to a gas condensing into a liquid at a given temperature as the volume of the system decreases.

\begin{figure}[t!]
	\centering
	\includegraphics[scale=0.28]{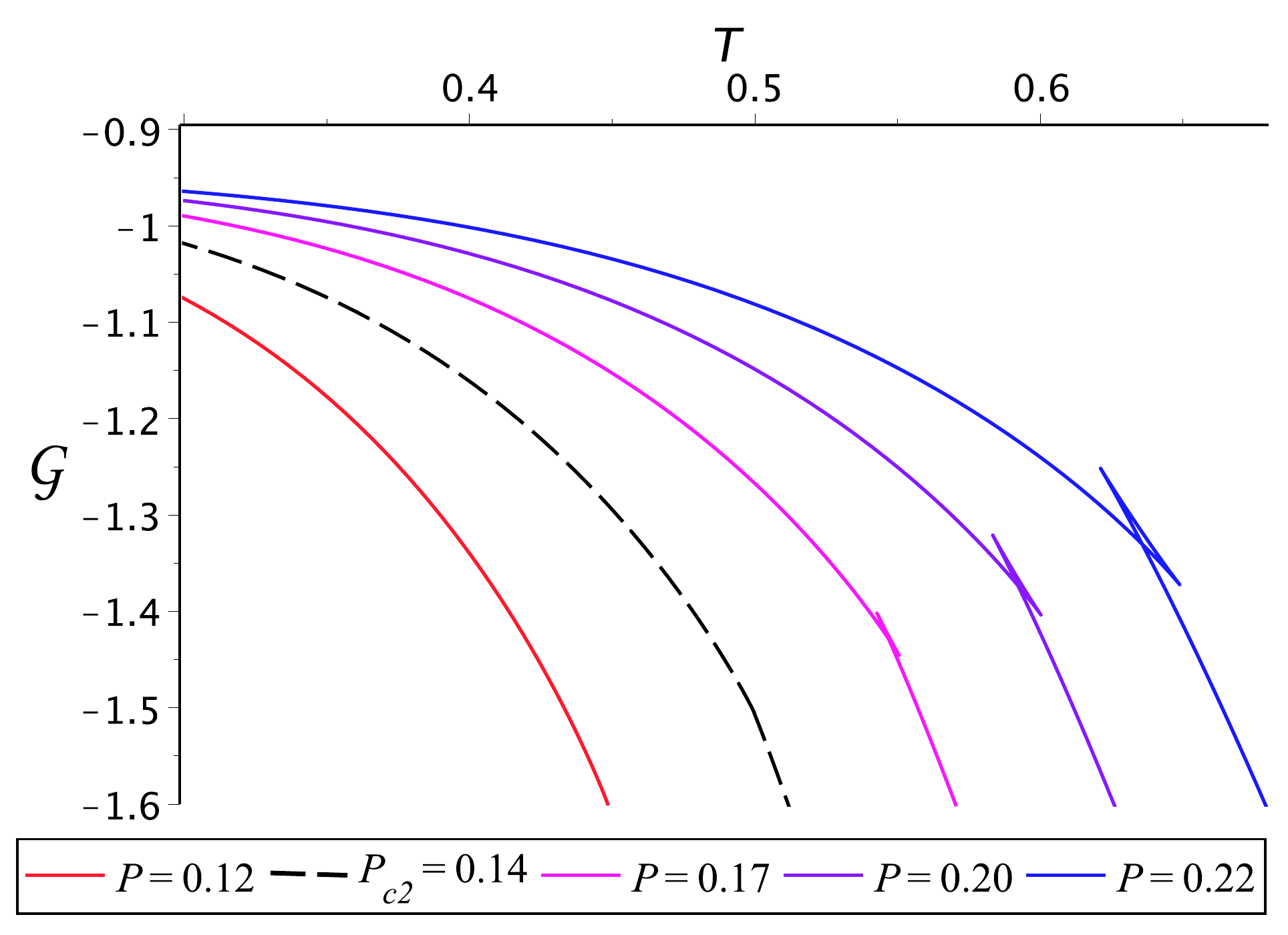}
	\includegraphics[scale=0.28]{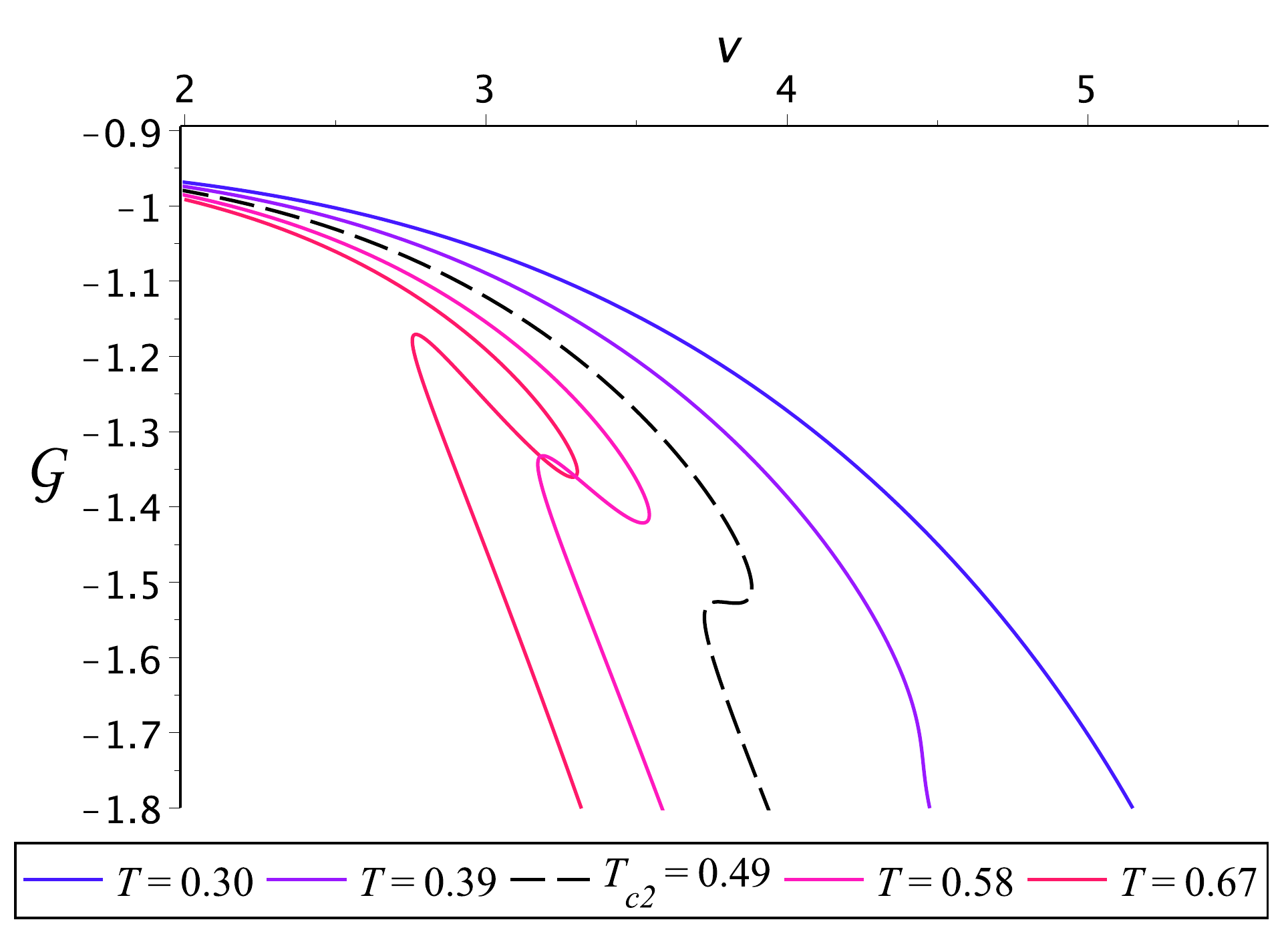}
	\includegraphics[scale=0.28]{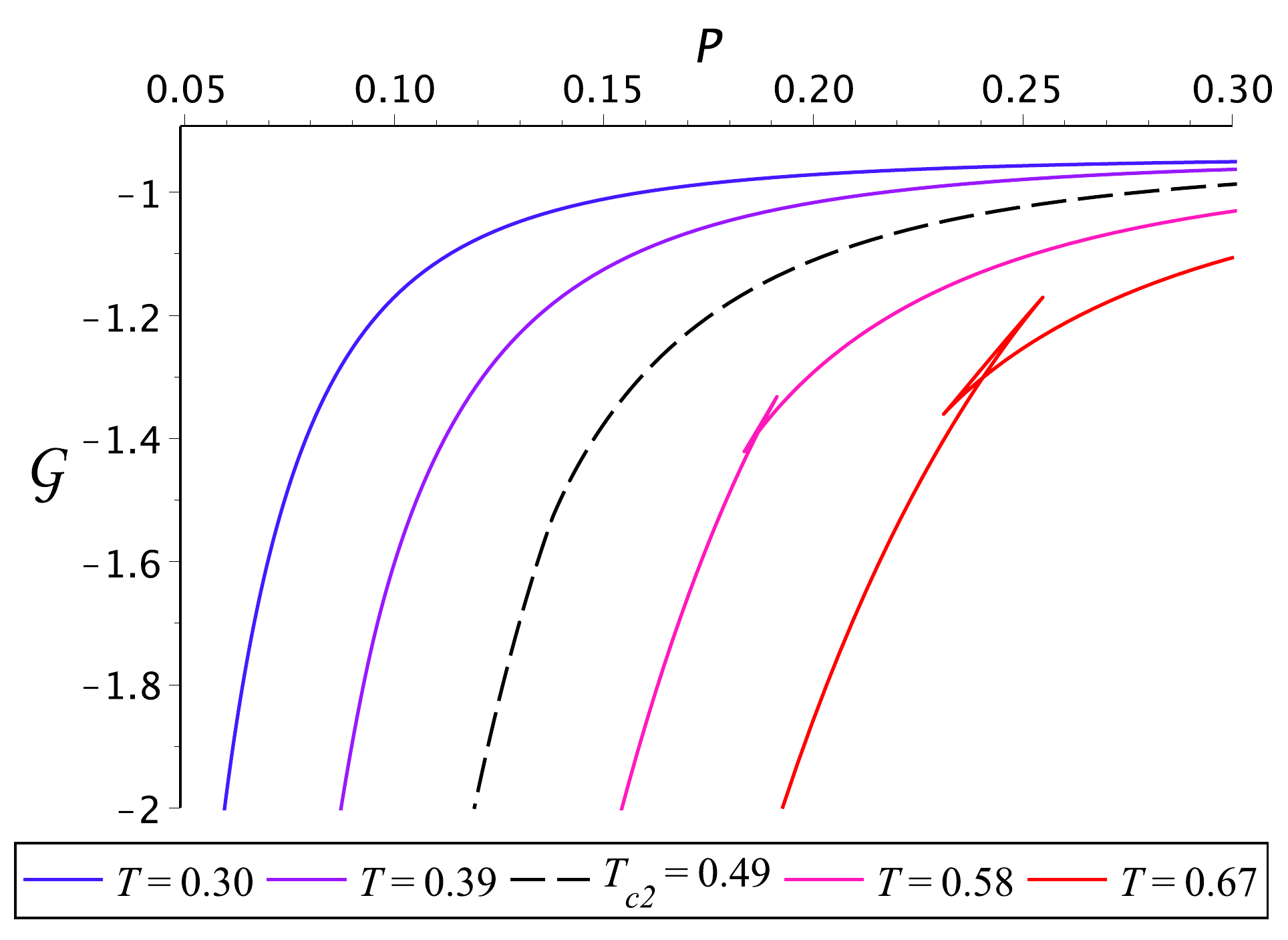}
	\caption{$\mathcal{G}-T$, $\mathcal{G}-v$ and $\mathcal{G}-P$ diagrams for the second critical point. The first order phase transition is discontinuous in $S=-(\pa\mathcal{G}/\pa{T})_{P}$ and $V=(\pa\mathcal{G}/\pa{P})_{T}$, like in the previous case.}
	\label{F14}
\end{figure}

Now, consider the second critical point, for which the corresponding diagrams are depicted in Fig.~\ref{F14}. Here the  swallowtails in the $\mathcal{G}-T$ and $\mathcal{G}-P$ planes exhibit opposite behaviour  compared to the previous case,  growing as $P$ increases (left diagram) and as $T$ increases (right diagram).  
The central panel is most instructive -- we see that neither $\mathcal{G}(v)$ nor $v(\mathcal{G})$  are single-valued functions above
$T_{c2}$.  For these temperatures the first order transition corresponds to what appears to be a form of `reverse condensation', in which black holes of smaller specific volume
condense into black holes of larger specific volume!

\begin{figure}[t!]
	\centering
	\includegraphics[scale=0.28]{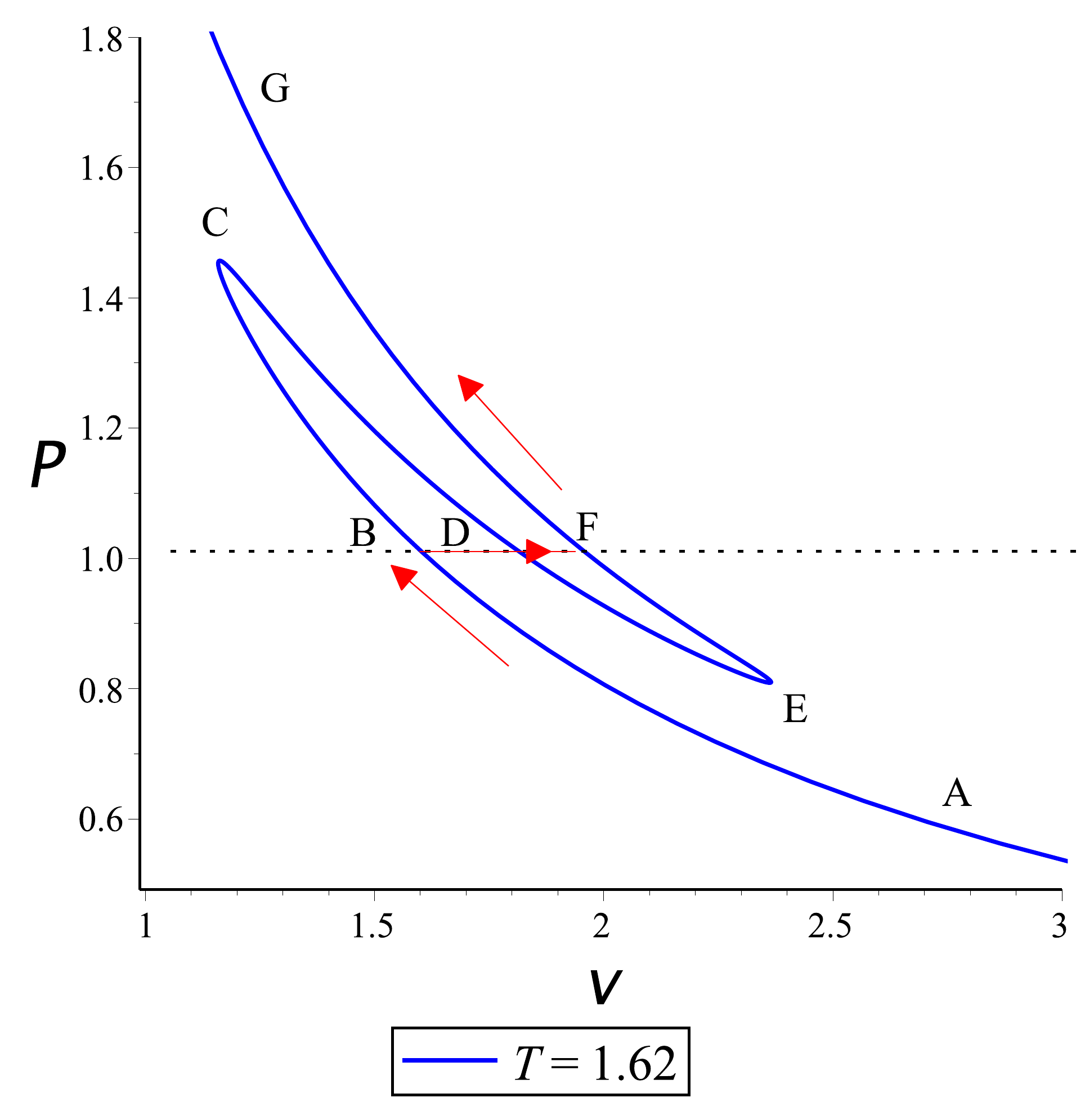}
	\includegraphics[scale=0.28]{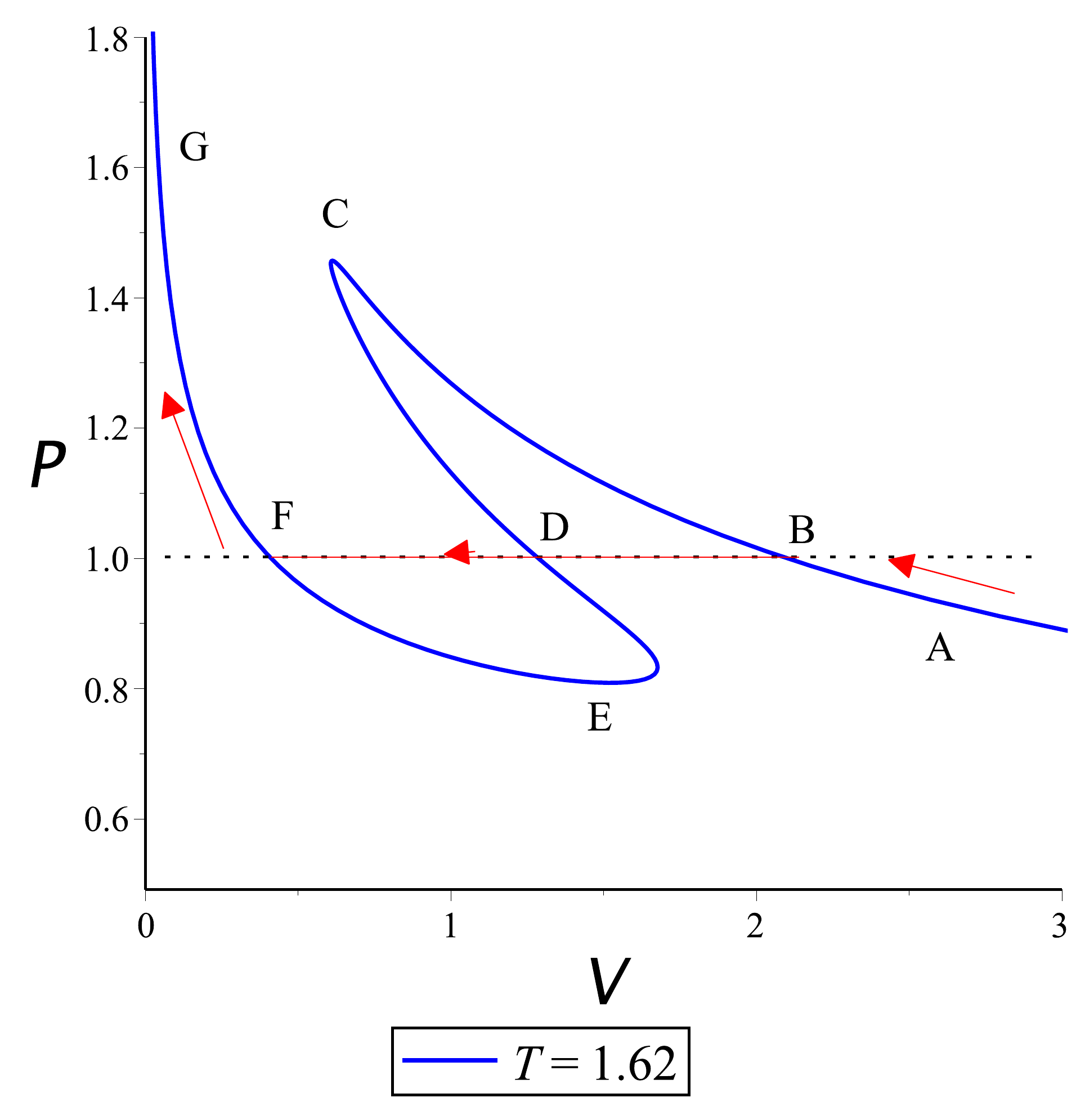}
	\includegraphics[scale=0.28]{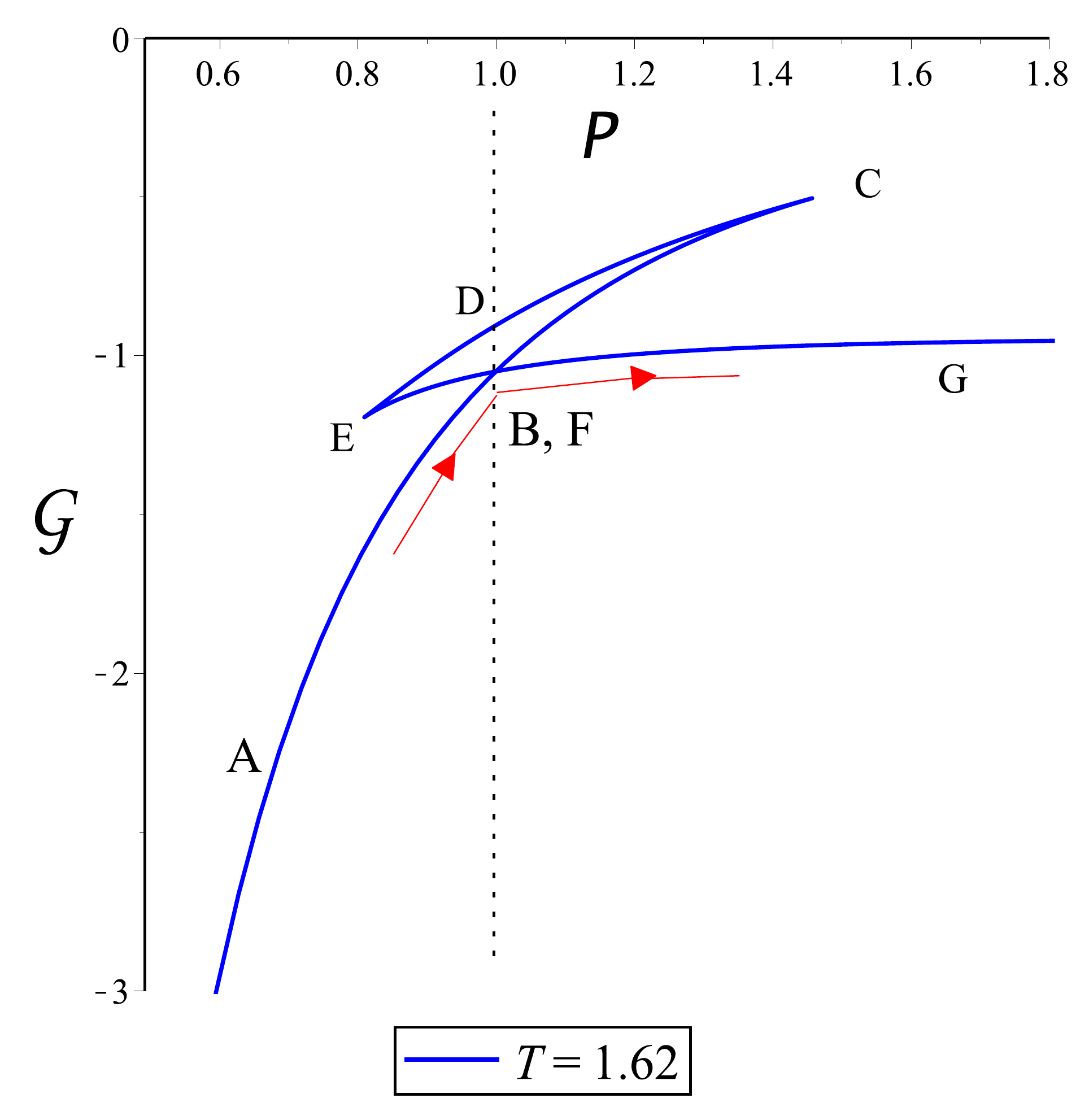}
	\caption{Equation of state $P-v$ and $P-V$, and the $\mathcal{G}-P$ diagram for $\Phi=0.85$ and $T=1.62$ in the model $\sigma=3$. The trajectory of points is compared side by side for the phase transition that occurs at $P\approx 1$. Point A corresponds to the large-$v$ and large-$V$ phase, respectively. Point B (and F) corresponds to the intersection point of the swallowtail. Points C and E correspond to the local maximum and minimum of $P(v)$ (and $P(V)$), respectively. Point G corresponds to the small-$v$ and small-$V$ phase, respectively.}
	\label{F15}
\end{figure}

We can understand this behaviour by considering the equation of state and its corresponding $\mathcal{G}-P$ diagram, both illustrated in Fig.~\ref{F15}, with $T=1.62$. For this choice of $T$, the transition takes place at $P=1$, corresponding to the intersection point of the swallowtail. 
The arrows indicate the novel transition from large-to-small specific volumes.  Consider the system at point A, corresponding to a large black hole of negative free-energy.  As the specific volume $v$ decreases, the free-energy and pressure both increase until point B is reached. At this point the system undergoes `reverse condensation', moving from B to D to F at constant $P$, with small-$v$ black holes condensing into larger-$v$ ones. This all takes place at the swallowtail crossover. 
After this, the system is at point F, corresponding to a larger value of $v$.  The equation of state then indicates that as $P$ increases, $v$ will again decrease (and the free energy will increase).   The net effect is a small-$v$-to-large-$v$ first order transition, despite the fact that increasing pressure corresponds to decreasing $v$. 
Note that points C and E correspond to the cusps in the swallowtail; the system does not actually transit through these points.

We emphasize that `reverse condensation' is a property of the specific volume $v$; the thermodynamic volume $V$ decreases at the first-order transition, as is clear from central and right panels of Fig.~\ref{F15}. Note that the equal-area law can be applied here, despite the fact that $P$ is not a single-valued function of $V$ and vice-versa.

\begin{figure}[t!]
\centering
\includegraphics[scale=0.28]{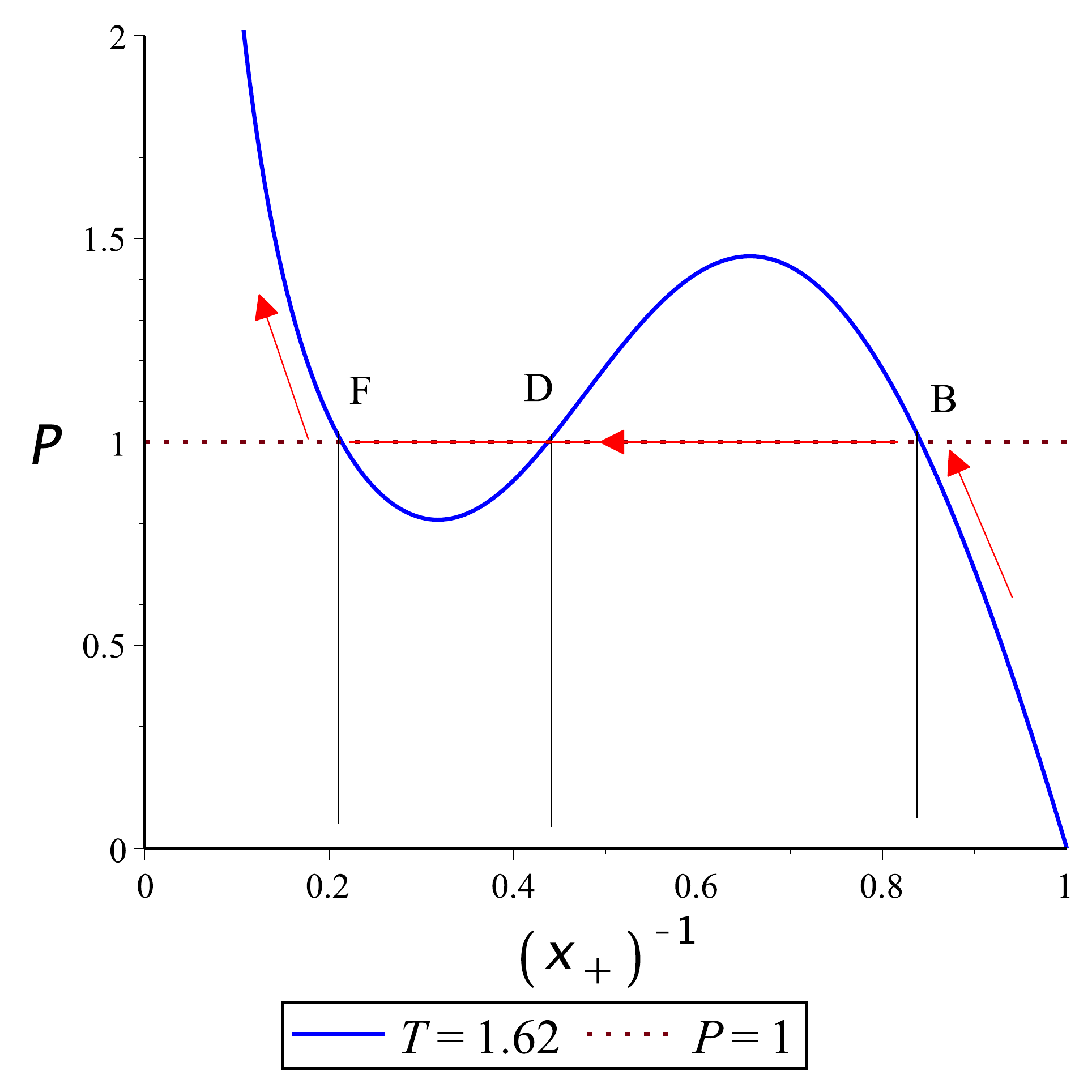}
\includegraphics[scale=0.28]{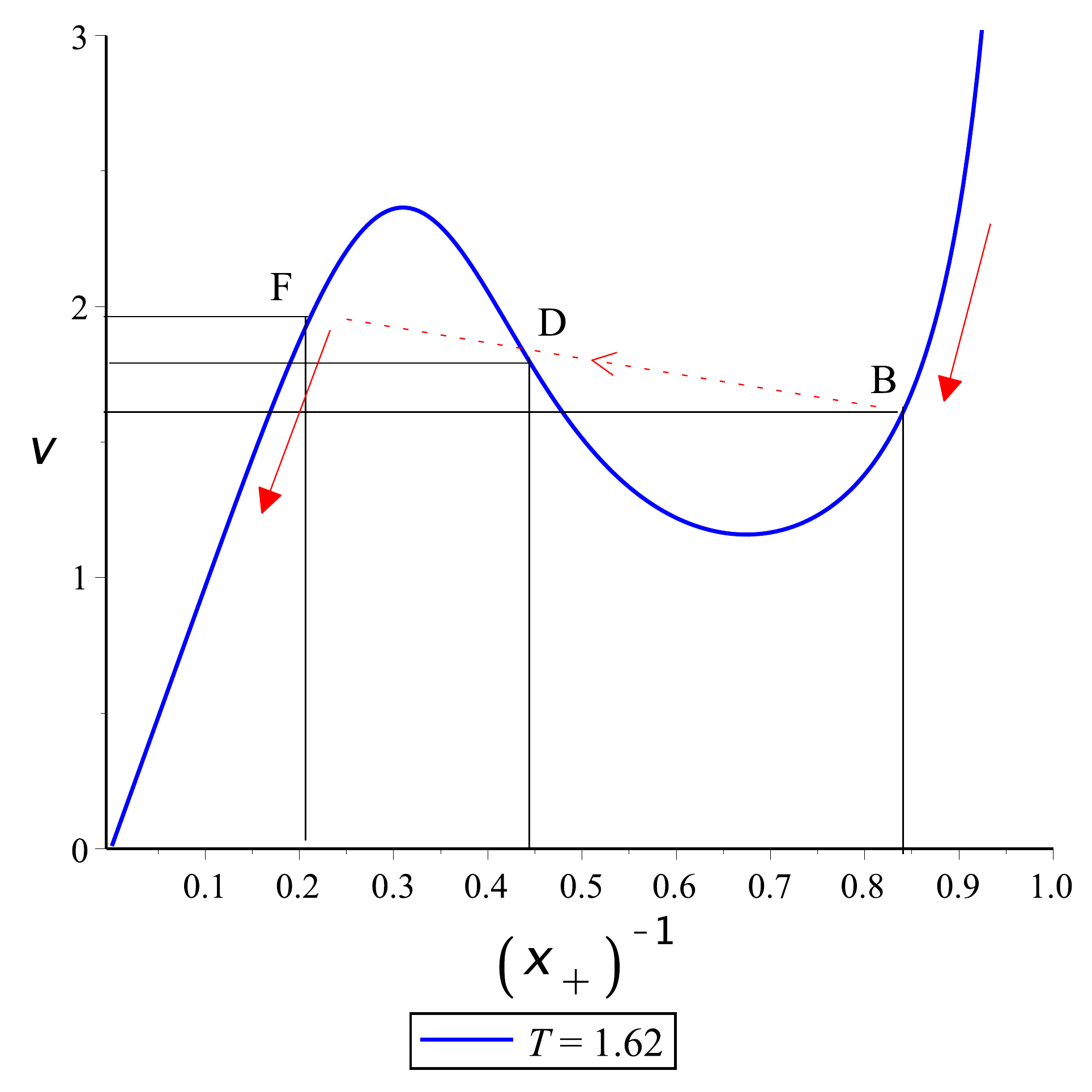}
\includegraphics[scale=0.28]{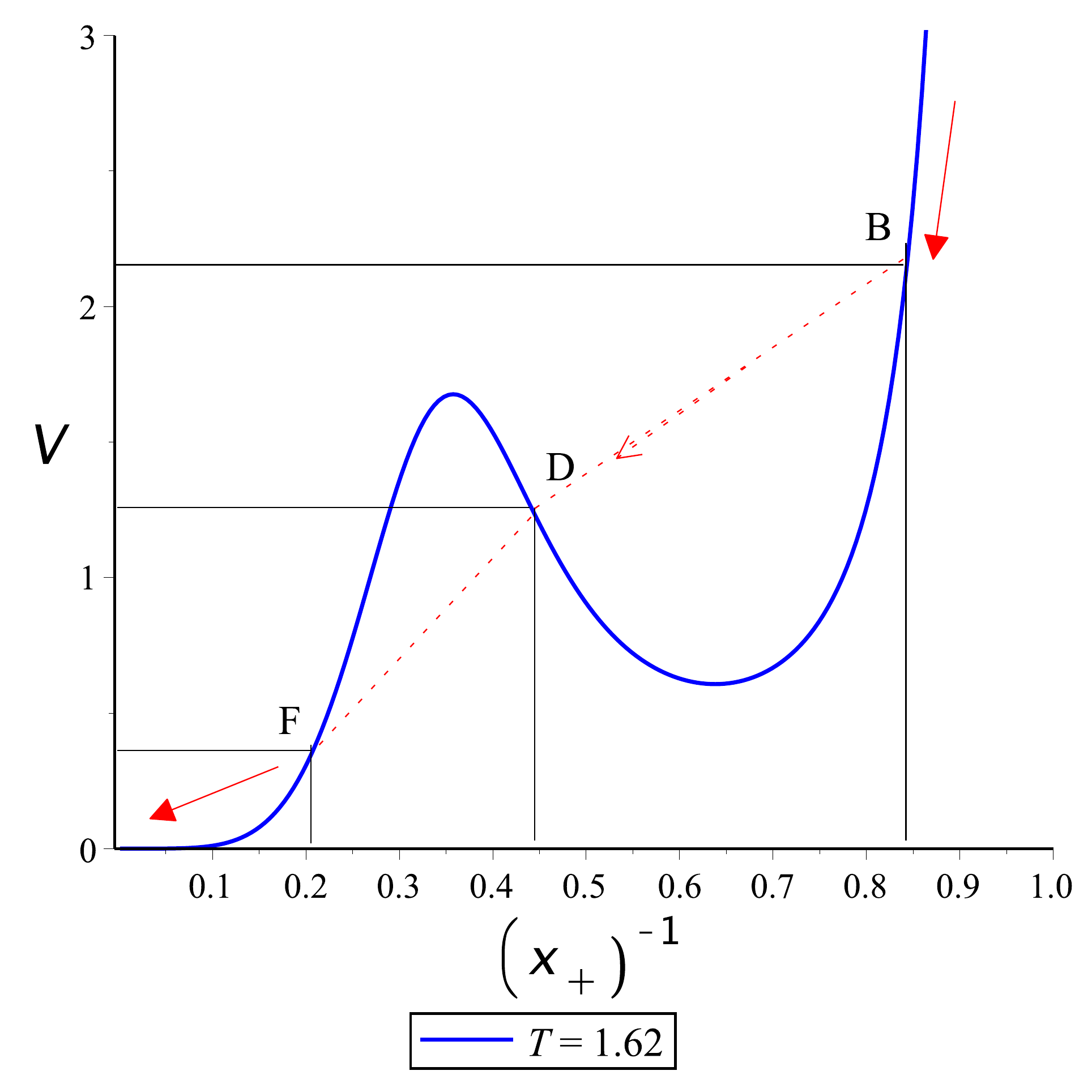}
\caption{$P-x_+^{-1}$, $v-x_+^{-1}$ and $V-x_+^{-1}$ for $T=1.62$ and $\Phi=0.85$, in the model $\sigma=3$. The limit $x_+^{-1}\rightarrow 1$ corresponds to the large black hole limit.}
\label{F16}
\end{figure}

This peculiar form of transition takes place because neither $v$ nor $V$ are monotonically increasing functions of the horizon size $1/x_+$  as shown in Fig.~\ref{F16}. An inspection of the central and right panels indicates the distinction between the two: the transition point B is smaller than the local maximum of $v(1/x_+)$ but larger than 
the local maximum of $V(1/x_+)$. Hence 
during the transition the thermodynamic volume $V$ and the
horizon radius of the black hole consistently decrease, as is clear from the left panel, whereas the specific volume increases.  During
condensation, the black hole gets smaller in size, but larger in specific volume, shown in the central panel.
Although it may seem counterintuitive that $V$ decreases while $v$ increases, this occurs because the entropy decreases considerably, and hence $v=3V/(2S)$ has a larger local maximum.

The case $\sigma\to \infty$ also exhibits similar behaviour, but  has a few special features that we discuss in the Appendix.

\section{Conclusions}\label{conc}
 
We have investigated the thermodynamics of a four-dimensional asymptotically $AdS$ family of exact hairy black hole solutions \cite{Anabalon:2013sra}, whose scalar field has a non-trivial self-interacting potential that can be embedded in supergravity \cite{Anabalon:2017yhv,Anabalon:2020pez}. The Euclidean action was regularized by using the counterterm method in the presence of a scalar field satisfying mixed boundary conditions \cite{Anabalon:2015xvl} and the conserved energy was obtained by using the quasilocal formalism of Brown and York \cite{Brown:1992br}. We have explicitly shown that the first law is satisfied and the Smarr relation holds as long as the two parameters of the theory ($\Lambda$ and $\alpha$) are included.

By considering the cosmological constant as a pressure term, we have analyzed the thermodynamics in the extended phase space. After revisiting the study of the special theory $\sigma\rightarrow\infty$ in Section \ref{sec2}, we studied the general case in Section \ref{sec3}. We have proved that the thermodynamic volume satisfies the Reverse Isoperimetric Inequality for the set of the hair parameter $\sigma>1$. One of the main new results is the existence of reentrant phase transition in both the canonical and grand canonical ensemble for suitable values of $Q$ and $\Phi$, respectively. As pointed out in \cite{Narayanan}, for a phase transition to be reentrant, it must involve the transformation of a system from one state into a macroscopically similar state via at least two phase transitions through the variation of a single thermodynamic parameter. In this case, this parameter is the temperature, while the pressure is held fixed. As $P$ is not required to be variable in order to have reentrant phase transition, these results do not restrict to the black hole chemistry approach.

For the case $\sigma\rightarrow\infty$, $P$ is double-valued in both, canonical and grand canonical, ensembles and no reentrant phase transition was observed. For finite $\sigma$, there is a region of the extended phase space where the pressure becomes triple-valued only in the grand canonical ensemble. This region, for which $\Phi>\Phi_c\equiv \sqrt{(\sigma-1)/(2\sigma)}$, is where the second critical point belongs to. We have analyzed in detail this multi-valuedness in pressure and the associated phase transition and the results are summarized in Fig. \ref{F15}.  We would like to point out that there also exist examples in ordinary chemistry systems where $P$ is a multi-valued function \cite{Fenton}. 

The results presented in this paper, i.e., the multiple critical points, the reentrant phase behaviours in both ensamble, and multi-valuedness of the pressure and its associated novel phase transition, are new features of charged hairy black holes in $d=4$ spacetime dimensions. They do not exist when either the scalar field or its self-interaction is turned off. Therefore, this study offers concrete evidence that a self-interacting scalar field can drastically change and enrich the thermodynamic behaviour of black holes.

In the context of string theory, it has recently been arguedw that variations of the cosmological constant can be interpreted as variations in the volume of the sphere of compactification \cite{Cvetic:2010jb,Astefanesei:2019ehu}. Therefore, in accordance with our results, namely, the existence of several critical isobars, the size of the sphere of compactification leads to a different thermodynamic behaviour of black holes.

\section*{Acknowledgments}
This work was supported in part by the Natural Sciences and Engineering Research Council of Canada. DA was supported during this work by the Fondecyt grant 1200986.
The work of RR was supported by the FONDECYT grant 3220663. The work of PC was supported by ANID Grant No. 21182145 and UTFSM's grant No. 072/2021 thanks to the DPP's support.  
	\bigskip
	
\appendix
\section{Novel Phase Behaviour for $\sigma=\infty$}

\begin{figure}[t!]
\centering
\includegraphics[scale=0.28]{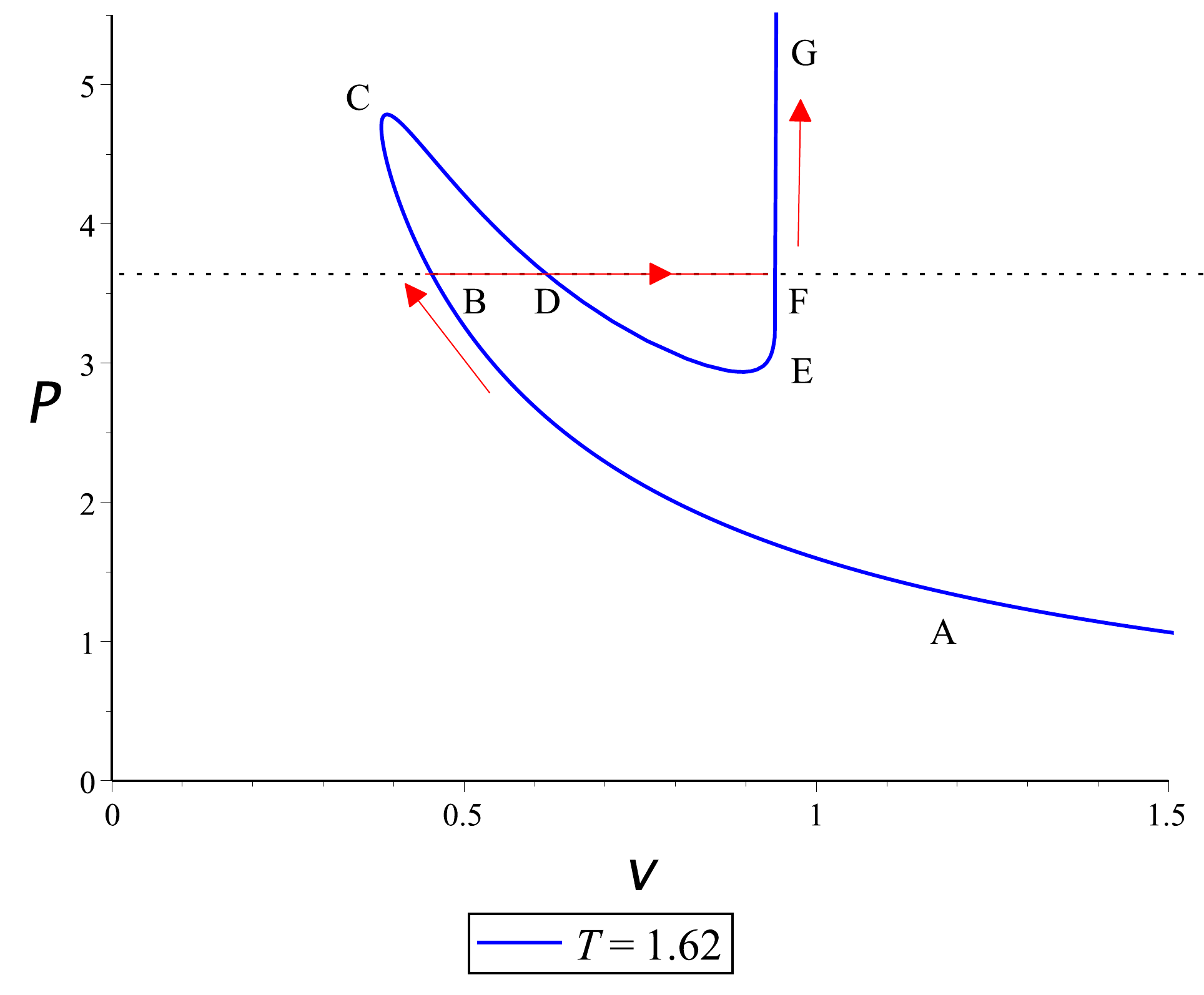}
\includegraphics[scale=0.28]{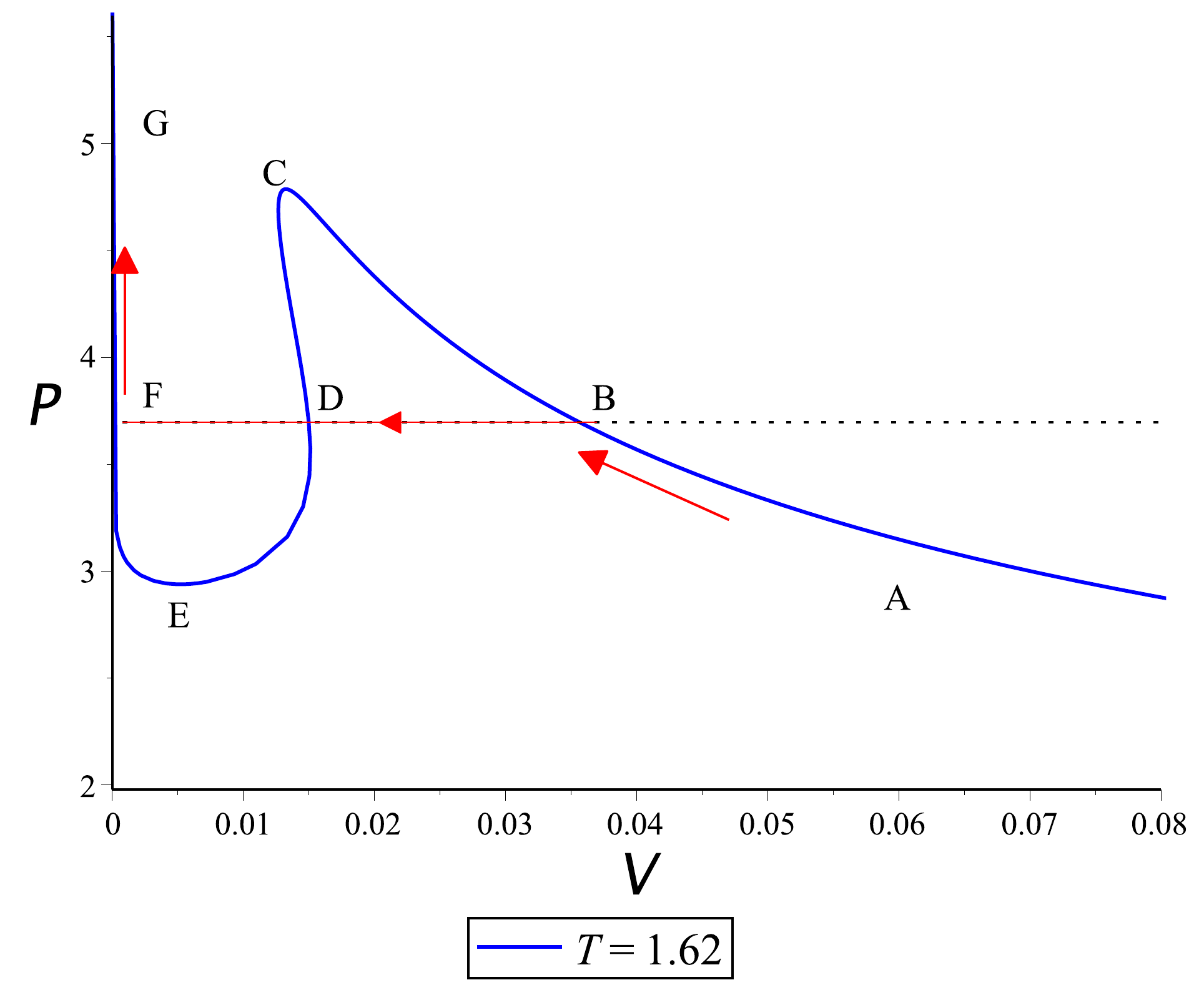}
\includegraphics[scale=0.28]{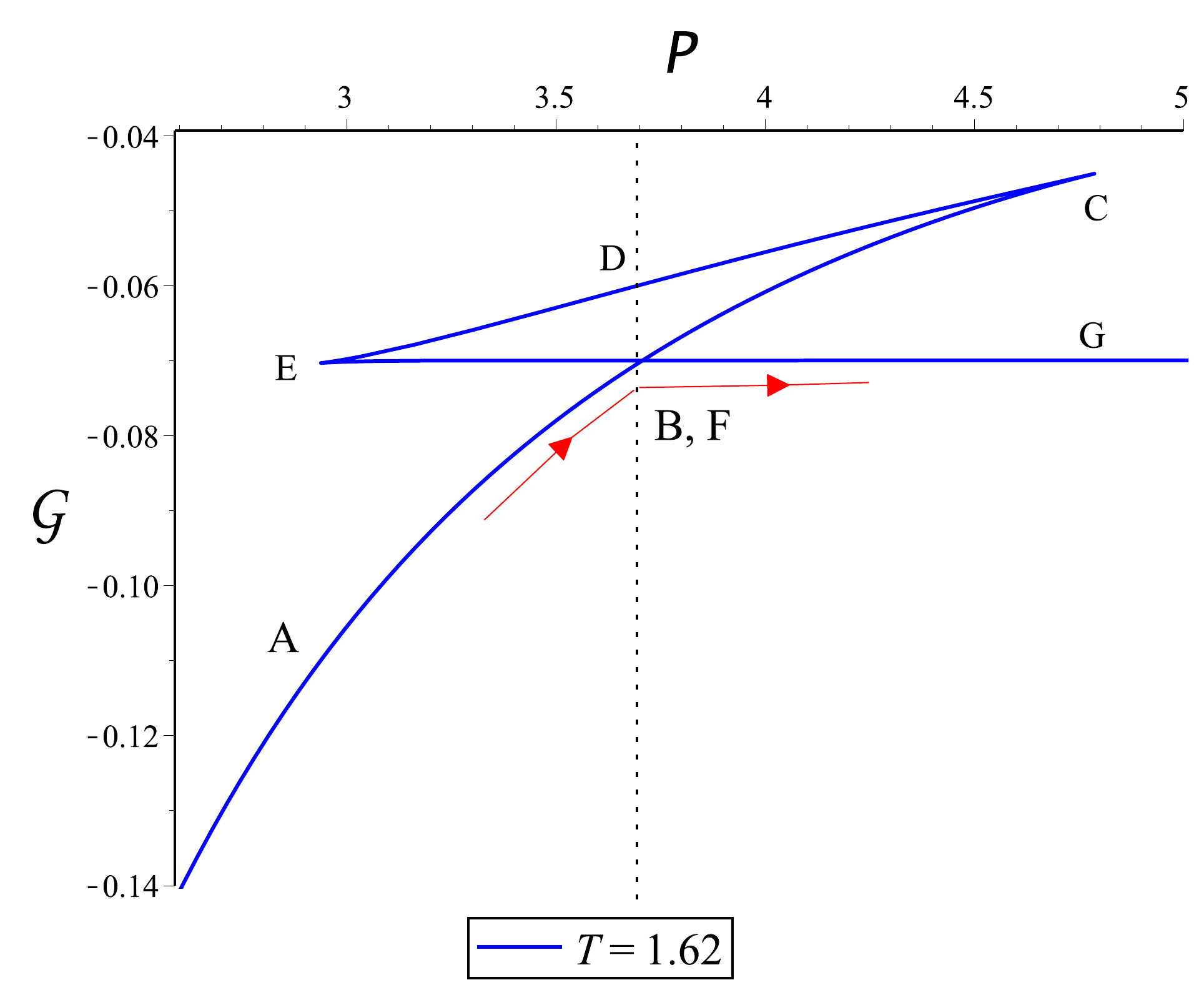}
\caption{Equation of state $P-v$, $P-V$ and the corresponding $\mathcal{G}-P$ diagram for $\Phi=0.85$, $T=1.62$ in the model $\sigma=\infty$. For this isotherm, the phase transition occurs at $P\approx 3.7$.}
\label{F17}
\end{figure}

For $\sigma=\infty$,   as long as $\Phi>1/\sqrt{2}$ (values of $\Phi$ for which the system exhibits critical phenomena), the specific volume 
\begin{equation}
\label{specificvol2}
v=\sqrt{2}\sqrt{2\Phi^2-1} -\frac{\sqrt{2}\(2\sqrt{2}\pi \sqrt{2\Phi^2-1}T-8\Phi^2+3\)}{\sqrt{2\Phi^2-1}x_+}+\mathcal{O}(x_+^{-2})
\end{equation}
goes to a constant value as $x_+\rightarrow\infty$.  The  equation of state consequently develops a branch characterized by an almost completely vertical line in $P-v$. One way to see this is that in the very tiny black hole limit, the entropy and the thermodynamic volume becomes proportional. Indeed, both decays in the same order of $x_+$,
\begin{equation}
S=\frac{2\pi(2\Phi^2-1)}{x_+}+\mathcal{O}(x_+^{,2})\,, \qquad V=\frac{4\sqrt{2}\pi(2\Phi^2-1)^{3/2}}{3x_+}+\mathcal{O}(x_+^{-2})
\end{equation}
Another way to see this proportionality is by writing $x_+=1/(\eta^2r_+^2)+2-\eta^2r_+^2+\mathcal{O}(r_+
^{4})$, obtained from $\Omega(x_+)=r_+^2$ in the limit $x_+\rightarrow\infty$, in the expression for $V$ given in (\ref{quantities2}). The entropy is simply $S=\pi r_+^2$ and the volume is
\begin{equation}
V=\frac{2\pi r_+^2}{3\eta}+\mathcal{O}(r_+^4)
\end{equation}
In any case, $V/S=2/(3\eta)$, where $\eta(x_+\rightarrow\infty)=1/(\sqrt{2}\sqrt{2\Phi^2-1})$ can be directly obtained by taking the limit $x_+\rightarrow\infty$ in the expression for $\eta$ obtained from the horizon equation $f(x_+)=0$. The equation of state near the second critical point and the $\mathcal{G}-P$ diagram are depicted in Fig.~\ref{F17} for $\Phi=0.85$. For this value of $\Phi$, $v(x_+\rightarrow\infty)\approx 0.9434$. Apart from the fact that the specific volume tends to a constant value  in the limit $x_+\rightarrow\infty$, the phase transition follows the same general features as for finite $\sigma$.

For completeness, we have plotted $P-(x_+)^{-1}$, $v-(x_+)^{-1}$ and $V-(x_+)^{-1}$.
\begin{figure}[t!]
	\centering
	\includegraphics[scale=0.25]{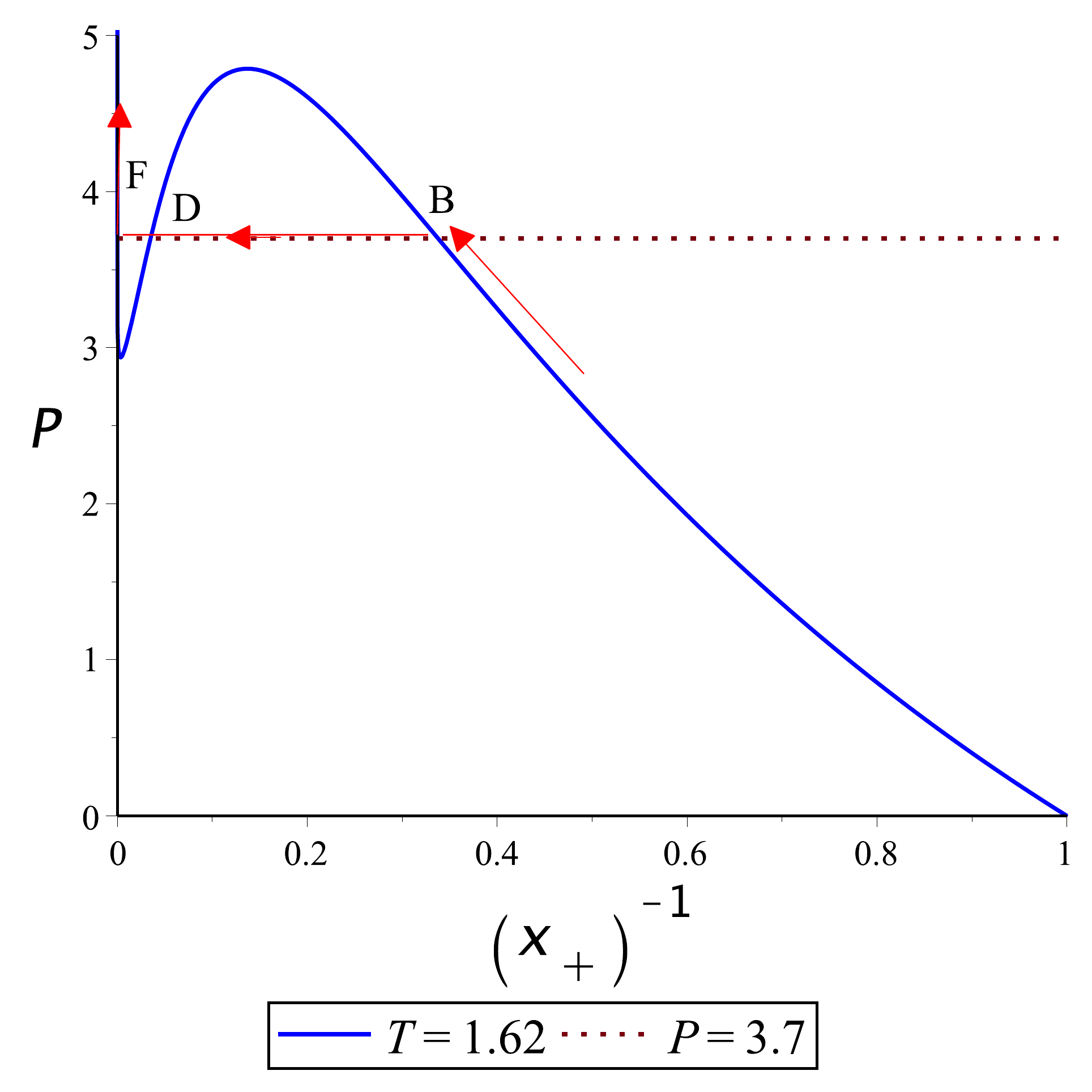}
	\includegraphics[scale=0.25]{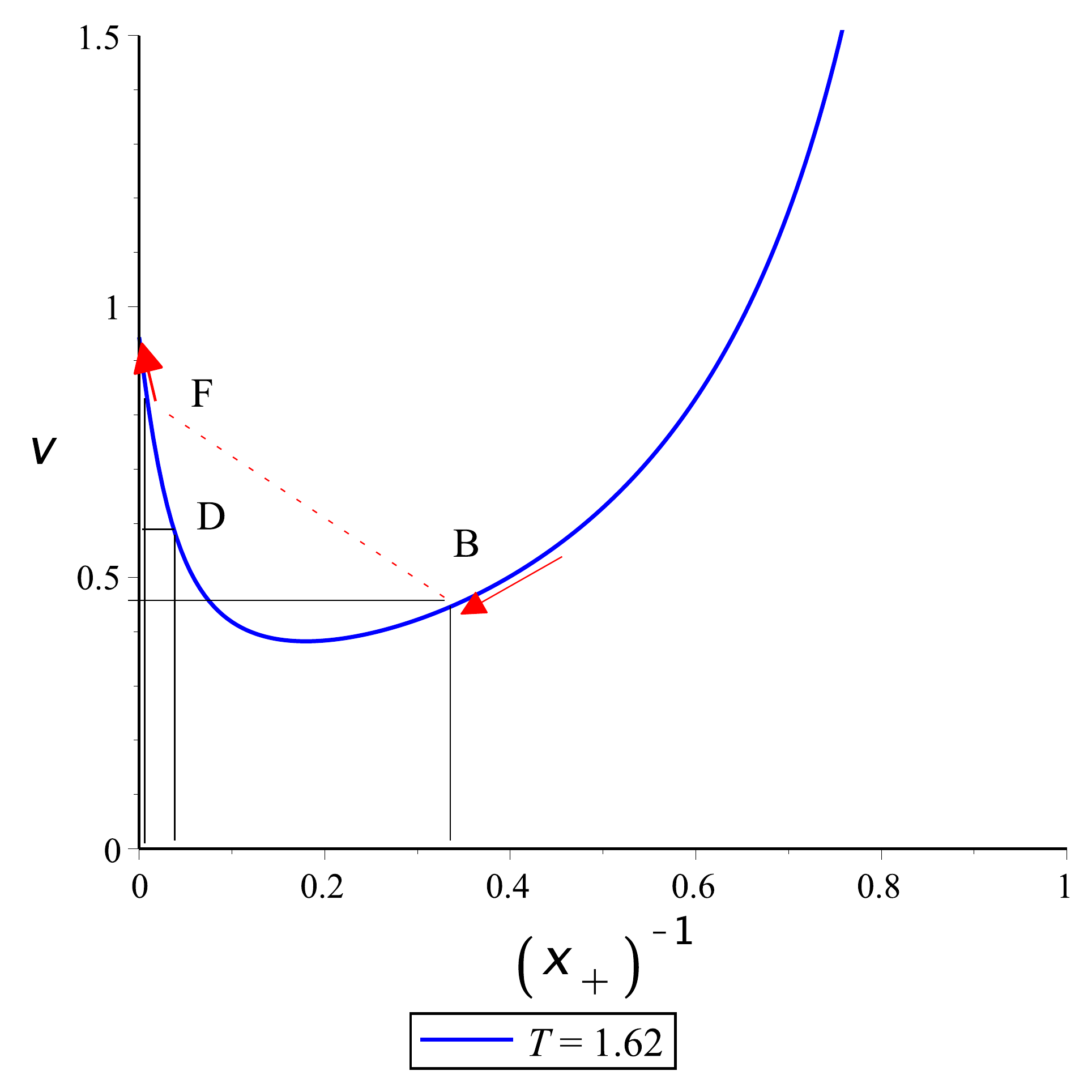}
	\includegraphics[scale=0.25]{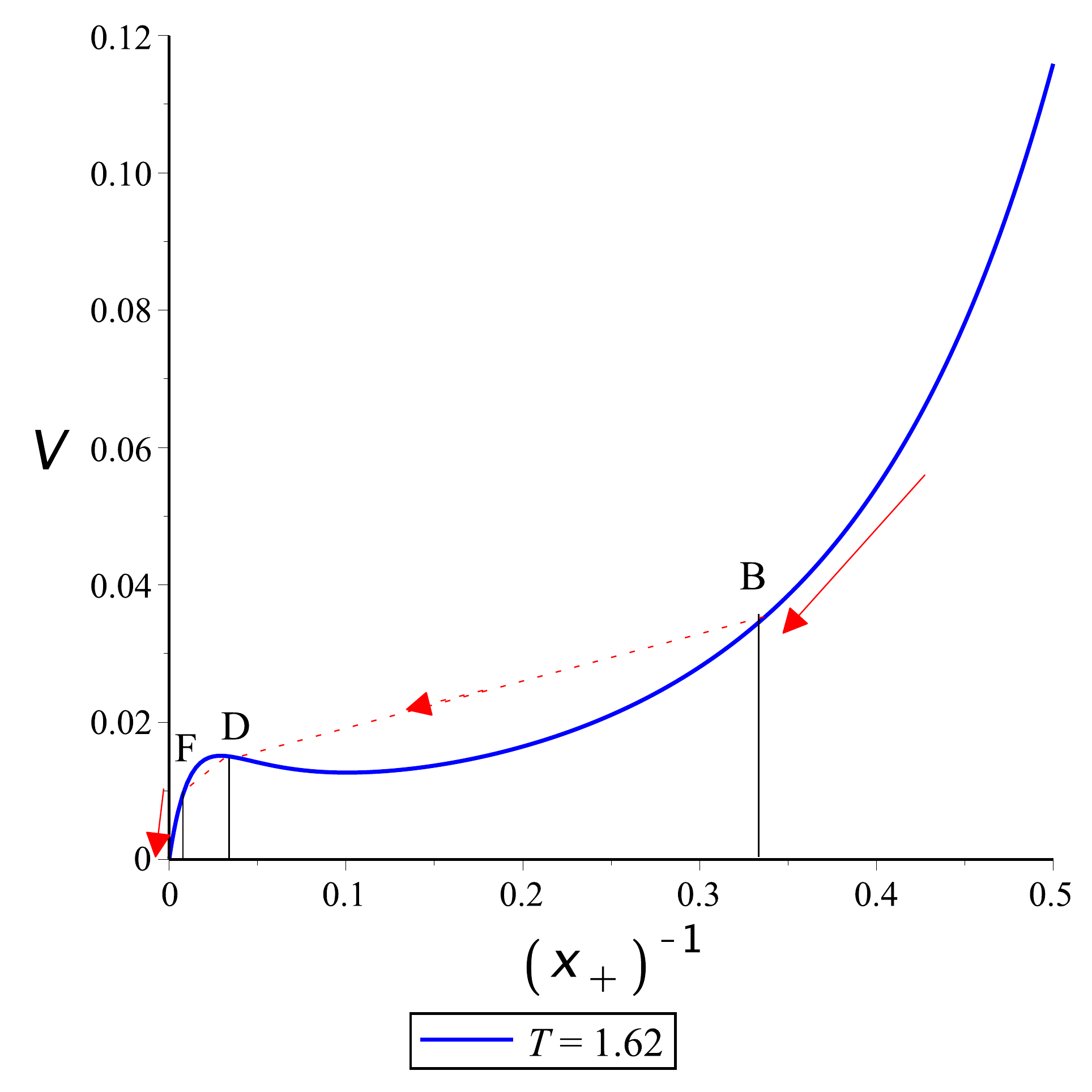}
	\caption{$P-x_+^{-1}$, $v-x_+^{-1}$ and $V-x_+^{-1}$ for $T=1.62$ and $\Phi=0.85$, in the model $\sigma=3$. The limit $x_+^{-1}\rightarrow 1$ corresponds to the large black hole limit.}
	\label{F18}
\end{figure}
A subtle difference in this case, compared with the finite $\sigma$ case, is observed in the second panel of Fig.~\ref{F18}. After the transition is taking place, from B to F, the specific volume $v$ still increases a little bit when moving from F to G. It follows from (\ref{specificvol2}) that, in the limit $x_+\rightarrow\infty$, $(\pa v/\pa x_+)_T$ is positive only provided
\begin{equation}
T>\frac{8\Phi^2-3}{2\sqrt{2}\pi\sqrt{2\Phi^2-1}}
\end{equation}
For $\Phi=0.85$, this inequality is $T>0.47$, which is fulfilled since $T_{c2}\approx 0.82>0.47$. Therefore, the first order phase transition associated with the second critical point $c2$ has the peculiarity that, for the small$-S$ phase, $v$ slightly increases as $S$ decreases, contrary to the case of finite $\sigma$.

\end{document}